\documentclass[a4paper,11pt]{article}
\usepackage{jcappub}
\usepackage{amsmath}
\usepackage{graphicx}
\usepackage{bayaspic}
\usepackage{eivars}

\usepackage{xspace}
\usepackage{lscape}

\makeatletter
\gdef\@fpheader{}
\makeatother

\newcommand{\ie}{i.e.\xspace}

\newlength{\wsingfig}
\newlength{\wdblefig}
\newlength{\wfull}
\newlength{\hfull}
\newcommand{\sss}[1]{{\scriptscriptstyle{#1}}}

\DeclareMathOperator{\expbeta}{\exp_{1-\beta}}

\newcommand{\uk}{\mathrm{k}}
\newcommand{\uuc}{\mathrm{uc}}
\newcommand{\uSZ}{\mathrm{SZ}}
\newcommand{\uML}{\mathrm{ML}}
\newcommand{\usssMC}{\sss{\mathrm{MC}}}
\newcommand{\usssPS}{\sss{\mathrm{PS}}}
\newcommand{\usssREF}{\sss{\mathrm{REF}}}
\newcommand{\usssCIB}{\sss{\mathrm{CIB}}}
\newcommand{\umod}{\sss{\mathrm{mod}}}
\newcommand{\utSZ}{\ut\sss{\uSZ}}
\newcommand{\ukSZ}{\uk\sss{\uSZ}}

\newcommand{\calP}{\mathcal{P}}

\newcommand{\calI}{\mathcal{I}}
\newcommand{\calN}{\mathcal{N}}
\newcommand{\calL}{\mathcal{L}}
\newcommand{\calV}{\mathcal{V}}
\newcommand{\calR}{\mathcal{R}}
\newcommand{\calE}{\mathcal{E}}
\newcommand{\calM}{\mathcal{M}}
\newcommand{\calMref}{\mathcal{M}_{\usssREF}}

\newcommand{\Refc}[1]{Ref.~\cite{#1}}
\newcommand{\Refcs}[1]{Refs.~\cite{#1}}
\newcommand{\Eq}[1]{Eq.~\eqref{#1}}
\newcommand{\Eqs}[1]{Eqs.~\eqref{#1}}
\newcommand{\Fig}[1]{Fig.~\ref{#1}}

\newcommand{\order}[1]{\mathcal{O}\!\left(#1\right)}

\newcommand{\logdec}{\log}

\DeclareMathOperator{\sech}{sech}

\newcommand{\dd}{\mathrm{d}} 
\newcommand{\ee}{e} 
\newcommand{\efold}{$\ee$-fold\xspace}
\newcommand{\efolds}{$\ee$-folds\xspace}

\newcommand{\MeV}{\mathrm{MeV}}
\newcommand{\GeV}{\mathrm{GeV}}
\newcommand{\GHz}{\mathrm{GHz}}
\newcommand{\TeV}{\mathrm{TeV}}

\newcommand{\Mpc}{\mathrm{Mpc}}

\newcommand{\OmegaCDM}{\Omega_\udm}
\newcommand{\OmegaB}{\Omega_\ub}

\newcommand{\thetaMC}{\theta_{\usssMC}}
\newcommand{\APSa}{A^{\usssPS}_{100}}
\newcommand{\APSb}{A^{\usssPS}_{143}}
\newcommand{\APSc}{A^{\usssPS}_{217}}
\newcommand{\rPSbc}{r^{\usssPS}_{143\times217}}
\newcommand{\ACIBb}{A^{\usssCIB}_{143}}
\newcommand{\ACIBc}{A^{\usssCIB}_{217}}
\newcommand{\rCIBbc}{r^{\usssCIB}_{143\times217}}
\newcommand{\gamCIB}{\gamma^{\usssCIB}}
\newcommand{\AtSZ}{A_{\utSZ}}
\newcommand{\AkSZ}{A_{\ukSZ}}
\newcommand{\xitSZCIB}{\xi^{\utSZ\times\usssCIB}}

\newcommand{\ca}{c_{100}}
\newcommand{\cc}{c_{217}}
\newcommand{\betaoo}{\beta_1^1}

\newcommand{\rlss}{r_{\mathrm{\ell ss}}}

\newcommand{\Clt}{C_{\ell}^{\mathrm{th}}}
\newcommand{\Clm}{C_{\ell}^{\mathrm{obs}}}

\newcommand{\almm}{a_{\ell m}^{\mathrm{obs}}}
\newcommand{\tetas}{\theta_{\mathrm{s}}}
\newcommand{\tetai}{\theta_{\uinf}}
\newcommand{\tetar}{\theta_{\ureh}}
\newcommand{\tetamin}{\theta_{\min}}
\newcommand{\tetamax}{\theta_{\max}}

\newcommand{\bartetamax}{\bar{\theta}_{\max}}

\newcommand{\ellmin}{\ell_\umin}
\newcommand{\ellmax}{\ell_\umax}

\newcommand{\mpl}{m_\usssPl}
\newcommand{\Mp}{M_\usssPl}
\newcommand{\vev}{\textit{vev}\xspace}

\newcommand{\Leff}{\ensuremath{\calL_\text{eff}}}

\newcommand{\urlaspic}{\url{http://cp3.irmp.ucl.ac.be/~ringeval/aspic.html}}
\newcommand{\CAMB}{\texttt{CAMB}\xspace}
\newcommand{\COSMOMC}{\texttt{COSMOMC}\xspace}
\newcommand{\ASPIC}{\texttt{ASPIC}\xspace}

\newcommand{\MULTINEST}{\texttt{MultiNest}\xspace}
\newcommand{\COBE}{COBE\xspace}
\newcommand{\CLIK}{\texttt{CLIK}}

\newcommand{\EI}{\textit{Encyclop{\ae}dia Inflationaris}\xspace}

\newcommand{\kstar}{k_*}
\newcommand{\etastar}{\eta_*}
\newcommand{\Pstar}{P_*}
\newcommand{\calPstar}{\calP_0}

\newcommand{\fnlloc}{f_\uNL^\uloc}
\newcommand{\fnleq}{f_\uNL^\ueq}
\newcommand{\fnlortho}{f_\uNL^\uortho}

\newcommand{\wrehbar}{\overline{w}_{\ureh}}

\newcommand{\epsstar}[1]{\epsilon_{#1}}

\newcommand{\AI}{A_{_{\mathrm{I}}}}

\newcommand{\Rrad}{R_\urad}

\newcommand{\vstar}{v_*}
\newcommand{\rhoreh}{\rho_\ureh}
\newcommand{\rhoend}{\rho_\uend}

\newcommand{\rhonuc}{\rho_\unuc}

\newcommand{\likeinf}{\Like_\text{eff}}

\newcommand{\BayesFactor}[2]{B^{#1}_{#2}}
\newcommand{\Bref}[1]{\BayesFactor{#1}{\usssREF}}
\newcommand{\Bhi}[1]{\BayesFactor{#1}{\hi}}
\newcommand{\Cb}{\Complexity}

\newcommand{\Nevid}{193}
\newcommand{\Nincon}{52}
\newcommand{\Nweak}{41}
\newcommand{\Nmoder}{34}
\newcommand{\Nstrong}{66}
\newcommand{\pourincon}{26}

\newcommand{\pourstrong}{34}

\newcommand{\Nmod}{N^{\umod}}
\newcommand{\Nparam}{\Params}
\newcommand{\Nuc}{N^{\uuc}}

\title{The Best Inflationary Models After Planck}

\author[a]{J\'er\^ome Martin,}
\author[b]{Christophe Ringeval,}
\author[c]{Roberto Trotta}
\author[a]{and Vincent Vennin}

\affiliation[a]{Institut d'Astrophysique de Paris, UMR
7095-CNRS, Universit\'e Pierre et Marie Curie, 98bis boulevard Arago,
75014 Paris (France)}

\affiliation[b]{Centre for Cosmology, Particle Physics and Phenomenology,
  Institute of Mathematics and Physics, Louvain University, 2 Chemin
  du Cyclotron, 1348 Louvain-la-Neuve (Belgium)}

\affiliation[c]{Imperial College London, Astrophysics \& Imperial
  Centre for Inference and Cosmology, Blackett Laboratory, Prince
  Consort Road, London SW7 2AZ (United Kingdom)}

\emailAdd{jmartin@iap.fr}
\emailAdd{christophe.ringeval@uclouvain.be}
\emailAdd{r.trotta@imperial.ac.uk}
\emailAdd{vennin@iap.fr}

\date{today}

\begin{document}

\abstract{We compute the Bayesian evidence and complexity of $\Nevid$
  slow-roll single-field models of inflation using the Planck 2013
  Cosmic Microwave Background data, with the aim of establishing which
  models are favoured from a Bayesian perspective. Our calculations
  employ a new numerical pipeline interfacing an inflationary
  effective likelihood with the slow-roll library $\ASPIC$ and the
  nested sampling algorithm $\MULTINEST$. The models considered
  represent a complete and systematic scan of the entire landscape of
  inflationary scenarios proposed so far. Our analysis singles out the
  most probable models (from an Occam's razor point of view) that are
  compatible with Planck data, while ruling out with very strong
  evidence $\pourstrong \%$ of the models considered. We identify
  $\pourincon\%$ of the models that are favoured by the Bayesian
  evidence, corresponding to $15$ different potential shapes. If the
  Bayesian complexity is included in the analysis, only $9\%$ of the
  models are preferred, corresponding to only $9$ different potential
  shapes. These shapes are all of the plateau type.}

\keywords{Cosmic Inflation, Slow-roll, Reheating, Cosmic Microwave
    Background, Aspic, Bayesian model comparison}

\maketitle

\section{Introduction}
\label{sec:intro}

The recent release of the Planck satellite data has had important and
profound consequences for our understanding of primordial
cosmology. These data clearly support the idea that inflation is the
correct description of the physical conditions that prevailed in the
early universe since they are in agreement with several important and
generic predictions made by the inflationary theory. For instance, a
basic property of inflation is that spatial curvature should
vanish. And one indeed finds that $100 \Omega
_K=-0.05{}^{+0.65}_{-0.66}$ by combining Planck with Wilkinson
Microwave Anisotropy Probe (WMAP) large-scale polarisation (denoted WP
in Ref.~\cite{Ade:2013zuv}) and Baryon Acoustic Oscillations (BAO)
measurements. Another important consequence of the Planck data is the
detection of a spectral tilt, $\ns=0.9603\pm 0.0073$ thus ruling out
scale invariance at more than $5\sigma$, a level of significance
predicted in \Refc{Trotta:2007hy}, and convincingly confirming a
crucial inflationary prediction. Moreover, the Planck data seem to
point to the simplest (but non-trivial) version of inflation. Indeed,
neither a significant running nor a significant running of the running
have been detected since it is found that $\dd \ns /\dd \ln
k=-0.0134\pm 0.009$ (Planck+WP) and $\dd ^2 \ns/\dd \ln^2 k=0.02\pm
0.016 $ (Planck+WP), with a pivot scale chosen at $\kstar=0.05
\Mpc^{-1}$. The data are also compatible with adiabaticity at $95\%$
CL. If one defines $\alpha_{ab}^{(\ellmin,\ellmax)}\equiv \left(\Delta
  T\right)_{ab}^2(\ellmin,\ellmax) /\left(\Delta
  T\right)_{\mathrm{tot}}^2(\ellmin,\ellmax)$, with $a,b=\calR,\calI$,
where $\calI$ stands for Cold Dark Isocurvature (CDI), Neutrino
Density Isocurvature (NDI) or Neutrino Velocity Isocurvature (NVI) and
$\left(\Delta T\right)_{X}^2(\ellmin,\ellmax)=
\sum_{\ell=\ellmin}^{\ell=\ellmax}\left(2\ell+1\right)C_{\ell,X}^{TT}$,
then one obtains $\alpha_{\calR \calR}^{(2,2500)}\in[0,98,1.07]$ and
$\alpha_{\calR \calI}^{(2,2500)}\in[-0.093,0.014]$ for
$\calI=\mbox{CDI}$, $\alpha_{\calR \calR}^{(2,2500)}\in[0,99,1.09]$
and $\alpha_{\calR \calI}^{(2,2500)}\in[-0.18,0.0]$ for
$\calI=\mbox{NDI}$, $\alpha_{\calR \calR}^{(2,2500)}\in[0,96,1.05]$
and $\alpha_{\calR \calI}^{(2,2500)}\in[-0.09,0.026]$ for
$\calI=\mbox{NVI}$. This implies that isocurvature modes are
compatible with zero although the analysis is done with one
isocurvature mode at a time only. A quite large non-adiabatic
contribution remains possible but, as discussed in
Ref.~\cite{Ade:2013uln}, this is in fact driven by the data in the
range $\ell \le 40$. The Planck data also imply that primordial
non-Gaussianity is compatible with zero, namely $\fnlloc=2.7\pm 5.8$,
$\fnleq=-42\pm 75$ and $\fnlortho=-25\pm 39$ \cite{Ade:2013ydc}. Some
anomalies or ``glitches'' have also been reported but the
corresponding statistical significance is unclear and, in any case,
not yet sufficient to claim a detection.

Therefore, the overall picture that emerges is that the inflationary
mechanism is non-trivial but, at the same time, ``non-exotic''. In
particular, the complicated scenarios that were considered, at some
point, as attractive are now disfavoured (but not necessarily ruled
out). Therefore, in accordance with an Occam's razor principle, that
the simplest viable explanation for the observations at hand ought to
be preferred, it is appropriate to consider -- at least for the moment
-- the simplest scenarios, namely single field slow-roll inflation
with a standard kinetic term. This type of scenarios is characterised
by one free function, the potential $V(\phi)$. Therefore, identifying
the ``best model of inflation'' boils down to determining the
potential $V(\phi)$ which fits the data the best with the smallest
number of free parameters and the least fine-tuning.

In order to achieve this task, it is first necessary to identify all
the scenarios belonging to the above-mentioned class. This
is not so easy since, even if restricted to a small part of the
inflationary landscape, the ``single-field region'' remains densely
populated. This was accomplished recently in the ``{\EI}'' of
\Refc{Martin:2013tda}.  Once all the single-field models have been
identified, one needs to quantify statistically whether a model is
``better'' than another.  This question can be addressed in the
framework of Bayesian model comparison, which requires the computation
of the Bayesian evidence, or global likelihood, \ie the integral of
the likelihood over the prior space for each model. The ratio of such
evidences then gives the Bayes factor, representing the degree by
which the Planck data have modified our a priori relative belief in
each pair of models. From the Bayes factors, one can then evaluate the
posterior probability for each model, and thus identify the ``best''
(in a Bayesian sense) model of inflation.  The calculation of the
Bayesian evidence of each of the {\EI} scenarios constitutes the main
subject of the present paper.

This article is organised as follows. In the next section,
section~\ref{sec:evidence}, we briefly present the theory of Bayesian
inference and how it can be used to perform model comparison. In
sub-section~\ref{subsec:bfactor}, we recall the definition of the
Bayesian evidence and, in sub-section~\ref{subsec:priorsens}, we
discuss how this quantity depends on the prior choices. In
sub-section~\ref{subsec:complexity}, we also introduce the Bayesian
complexity and explains its meaning. In section~\ref{sec:fastcalcul},
we discuss how the Bayesian evidences and complexities can be
calculated efficiently and rapidly from the $\ASPIC$\footnote{\urlaspic} library. In
sub-section~\ref{subsec:srshortcut}, we present the idea behind the
method introduced in \Refc{Ringeval:2013lea} (and used in the present
article) and, in sub-section~\ref{subsec:effectiveLike}, we detail how
the effective likelihood, which is the crucial tool of the method of
\Refc{Ringeval:2013lea}, can be determined from the Planck 2013 data. In
sub-section~\ref{subsec:computingEvid}, we describe the numerical
methods used in order to calculate the evidences from the effective
likelihood. We also specify the priors chosen on the non-primordial
parameters. In sub-section~\ref{subsec:finetuning}, we briefly discuss
the accuracy of our calculations and its limitations. Then, in
section~\ref{sec:res}, we present our results, namely the numerical
values of the evidence and complexity for all the models considered
and we discuss the physical implications of our calculations. In
section~\ref{sec:conclusion}, we summarise our findings and present
our conclusions. Finally, in appendix~\ref{sec:priors}, we review in
detail how the priors, for each model, have been chosen. Special
attention has been paid to their physical origin and we discuss how
the Bayesian evidence would be modified if the priors were changed.

\section{Bayesian inference and model comparison}
\label{sec:evidence}

In this section, we briefly review Bayesian inference theory and
Bayesian model comparison, which we adopt to compare the performance
of the {\EI} scenarios.

\subsection{Bayes factor and posterior model probability}
\label{subsec:bfactor}

Let $\calM_i$ be a collection of $\Nmod$ models ($i=1,\cdots, \Nmod$)
describing a given physical situation. In this paper, we will denote
by ``model'' a choice of inflationary potential, together with the
specification of a prior distribution for its parameters. A given
shape of the potential can support different prior choices, and we
call the selection of a potential shape (without specification of a
prior for its parameters) a ``scenario''. Thus within a given
inflationary scenario there can be multiple models. The following
considerations are however fully general. A model $\calM_i$ is
specified by a set of $\Nparam_i$ parameters $\theta_{ij}$ (with $j=1,
\cdots, \Nparam_i$) and by the prior probability distribution of each
of its parameters, namely $\pi(\theta_{ij}\vert \calM_i)$. In the
context of inference on the model's parameter (where the model is
assumed to be correct), the prior can be set from the posterior of a
previous observation. However, if one is interested in assessing a
model's performance via Bayesian model comparison, it is preferable to
understand the priors in terms of the a priori available parameter
space under the theory represented by model $\calM_i$ (see e.g.~
\Refcs{Cox:1946, Jeffreys:1961, deFinetti:1974, Bernardo:1994,
  Box:1992, JaynesBook, Berger:2003, Trotta:2005ar, Trotta:2008qt} for
further details).

Bayesian inference uses Bayes' theorem to update our degree of belief
in hypotheses when some new data $D$ becomes available (here, we think
of $D$ as the Cosmic Microwave Background - CMB - Planck data but the
formalism is generic). Assuming that model $\calM_i$ is true, from
Bayes' theorem, the posterior probability of its parameters
$\theta_{ij}$'s can be expressed as
\begin{equation}
  p\left(\theta_{ij}\vert D,\calM_i\right)=\frac{1}{\calE(D\vert \calM_i)}
  \calL \left( \theta_{ij} \right)
\pi\left(\theta_{ij}\vert \calM_i\right),
\end{equation}
where $\calL(\theta_{ij}) = p\left(D\vert \theta_{ij},\calM_i\right)$
is the likelihood function for the parameters of model $\calM_i$. The
quantity $\calE(D\vert\calM_i)$ is just a normalisation factor, called
the Bayesian evidence or model likelihood, and it is given by
\begin{equation}
\label{eq:generaldefevidence}
\calE\left(D\vert\calM_i\right)=\int \dd \theta_{ij}
\calL(\theta_{ij})
\pi\left(\theta_{ij}\vert \calM_i\right).
\end{equation}
If we are only interested in constraining the parameters $\theta_{ij}$
of the model, then the Bayesian evidence can be neglected. However, in
the following we shall focus on the question of assessing the
posterior model's probability, for which the Bayesian evidence plays a
central role.

Using again Bayes' theorem, one obtains the posterior probability
of the model $\calM_i$, which is given by
\begin{eqnarray} 
\label{eq:postM}
 p(\calM_i|D)&=&\frac{\calE(D|\calM_i)\pi(\calM_i)}{p(D)}\, ,
\end{eqnarray}
where $\pi(\calM_i)$ is the prior belief in model $\calM_i$. The
quantity $p(D)$ is a normalisation factor (which only depends on the
data but not on the model under consideration), given by
\begin{equation}
p(D) = \sum_i \calE(D|\calM_i)\pi(\calM_i).
\end{equation}
When comparing two models against each other, this factor cancels.  If
one defines a ``reference model'', $\calMref$, against which all other
models are compared, the posterior odds between a model $\calM_i$ and
the reference model are given by
\begin{equation}
  \frac{p(\calM_i| D)}{p(\calMref | D)} =
  \Bref{i}\frac{\pi(\calM_i)}{\pi(\calMref)}\,.
\end{equation}
Here, we have introduced the Bayes factor $\Bref{i}$ which can be
expressed as the ratio of the evidences, namely
\begin{eqnarray}
\Bref{i}  & \equiv & \frac{\calE(D|\calM_i)}{\calE(D|\calMref)}\, .
\end{eqnarray}
Under the principle of indifference, we can assume non-committal model
priors, i.e. we give all models the same a priori probability,
$\pi(\calM_i) = 1/\Nmod$, in which case the Bayes factor
becomes identical with the posterior odds.  With this assumption, a
Bayes factor larger (smaller) than one means a preference for the
model $\calM_i$ over the reference model (a preference for the
reference model over $\calM_i$).  The ``Jeffreys' scale'', see
Table~\ref{tab:Jeff}, gives an empirical prescription for translating
the values of $\Bref{i}$ into strengths of belief.
\begin{table}[t]
\begin{center}
\begin{tabular}{l l  l} \hline 
  $|\ln \Bref{i}|$ & Odds  & Strength of evidence \\\hline 
 $<1.0$ & $\lesssim 3:1$ &  Inconclusive \\
 $1.0$ & $\sim 3:1$ &  Weak evidence \\
 $2.5$ & $\sim 12:1$ & Moderate evidence \\
 $5.0$ & $\sim 150:1$ &  Strong evidence \\
\hline
\end{tabular}
\caption{Jeffreys' scale for evaluating the strength of evidence when
  comparing two models, $\calM_i$ versus a reference model $\calMref$,
  here slightly modified following the prescriptions given in
  Refs.~\cite{Gordon:2007xm,Trotta:2008qt}.}
\label{tab:Jeff} 
\end{center}
\end{table}

With non-committal model priors, the posterior probability for model
$\calM_i$ is then given by
\begin{equation}
p(\calM_i|D) = \frac{\Bref{i}}{\sum_j \Bref{j}}\,.
\end{equation}
This implicitly further assumes that the list of $\Nmod$ is
reasonably complete -- i.e. that there isn't a yet undiscovered better
models that have not been considered a priori (see~\Refc{March:2010ex}
for a Bayesian method leading to the discovery of such unknown
models).

The fundamental idea underpinning Bayesian model comparison is that
``economic'' models that fit well the data while exhibiting strong
predictivity are rewarded, while models with a large number of free
parameters that turn out not to be required by the data are penalised
for the wasted parameter space. Therefore, in a Bayesian sense, the
``best'' model is the one that achieves the best compromise between
quality of fit and simplicity (see~\Refc{Trotta:2008qt,Berger:1987}
for further details and \Refc{Berger:2004,Johnson:2013} for a
discussion of issues in Bayesian-frequentist calibrations). One of the
attractive features of Bayesian model comparison is that it
automatically embodies a quantitative version of Occam's razor, that
is to say, the principle of simplicity (see \Refc{Cousins2013} for a
critical discussion and comparison with frequentist methods). The
price to pay is that the Occam's razor effect depends in an
irreducible way on the choice of prior (and particularly on its range)
hence the latter must be set according to physical considerations
stemming from the model. We now turn to the crucial question of prior
sensitivity.

\subsection{Prior sensitivity considerations}
\label{subsec:priorsens}

As mentioned above, since the priors $\pi(\theta_{ij}\vert \calM_i)$
play a crucial role, a detailed description on how they have been
chosen is provided for each model in appendix~\ref{sec:priors}. We also
discuss how the evidence is affected by alternative prior choices
within various theoretical scenarios. For this reason, the number of
evidences presented in this paper is much larger than the number {\EI}
scenarios. Indeed, a given field potential can support several prior
choices motivated by different theories, each of them leading to
different evidences. We thus consider them as different models.

For each field potential, physical considerations have been used to
determine the shape of the prior. If a parameter is small but its
order of magnitude is unknown, as it is typically the case for a
coupling constant used in a perturbative expansion, then a Jeffreys'
prior (uniform in the logarithm of the parameter) is the most
uninformative. If, on the contrary, we deal with a parameter whose
order of magnitude is known, then this is a scale parameter and a
uniform prior on the parameter itself is appropriate. As priors must
be proper (i.e., normalised), the support of the prior
$[\tetamin,\tetamax]$ must also be chosen according to the natural
values allowed by the underlying physical scenario. Indeed, the
strength of the Occam's razor effect depends on this range, as
generically the Bayesian evidence scales as (for uniform priors)
\begin{equation}
\calE(D|\calM_i) \propto \frac{1}{\tetamax - \tetamin}\,,
\end{equation}
for cases where the support of the likelihood is much smaller than the
support of the prior. However, since the Jeffreys' scale is
logarithmic in the Bayes factor, the dependence on the prior range is
relatively mild.  Still, there are many cases in which $\tetamin$ and
$\tetamax$ remain unspecified by the model. When this happens,
attention has been paid on how the evidence is affected when this
range is modified. 

From the above argument it follows that one can
estimate the variation in the evidence that one would get from a
change of the range of the prior simply by rescaling it proportionally
to the ratio of the prior volumes in the parameter space. This holds
approximately true as long as the support of the likelihood is well
within that of the prior. This is more detailed in
appendix~\ref{sec:priors} where, if necessary, we discuss for each
model how this calculation can be done in practice.

Another often-encountered situation is when the likelihood is flat
along the $\theta_{ik}$ direction, \ie the data are insensitive to one
of the parameters of the model under consideration. In this case, the
posterior for that parameter is identical to the prior and the Bayes
factor reduces to unity -- the Bayesian evidence is insensitive to the
number of \emph{unconstrained} parameters in a model. For such flat
directions in parameter space, the prior boundary does not matter (as
long as the likelihood stays flat), and the evidence is unchanged by a
rescaling of the boundaries of the prior. A second quantity is thus
required to measure the number of effective parameters that the data
can constrain in a given model.  This can be implemented in various
way, as for instance by using Kullback-Leiber divergence between the
prior and the posterior, leading to the notion of model complexity
that we now discuss~\cite{Kullback:1951, Kunz:2006mc, Trotta:2008qt}.

\subsection{Bayesian complexity}
\label{subsec:complexity}

The number of parameters in a model is a poor description of its
``complexity'', as parameters that are not constrained by the data
should not be counted. A better evaluation of complexity (in a
Bayesian sense) has been introduced by \cite{Spiegelhalter}, who
advocates using the relative entropy between the prior and the
posterior distribution (\ie , the Kullback-Leibler divergence) as a
better suited measure of the number of free parameters in a model that
the data can actually constrain.

As shown in~\Refc{Kunz:2006mc}, such an effective number of
parameters, or Bayesian complexity, $\Cb$, can be written as
\begin{equation}
\label{def:Cb}
\Cb_i = \langle -2 \log \calL\left(\theta_{ij}\right) \rangle 
+ 2 \log \calL\left({\theta}_{ij}^\uML\right),
\end{equation}
where $\langle \cdot \rangle$ denotes averaging over the posterior
$p(\theta_{ij} | D, \calM_i)$ and ${\theta}_{ij}^\uML$ is the
maximum-likelihood estimate of the model's parameters which can be
approximately obtained from the posterior samples used to map out the
posterior distribution\footnote{See however \Refc{Feroz:2011bj} for
  the caveats that apply when one wants to derive maximum likelihood
  estimates from Bayesian posterior maps.}. The Bayesian complexity is
thus not an absolute measure of the number of constrained parameters --
rather it assesses the constraining power of the data with respect to
the measure provided by the prior.

The use of model complexity together with the Bayesian evidence allows
us to distinguish between cases where $\calE(D|\calM_i) \simeq
\calE(D|\calM_j)$ (\ie, two models exhibiting approximately the same
Bayesian evidence) but $\Cb_i \simeq \Cb_j$, in which case the data is
insufficient to distinguish between the two models (as their effective
complexities are the same); or the case where $\Cb_i > \Cb_j$, which
means that the data are sufficient to measure extra parameters of
model $i$ but that those parameters are not required by the evidence,
in which case we ought to prefer model $j$, as the one with the
smallest (measured) complexity.

\section{Fast Bayesian evidence calculation}
\label{sec:fastcalcul}

The computation of the Bayesian evidence can be a numerically
demanding task, as it requires the evaluation of the multi-dimensional
integral of Eq.~\eqref{eq:generaldefevidence}. This is particularly
computationally intensive for Markov Chains Monte-Carlo (MCMC)-based
methods. In recent years, a powerful tool has emerged in the shape of
nested sampling, and its implementation in the \MULTINEST
code~\cite{Feroz:2007kg, Feroz:2008xx}. Even with such a highly
efficient algorithm, the Bayesian evidence requires hundreds of
thousands of likelihood evaluations for each model. A typical analysis
based on the Planck likelihood coupled with an exact inflationary code
to integrate the perturbations requires roughly $3\times10^5$ CPU
hours (or 3.4 CPU years) of computing time on modern
$\texttt{x86\_64}$ processors. Performing this for each model
considered here would become prohibitively time consuming, even with
high-performance computing.

In this section, we briefly describe the method introduced in
Ref.~\cite{Ringeval:2013lea} which allows us to calculate the Bayesian
evidences in a fraction of the time that would be required using
conventional tools. We also mention the limitation of the method,
especially the fact that the very low evidences may be poorly
approximated.

\subsection{Effective likelihood via slow-roll reparameterisation}
\label{subsec:srshortcut}

Let us denote by $\almm$ the CMB temperature map recently observed by
the Planck satellite. From this map, one can estimate the measured
multipole moments $\Clm=\langle \almm \almm {}^\star\rangle$. From the
$\Lambda$CDM model (or any other post-inflationary history) and the
scenario of inflation, one can compute the theoretical prediction for
those multipole moments, $\Clt(\tetas,\tetar,\tetai)$ as a function of
the parameters in the model. Here, $\tetas$ represents a set of
parameters describing post-inflationary physics, see
Eq.~\eqref{eq:camparams} for a precise definition, $\tetar$ are the
parameters of reheating and $\tetai$ describe the shape of the
potential $V(\phi)$.  The reheating epoch can be described either with
$\tetar= (\rhoreh,\wrehbar)$, namely the energy density of the
universe at the end of reheating and the mean equation of state
parameter during reheating; or with the completely generic rescaled
reheating parameter $\tetar=\ln(R)$, defined by
\begin{equation}
R \equiv \Rrad \dfrac{\rhoend^{1/4}}{\Mp}, \qquad \Rrad \equiv
\dfrac{\aend}{\areh} \left(\dfrac{\rhoend}{\rhoreh} \right)^{1/4}.
\end{equation}
Here the indices ``$\uend$'' and ``$\ureh$'' denote the end of
inflation and end of the reheating era (\ie the beginning of the
radiation dominated era, see Ref.~\cite{Martin:2013tda} for further
details), $\rho$ and $a$ being the energy density of the universe and
the FLRW scale factor, respectively. Here, we have chosen to sample
over the same optimised set discussed in Refs.~\cite{Martin:2006rs,
  Martin:2010kz, Martin:2010hh, Ringeval:2013lea}, see also
\Refcs{Easther:2011yq, Norena:2012rs}. All possible reheating
histories are sampled using the rescaled reheating parameter and with
a prior uniform in its logarithm,
\begin{equation}
\pi(\tetar) = \pi[\ln(R)] = U(-46,15).
\end{equation}
The boundaries of the prior support encompass all
reheating histories satisfying the constraints that the mean equation of
state during reheating verifies $-1/3<\wrehbar<1$, and
$\rhonuc<\rhoreh<\rhoend$. The last inequality enforces that reheating
takes place after inflation and before Big-Bang Nucleosynthesis
(BBN). Practically, we have chosen $\rhonuc^{1/4} \equiv
10\,\MeV$. More details can be
found in Refs.~\cite{Martin:2006rs, Ringeval:2007am, Martin:2007ue,
  Martin:2010kz, Martin:2010hh, Demozzi:2012wh, Kuroyanagi:2013ns,
  Ringeval:2013hfa}.

The expression for $\Clt$ can be written as
\begin{align}
\label{eq:defClt}
\Clt\left(\tetas,\tetar,\tetai\right)=\int_0^{+\infty}
\frac{\dd k}{k}& j_{\ell}(k\rlss)T(k;\tetas)
\calP_\zeta(k;\tetar,\tetai),
\end{align}
$j_{\ell}$ being a spherical Bessel function, $\rlss$ the comoving
radial distance to the last scattering surface, $T(k;\tetas)$ the
transfer function which describes the evolution of cosmological
perturbations during the standard Friedmann-Lema\^{\i}tre eras and
$\calP_\zeta$ the inflationary power spectrum.

The posterior
distribution for the parameters of interest is given by
\begin{eqnarray}
\label{eq:postdistri}
p\left(\tetas,\tetar,\tetai \vert \almm\right) &=&
\frac{1}{\calE}\calL\left(\tetas,\tetar,\tetai \right)
\pi\left(\tetas,\tetar,\tetai\right),
\end{eqnarray}
where $\calL\left(\tetas,\tetar,\tetai \right)= p \left(\almm\vert
  \tetas,\tetar,\tetai \right) \propto
\ee^{-\chi^2\left(\tetas,\tetar,\tetai\right)/2}$ is the likelihood
function (and the normalisation constant in front is irrelevant),
$\chi^2$ being the effective chi-squared. The prior distribution
$\pi\left(\tetas,\tetar,\tetai\right)$ describes our a priori state of
knowledge about the values of the parameters before our information is
updated. Notice that, for clarity, we have dropped the dependence on
the model $\calM$ under scrutiny.  In \Eq{eq:postdistri}, $\calE$ is
the Bayesian evidence discussed in the previous section and reads
\begin{equation}
\label{eq:defevidence}
\calE=\int  \dd \tetas \dd \tetar \dd \tetai \calL\left(
\tetas,\tetar,\tetai \right) \pi\left(\tetas,\tetar,\tetai\right).
\end{equation}
It is the quantity we need to calculate for the $\Nevid$ models
considered here.

The effective chi-squared, and, therefore, the likelihood function, is
a function of $\Clt$ and of the data, namely
\begin{equation}
\chi^2\left(\tetas,\tetar,\tetai \right)=
\chi^2\left[\Clt\left(\tetas,\tetar,\tetai \right) ,\almm,\Sigma \right],
\end{equation}
where $\Sigma$ is the noise covariance matrix of the measurement.  The
above expression is only illustrative -- in practice one has to deal
with more complex issues, including foregrounds, instrumental
systematics and the measurements of polarisation in addition to
temperature~\cite{Planck:2013kta}.  Assuming that the
post-inflationary physics is the same for all inflationary
scenarios, different models have different evidences
because they have a different power spectrum
$\calP_\zeta(k;\tetar,\tetai)$. In order to calculate the evidence of
a given inflationary model, one must therefore evaluate
$\calP_\zeta(k;\tetar,\tetai)$ for the sampled values of $\tetar$ and
$\tetai$, then perform the integral~(\ref{eq:defevidence}). In
general, $\calP_\zeta(k;\tetar,\tetai)$ is only known numerically and
this procedure is computationally intensive.

It is, however, possible to speed up dramatically  this calculation if
one uses the fact that the inflationary models under consideration
here are all slow-roll models. In that case, there exists a
general parametrisation of the power spectrum which is given by
($\kstar$ is the pivot scale)
\begin{equation}
\begin{aligned}
\label{eq:pzeta}
  \calP_\zeta(k) & = \calPstar \left[
    a_0\left(\epsilon_n\right) + a_1\left(\epsilon_n\right) \ln
    \left(\dfrac{k}{\kstar}\right) +
    \frac{1}{2}a_2\left(\epsilon_n\right)
    \ln^2\left(\dfrac{k}{\kstar}\right) + \dots \right] \\ & = \Pstar
  \left[1 + \dfrac{a_1(\epsilon_n)}{a_0(\epsilon_n)} \ln
    \left(\dfrac{k}{\kstar}\right) +
    \dfrac{a_2(\epsilon_n)}{a_0(\epsilon_n)}
    \ln^2\left(\dfrac{k}{\kstar}\right) + \dots \right] ,
\end{aligned}
\end{equation}
where $\epsilon_n$ are the Hubble-flow parameters evaluated at Hubble
exit and $\calPstar$ represents the overall
normalisation~\cite{Schwarz:2001vv, Schwarz:2004tz}. We have rendered
explicit the well-measured quantity $\Pstar = a_0(\epsilon_n)
\calPstar = \calP_\zeta(\kstar)$ which fixes the amplitude of the CMB
anisotropies. The explicit form of the $a_i$'s as functions of
$\epsilon_n$ is known~\cite{Martin:2013uma}.

Furthermore, one can express the Hubble flow parameters as a function
of the more fundamental inflationary parameters for every
scenario. The explicit functionals $\epsilon_n(\tetar,\tetai)$ are all
provided in the {\ASPIC} library and in the {\EI}.

The central idea, introduced in~\cite{Ringeval:2013lea}, is that the
likelihood function entering the evidence is invariant under a
reparameterisation of the primordial power spectrum parameters.  We
can thus rewrite the multipole moments (and hence the likelihood
function which depends on them) as $\Clt(\tetas, \tetar, \tetai) =
\Clt\left[\tetas,\Pstar(\tetar, \tetai),\epsilon_n(\tetar,
  \tetai)\right]$. The evidence of Eq.~\eqref{eq:defevidence} becomes
\begin{align}
\label{eq:evidence2}
\calE&=\int  \dd \tetas \dd \tetar \dd \tetai \calL\left[
\tetas,\Pstar(\tetar, \tetai),\epsilon_n(\tetar, \tetai) \right] \pi(\tetas)\pi(\tetar,\tetai) \\
& = \int \dd \tetar \dd \tetai 
\Leff\left[\Pstar(\tetar, \tetai),\epsilon_n(\tetar, \tetai)\right]
\pi(\tetar)\pi(\tetai),
\label{eq:evidence3}
\end{align}
where we have defined the effective likelihood, marginalised over the
post-inflationary parameters, $\tetas$, as
\begin{equation} \label{eq:effectivelike}
 \Leff \left[\Pstar(\tetar, \tetai),\epsilon_n(\tetar, \tetai)\right] \equiv \int \dd
\tetas \ee^{-\frac{1}{2}
  \chi^2\left[\Clt\left(\tetas,\Pstar,\epsilon_n \right) ,
    \almm,\Sigma \right]} \pi\left(\tetas\right).
\end{equation}
In Eq.~\eqref{eq:evidence3} we have made the reasonable assumption
that the prior on the post-inflationary, reheating and primordial
parameters are separable\footnote{More precisely, it is sufficient to
  require that $\pi(\tetas, \tetar,\tetai) = \pi(\tetas)\pi(\tetar,
  \tetai)$. However, it is sensible to assume that the reheating and
  inflationary parameters are separable, too, thus leading to
  Eq.~\eqref{eq:separablepriors}. }, i.e.
\begin{equation}
\label{eq:separablepriors}
\pi(\tetas, \tetar,\tetai) = \pi(\tetas)\pi(\tetar)\pi(\tetai).
\end{equation}
The effective likelihood, Eq.~\eqref{eq:effectivelike}, can be
computed as a function of the slow-roll parameters, $\Pstar,
\epsilon_n$, using machine-learning algorithms to interpolate the
functional form of $\Leff(\Pstar, \epsilon_n)$. Seen as a function of the
slow-roll parameters, $\Leff$ needs only to be computed once for all
inflationary models considered here. To then use it for a specific
inflationary model, it is sufficient to map its potential parameters
$\tetai$ and reheating parameters $\tetar$ onto the corresponding
functionals, $\Pstar(\tetai, \tetar)$, $\epsilon_n(\tetai, \tetar)$.

The computational advantages of our method are twofold. First, the
evaluation of the effective likelihood is very fast, since it is
obtained as the output of a neural network interpolator (typically,
one evaluation requires less than a $\mu\us$ of CPU-time on standard
\texttt{x86\_64} processor). Second, by integrating out once and for
all the post-inflationary parameters from the likelihood, we are left
with a much reduced parameter space over which the Bayesian evidence
integral has to be computed. The dimensionality of $\tetai$ is at most
three, while the reheating is described by just one parameter, so that the
Bayesian evidence integral is at most four-dimensional. Thanks to this
vastly increased efficiency, we were able to compute a large number of
Bayesian evidences with a much reduced numerical effort. More details
about the method can be found in Ref.~\cite{Ringeval:2013lea}.

\subsection{Effective likelihood from Planck 2013}
\label{subsec:effectiveLike}

In order to determine $\Leff$, we have used the Planck 2013
data~\cite{Ade:2013ktc} together with the second order slow-roll
expansion of the primordial power spectra for both the scalar and
tensor perturbations. The full Planck likelihood is provided by the
Planck collaboration~\cite{Planck:2013kta}. Concerning the
post-inflationary universe, it is assumed to be a flat $\Lambda$CDM
model such that the parameters $\tetas$ are:
\begin{equation}
\begin{aligned}
\tetas & = \left(\OmegaB h^2, \OmegaCDM h^2, \tau,
100\thetaMC, \APSa, \APSb, \APSc, \rPSbc, \ACIBb,
\ACIBc, \rCIBbc, \gamCIB, \right. 
\\& \left. \AtSZ, \AkSZ, \xitSZCIB,
\ca, \cc, \betaoo \right).
\label{eq:camparams}
\end{aligned}
\end{equation}
The usual $\Lambda$CDM parameters are the density of baryons
$\OmegaB$, of cold dark matter $\OmegaCDM$, the reduced Hubble
parameter today $h$, the Thompson optical depth $\tau$ to last
scattering and an angle, $\thetaMC$, related to the angular size of
the sound horizon on the last scattering
surface~\cite{Lewis:2002ah}. The remaining parameters describe
astrophysical signals on top of the CMB and any relevant instrumental
distortions, as they have been modelled by the Planck
collaboration~\cite{Ade:2013zuv}. They are the power contribution at
$\ell=3000$ of unresolved point sources at $100\,\GHz$, at
$143\,\GHz$, at $217\,\GHz$ and their cross correlation ($\APSa$,
$\APSb$, $\APSc$, $\rPSbc$). The next are their equivalent for the
Cosmic Infrared Background (CIB), namely $\ACIBb$, $\ACIBc$,
$\rCIBbc$, and $\gamCIB$ stands for the spectral index of the CIB
angular power spectrum. The Sunyaev-Zel'dovich (SZ) signals, either
thermal or kinetic, and their correlations with the CIB are encoded in
the parameters $\AtSZ$, $\AkSZ$, $\xitSZCIB$. Finally, calibration and
beam uncertainties are taken into account in the last three
parameters. More details on how these signals are accounted for can be
found in Ref.~\cite{Planck:2013kta}.

Using the Planck likelihood and its associated public code $\CLIK$, we
have performed a MCMC exploration of the parameter space
$(\tetas,\Pstar,\epsilon_1,\epsilon_2,\epsilon_3)$. In order to do so,
we have used the public code $\COSMOMC$~\cite{Lewis:2002ah}
complemented by a modified version of the $\CAMB$
code~\cite{Lewis:1999bs} in order to implement as initial conditions
the slow-roll primordial power spectra discussed above. All
$\epsilon_n$ in these equations are evaluated at the conformal time
$\etastar$ defined by $\kstar \etastar = -1$, $\kstar =
0.05\,\Mpc^{-1}$ being the pivot scale.

\begin{figure}
\begin{center}
\includegraphics[width=\wsingfig]{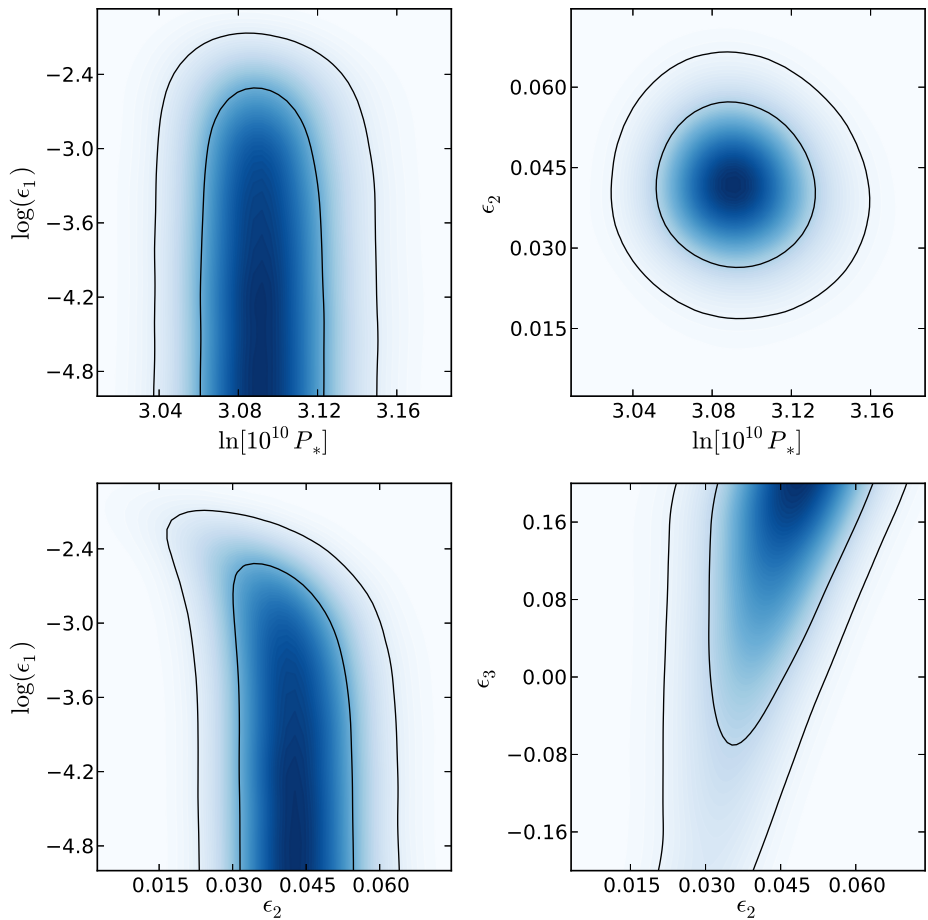}
\caption{Two-dimensional marginalised posterior distributions of the
  slow-roll parameters ($\Pstar$, $\epsstar{1}$, $\epsstar{2}$,
  $\epsstar{3}$) using the Planck 2013 data.}
\label{fig:sr2ndlog}
\end{center}
\end{figure}

The prior choices for the parameters $\tetas$ have been chosen as in
Ref.\cite{Ade:2013zuv}. For the primordial parameter space, we have
chosen a Jeffreys' prior for $\Pstar$ such that $\ln(10^{10} \Pstar)
\in[2.7,4.2]$, \ie centred around its well-measured value. The order
of magnitude of the tensor-to-scalar ratio being unknown, we have
chosen a wide Jeffreys' prior on $\epsilon_1$ as $\log(\epsilon_1) \in
[-5,-0.7]$, the upper bound being such that $\epsilon_1<0.2$ to be
within the slow-roll approximation. Finally, for $\epsilon_2$ and
$\epsilon_3$ we have chosen uniform priors in $[-0.2,0.2]$. The MCMC
exploration has been stopped once the total number of samples reached
two millions, which corresponds to the $R$-statistics convergence of
$\COSMOMC$ (the Gelman-Rubin criterion) to be less than $10^{-3}$ (see
Ref.~\cite{Lewis:2002ah}). The thus obtained two-dimensional
marginalised posterior probability distributions for the slow-roll
parameters are shown in figure~\ref{fig:sr2ndlog}. More details on the
analysis can be found in \Refc{Ringeval:2013lea}. In particular, all the
posteriors are compatible with those obtained by the Planck
Collaboration in Refs.~\cite{Ade:2013zuv,Ade:2013uln}.

These MCMC samples have then be used to determine the effective
likelihood for inflation $\likeinf$ according to
\Eq{eq:effectivelike}, \ie by marginalisation over all the
$\tetas$. However, as shown in figure~\ref{fig:sr2ndlog},
$\epsstar{3}$ is not well constrained. Therefore, following
Ref.~\cite{Ringeval:2013lea}, it is more convenient to fit a
three-dimensional likelihood
$\likeinf(\Pstar,\epsstar{1},\epsstar{2})$ by additionally
marginalising over $\epsstar{3}$. Notice that doing so renders our
analysis robust with respect to any uncertainties that are associated
with the unconstrained second order terms. The fit itself have been
implemented by a multivariate interpolation using a modified quadratic
Shepard's method~\cite{Shepard:1968, Thacker:2010}. Discussions on the
method's accuracy can be found in Ref.~\cite{Ringeval:2013lea} and we
emphasise that the effective likelihood is only well approximated
within the bounds $\ln(\likeinf^{\min}/\likeinf^{\max})=-10$. Lower
values of the likelihood have been extrapolated by assuming
Gaussian tails. As a result, for a given model, the contribution to
the Bayesian evidence from regions in parameter space where the
likelihood is smaller than this value are not reliable. In practice,
this is unlikely to be problematic because the contribution of
regions with exceedingly small likelihood values to the evidence
integral is minimal. Furthermore, models that never achieve a large
value of the likelihood are in any case clearly ruled out, even though
the value for their Bayesian evidence is only approximate.

Let us also stress that, for our purpose,
$\likeinf(\Pstar,\epsstar{1},\epsstar{2})$ is now numerically known
for any input values of $\Pstar$, $\epsstar{1}$ and $\epsstar{2}$
within the prior bounds mentioned earlier. As can be seen in the
posterior of $\epsstar{1}$ (see figure~\ref{fig:sr2ndlog}), $\likeinf$
has a flat direction for very small values of $\epsstar{1}$. As a
result, and only for $\epsstar{1}$, $\likeinf$ has been extrapolated
by a constant along its flat direction for $\log(\epsstar{1}) < -5$,
without loss of accuracy.

\subsection{Computing the evidences}
\label{subsec:computingEvid}

From the effective likelihood, and within a given model of inflation,
we have used the nested sampling algorithm
$\MULTINEST$~\cite{Feroz:2008xx, Feroz:2007kg} to perform the
multidimensional integral of \Eq{eq:evidence3}. For each slow-roll
scenario of the {\EI}, the analytic form of the functionals
$\epsstar{n}(\tetar,\tetai)$ have been derived in
Ref.~\cite{Martin:2013tda} and they have been numerically evaluated
using the public code
$\ASPIC$. The
evidences reported below have been obtained by requiring a
$\MULTINEST$ target accuracy of $10^{-4}$ on the evidence and a number
of live points equals to $30000$. Typically, this amounts to a few
hundred thousand samples for each model and around one hour of CPU
time. We have not reported any numerical error on the evidences
because, with such a target accuracy, they remain completely
negligible with respect to the prior sensitivity effects.

Moreover, for all of the models, we have traded the parameter $M$,
namely the mass scale giving the normalisation of the potential
$V(\phi)$, by the amplitude $\Pstar$ of the scalar primordial power
spectrum at the pivot wavenumber. Both of these parameters are indeed
in one-to-one correspondence once the functionals
$\epsstar{n}(\tetar,\tetai)$ are given, but using $\Pstar$ instead of
$M$ has the advantage of minimising superfluous degeneracies in the
parameter space, as does the choice of using the rescaled parameter
$R$ instead of $\Rrad$. From the Friedmann-Lema\^{\i}tre equation, one
indeed has~\cite{Ringeval:2013lea}
\begin{equation}
M^4 = 24 \pi^2 \dfrac{\epsstar{1}}{\vstar} \Pstar,
\end{equation}
at first order in slow-roll. Here $\vstar \equiv V(\phistar)/M^4$, and
$\phistar = \phi(\etastar)$.

These prior choices have important consequences for the evidence
calculation. They imply that, for all models tested, the prior space
on both the reheating, and the potential normalisation are the
same. As a result, the Occam's razor factors for those parameters
cancel out when computing the Bayes' factor between two models (this
can be seen at once by employing the Savage-Dickey density ratio,
see~\cite{Trotta:2005ar,Trotta:2008qt}). In other words, we assume
that all models have the same ability to reheat the universe after
inflation and to produce the observed amplitude of the CMB
anisotropies. As definite reheating predictions are almost absent in
all the models we have explored, and those same models do not predict
definite values of $M$, this is a fair assumption.

However, if one imagines a situation in which $M$ is an actual output
of the model under scrutiny, its evidence should be reviewed. One may
envisage two cases. Either the predicted values for $M$ (and
$\epsstar{1}$) yield a prior on $\Pstar$ whose support is outside the
range we have used, {\ie} $\ln (10^{10} \Pstar) \in [2.7,4.2]$ (see
figure.~\ref{fig:sr2ndlog}), which is compatible with the data -- in
which case such a model would be ruled out; or it overlaps with it and
the evidence should be recomputed by sampling the parameter space
directly over $M$. In the situation for which the model's predictions
for $M$ would actually match very well the observed amplitude of the
CMB anisotropies, one should expect the Bayesian evidence of that
precise model to be boosted in accordance with the Occam's razor
principle. The same remarks hold concerning the reheating
parameter~\cite{Mazumdar:2013gya}. Let us stress, however, that we
have not encountered such a situation in all the models tested here.

\subsection{Fine-tuning issues}
\label{subsec:finetuning}

For some of the models presented here, the slow-roll regime of
inflation takes place only for a very limited range of values for some
of their parameters. Such ``fine-tuning'' of parameters which have, a
priori, no reason to take exactly such specific values, is disfavoured
by the Occam's razor penalty in-built into the Bayesian evidence. From
a technical point of view, the likelihood can reliably be worked out
only in regimes where the slow-roll is (at least roughly)
valid. Otherwise, the inflationary dynamics is very difficult to track
and not described by our modelisation. On the other hand, when the
slow-roll is completely violated, one knows that the associated
predictions are ruled out by observations, and that the likelihood in
this region of parameter space, being essentially 0, does not
contribute to the the total evidence. Therefore such situations result
into an Occam's razor effect which suppresses the evidence computed
over ``compatible'' parameters (the ones for which slow-roll inflation
exist) by a factor equal to the ratio of the volume of compatible
parameters over the whole prior volume. For the models in which this
occurs, we have added some discussions in the appendix.

\begin{landscape}
\begin{figure}
\begin{center}
\includegraphics[height=\wfull]{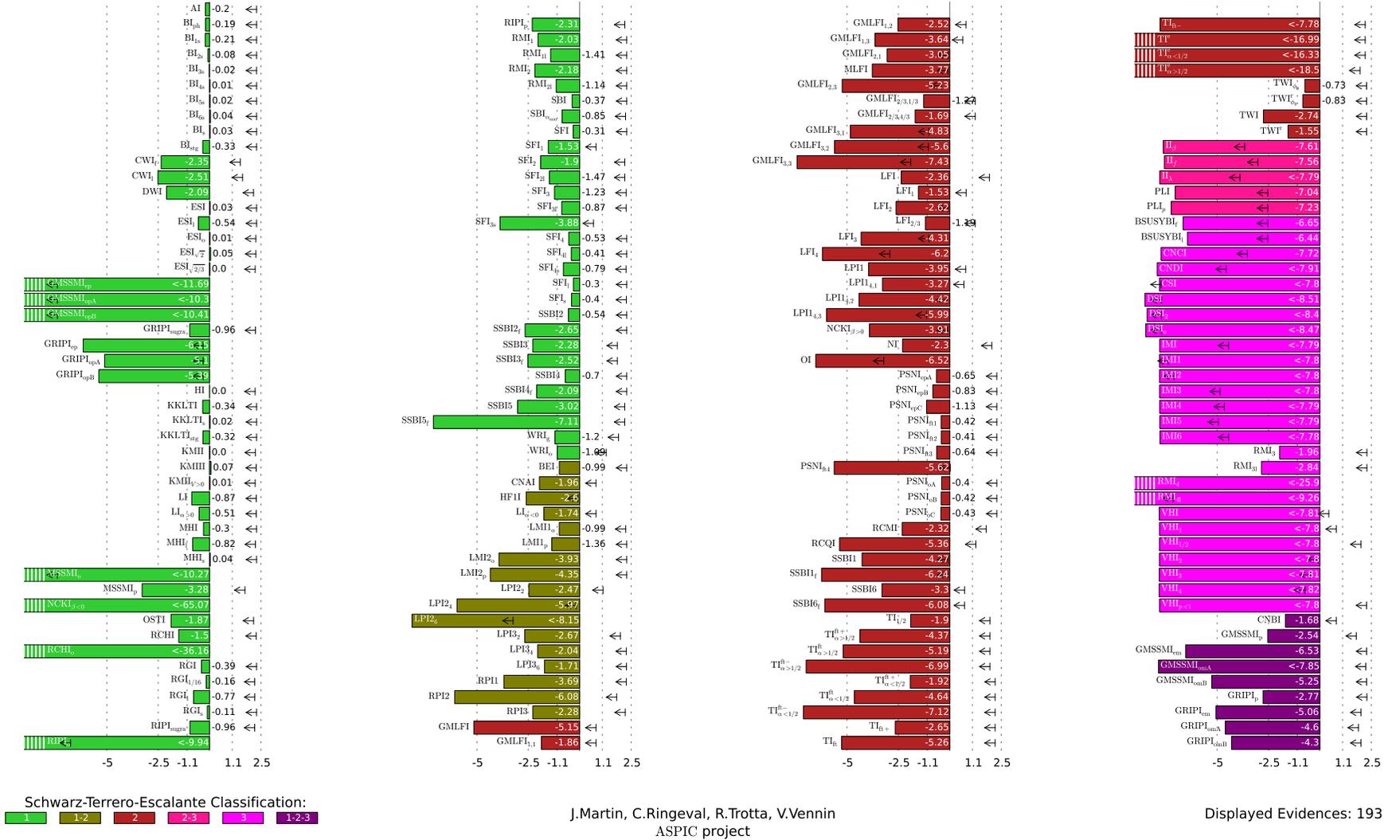}
\caption{Bayes factors (bars) and absolute upper bound to the Bayes
  factors (arrows) for the \EI inflationary scenarios, with Higgs
  inflation as the reference model (see the text for a more accurate
  description).}
\label{fig:evid}
\end{center}
\end{figure}
\end{landscape}

\section{Results and discussion}
\label{sec:res}


For all the models listed in the appendix~\ref{sec:priors}, \ie
$\Nmod=\Nevid$, we have computed the Bayes factors $\Bhi{i}$ with
respect to the Starobinsky model~\cite{Starobinsky:1980te,
  Ellis:2013nxa, Ellis:2013xoa} or Higgs Inflation ($\hi$), which is
our reference model. We have also evaluated each model's Bayesian
complexity $\Cb_i$.

Our main results are displayed in figure~\ref{fig:evid}, which
represents all the Bayes factors. Each model is represented by a
horizontal bar indicating the value of $\ln \Bhi{i}$. A bar extending
to the left corresponds to $\ln \Bhi{i}<0$ and the model under
consideration is disfavoured with respect to the the reference
model. If, on the contrary, the bar extends to the right, then $\ln
\Bhi{i}>0$ and the model is preferred to Higgs inflation. Obviously,
the Bayes factor of the reference model is one and, therefore, its
logarithm vanishes: this is why there is no bar for HI. In front of
(or inside) each bar, we have reported the exact numerical value of
$\ln \Bhi{i}$. We have also included the Jeffreys' scale of
Table~\ref{tab:Jeff}, as dashed vertical lines, as an indication of
the viability of a given model compared to $\hi$.

Bars are colour-coded according to the Schwarz--Terrero-Escalante (STE)
classification associated with the slow-roll parameters of the model
under consideration~\cite{Schwarz:2004tz}. Following the notation used
in \Refc{Martin:2013tda}, region $1$ are models predicting
$\epsstar{2}> 2 \epsstar{1} > 0$, \ie the kinetic energy increases
during inflation as well as the ratio of the kinetic energy to the
total energy. Region $2$ stands for potentials associated with
$0<\epsstar{2} < 2 \epsstar{1}$ for which the kinetic energy decreases
while the ratio of the kinetic energy to the total energy still
increases. Finally, region $3$ is such that both quantities decrease
during inflation. As shown in \Refc{Martin:2013tda}, the Planck 2013
results disfavour models living in regions $2$ and $3$ and the Bayes
factors also reflect this. Let us stress that the parameter space of
some models may span more than one region, \ie for some values of its
parameters the predictions of a model can fall in region 1 (say)
while, for some other regime, they can be in region 2. It is
referenced in the captions of figure~\ref{fig:evid} where the colour
code takes this fact into account.

Finally, for each model, we have also calculated the maximum value of
the evidence, that is to say the value that is obtained when all the
prior mass for the model under consideration is concentrated in a
delta-function centred at the maximum likelihood location. Clearly, in
that case, one has $\Evid_\mathrm{max}=\Like_\mathrm{max}$. It
represents an absolute upper bound on the evidence: any choice of
priors necessarily leads to a value of the evidence smaller than
$\Evid_\mathrm{max}$. This upper bound is represented by black
left-pointing arrows in figure~\ref{fig:evid}. Let us also remark that
this quantity would be relevant in a frequentist analysis where the
$p$-value obtained from the maximum likelihood ratio would be used to
compare the performances of different models.

Let us now analyse our results in more detail. Firstly, the answer to
the central question of this paper, namely ``what is the best model of
inflation given the Planck 2013 data?'' is $\kmiii$
inflation~\cite{Conlon:2005jm,Krippendorf:2009zza,Burgess:2013sla},
whose Bayes factor with respect to Higgs inflation is $\ln\Bhi{\kmiii}
= 0.07 > 0$. However, the preference is extremely mild, so much so
that it is within the margin of uncertainty of our analysis, and for
all practical purposes $\kmiii$ inflation has to be regarded as being
on the same footing with Higgs inflation, from the point of view of
the Planck data.

We can use the Jeffreys' scale as an indication for which of the
models remain viable, and which are disfavoured at various levels of
evidence with respect to the best models. We find $\Nincon$ models in
the ``inconclusive'' region (with respect to the best model), $\Nweak$
in the ``weakly disfavoured'' region, $\Nmoder$ in the ``moderately
disfavoured'' region and $\Nstrong$ in the ``strongly
disfavoured''. Therefore, our analysis concludes that surviving models
(\ie those in the ``inconclusive'' region) represent $\pourincon \%$
of the total. On the contrary, the number of models that are
conclusively ruled out (\ie those in the ``strong'' region) represent
$\pourstrong \%$ of the total numbers of models. The models in the
``inconclusive region'', which are to be considered the best models of
inflation after the Plank data, are (in alphabetical
order\footnote{The meaning of the different acronyms and the precise
  definition of the corresponding models can be found in
  appendix~\ref{sec:priors}.}): $\ai$, $\bi$, $\bis$, $\biONEs$,
$\biTWOs$, $\biTHREEs$, $\biFOURs$, $\biFIVEs$, $\biSIXs$, $\bistg$,
$\esi$, $\esil$, $\esisqrtTWOTHREE$, $\esisqrtTWO$, $\esio$, $\hi$,
$\kklti$, $\kkltis$, $\kkltistg$, $\kmiii$, $\kmii$, $\kmiivp$, $\li$,
$\lip$, $\mhi$, $\mhil$, $\mhis$, $\psniftONE$, $\psniftTWO$,
$\psniftTHREE$, $\psnioA$, $\psnioB$, $\psnioC$, $\psniepA$,
$\psniepB$, $\rgi$, $\rgis$, $\rgil$, $\rgiONEONESIX$, $\sbi$,
$\sbialphamin$, $\sfi$, $\sfiTHREEl$, $\sfiFOUR$, $\sfiFOURl$,
$\sfiFOURs$, $\sfil$, $\sfis$, $\ssbiTWO$, $\ssbiFOUR$, $\twiAONE$ and
$\twiATWO$. As explained above, there are more models than potential
shapes because a given potential can support different priors, which
are considered as separate model choices. As a consequence, the above
$\Nincon$ models in the ``inconclusive region'' encompass only $15$
different potentials or scenarios.

\begin{figure}
\begin{center}
\includegraphics[width=\wdblefig]{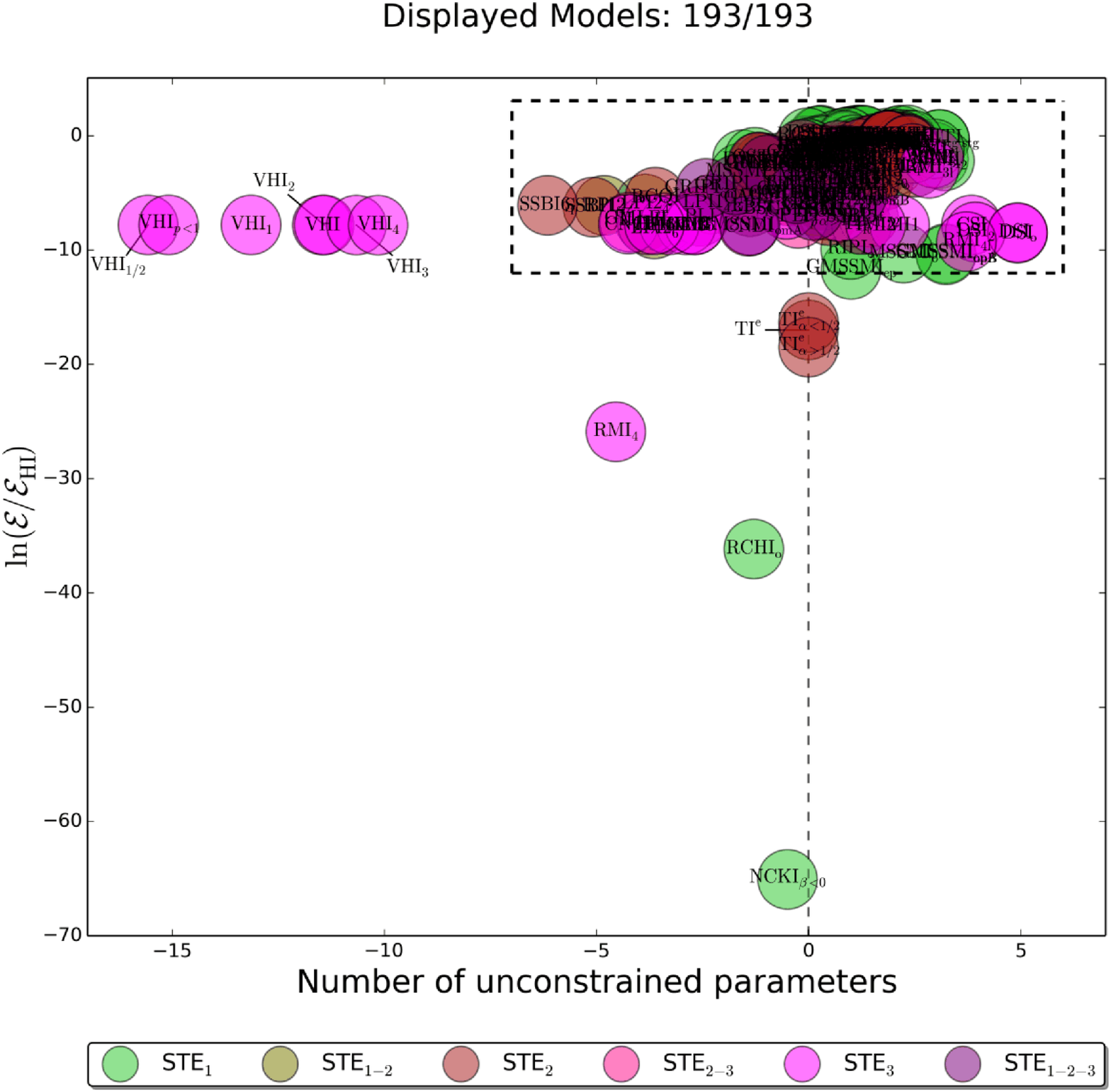}
\includegraphics[width=\wdblefig]{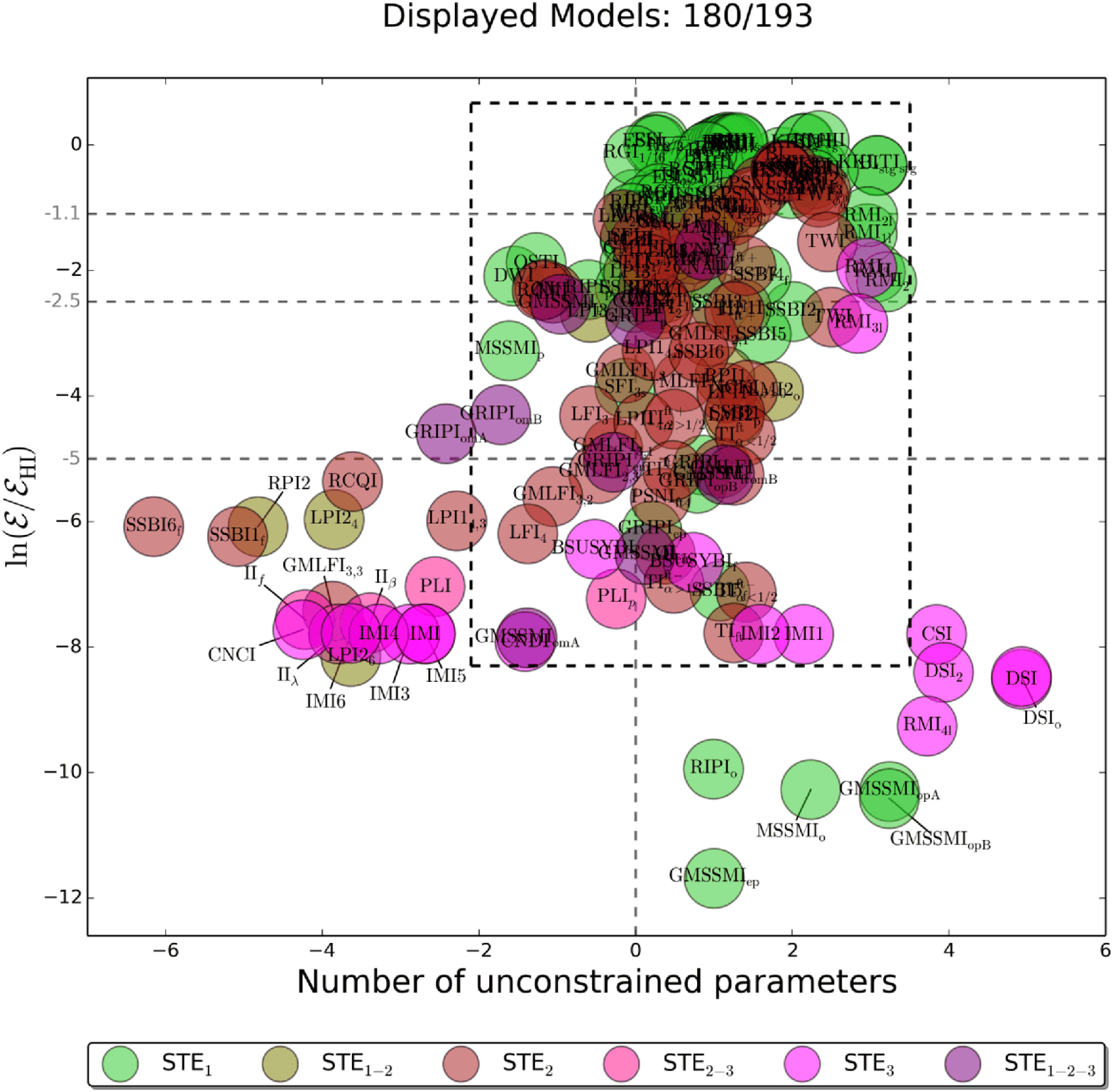}
\includegraphics[width=\wdblefig]{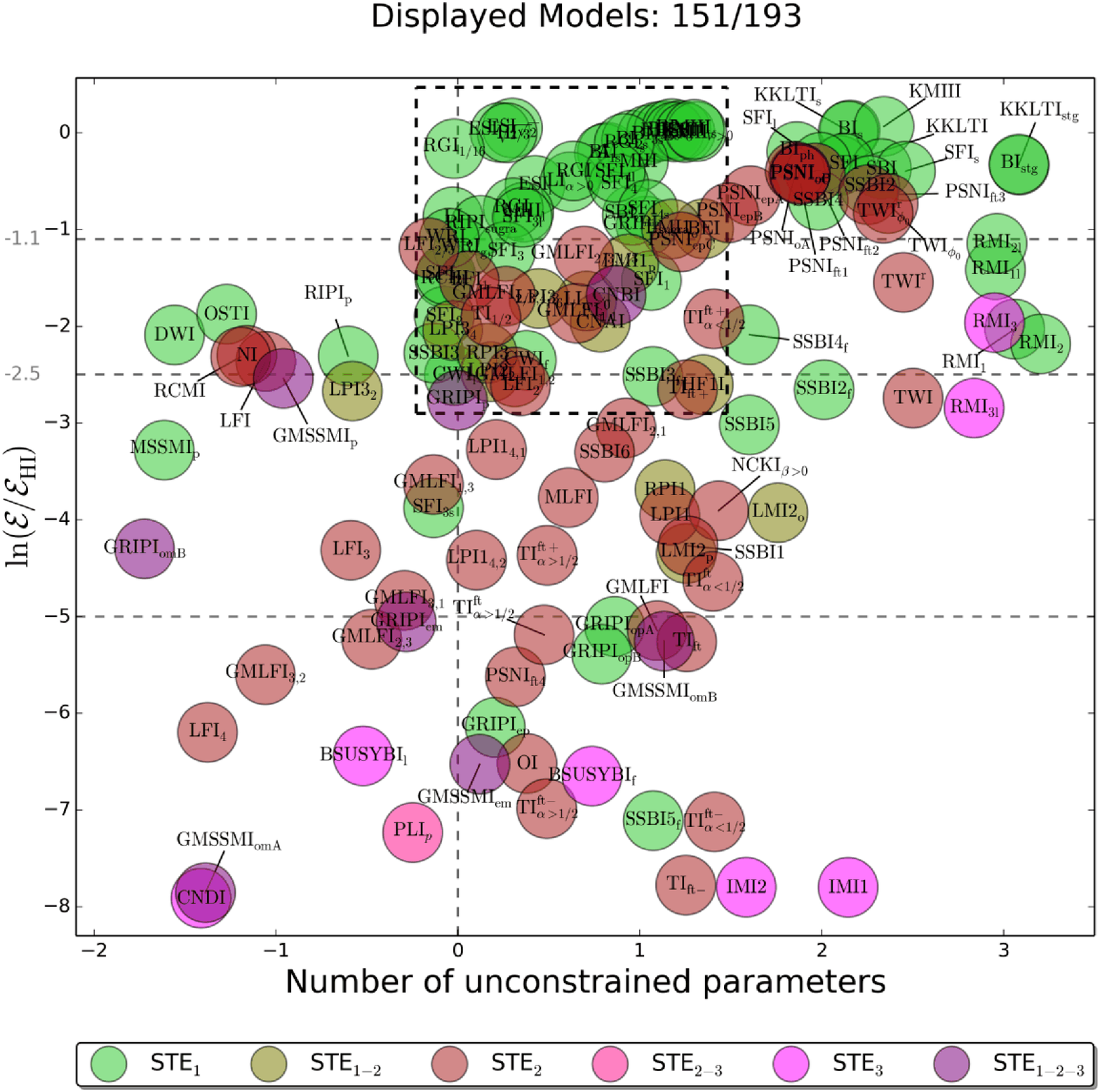}
\includegraphics[width=\wdblefig]{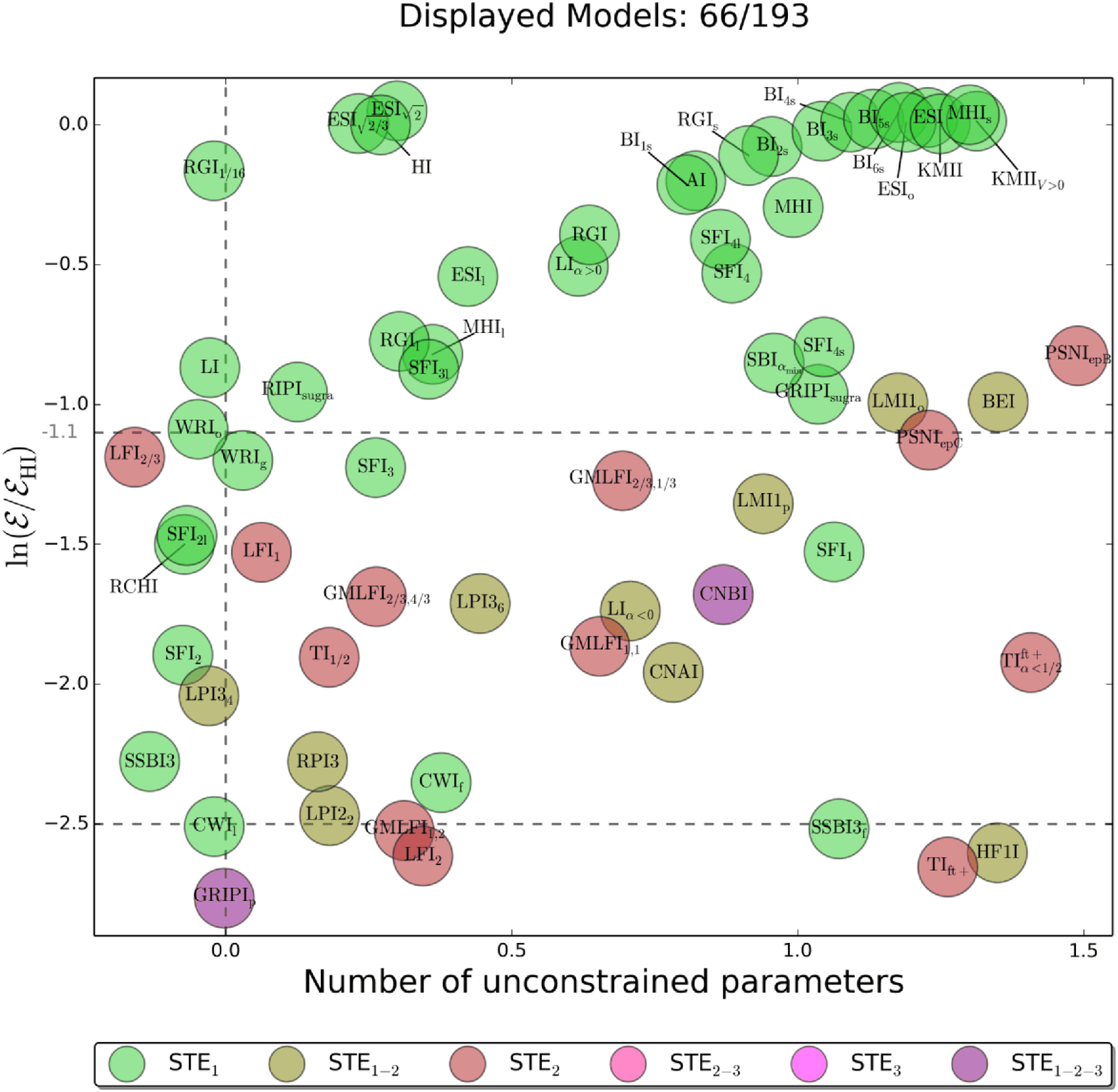}
\caption{Logarithm of the Bayes factor versus the number of
  unconstrained parameters $\Nuc$ for all the inflationary models
  investigated. The $\Nuc$ dimension allows us to disambiguate models
  with the same evidence, by preferring those with the smallest number
  of unconstrained (\ie, unnecessary) parameters. Optimal models are
  clustered around Higgs Inflation and have $\Nuc \simeq 0$ together
  with $\Bhi{} \gtrsim 0$. The four plots (from upper left to bottom
  right) increasingly zoom into the ``best region''. Each model is
  represented by a filled circle for illustration purposes only, and
  the radius of a circle has no meaning.}
\label{fig:evidcomp}
\end{center}
\end{figure}

Further insight can be garnered by considering the Bayesian complexity
for each {\EI} model. We are particularly interested in evaluating the
number of unconstrained parameters for the best models identified via
the Bayesian evidence, \ie the ones that are in the ``inconclusive
region''. Since the Bayesian complexity measures the number of
effective parameters supported by the data, one can define a measure
of the number of unconstrained parameters by
\begin{equation}
  \Nuc_i \equiv  \Nparam_i-\Cb_i,
\end{equation}
where $\Nparam_i$ is the total number of free parameters of the model
under consideration, {\ie} the inflationary potential parameters, plus
the reheating parameter. For models providing a reasonable good fit to
the Planck data, one expects $\Nuc_i \geq 0$. However, if the best-fit
log-likelihood of a given model is very poor, then the Bayesian
complexity can be arbitrary large, as the second term in
Eq.~\eqref{def:Cb} is large. This means that for such models $\Nuc_i <
0$. So we expect a negative measure of the number of unconstrained
parameters to be correlated with a small value of the Bayes factor.

In figure~\ref{fig:evidcomp}, we have plotted the location of all models
in the two-dimensional plane $(\Nuc,\ln\Bhi{})$. Models appearing
along the same horizontal lines have thus the same Bayes factor but
different number of unconstrained parameters $\Nuc$. Models with the
smallest, non-negative number of unconstrained parameters are to be
preferred in that they can be deemed to be simpler, even if they have
the same evidence as other models with a larger value of $\Nuc$.

We can observe in figure~\ref{fig:evidcomp} that models with $\Nuc<0$
do have poor values of the evidence as well ($\ln\Bhi{i} \ll 0$), as
expected from the above argument. Focusing on the models having the
best evidences together with a minimal number of unconstrained
parameters, \ie $0<\Nuc_{i}<1$ narrows down the slow-roll landscape to
a few preferred models: $\ai$, $\biONEs$, $\biTWOs$, $\esil$,
$\esisqrtTWOTHREE$, $\esisqrtTWO$, $\hi$, $\lip$, $\mhi$, $\mhil$,
$\rgi$, $\rgis$, $\rgil$, $\sbialphamin$, $\sfiTHREEl$, $\sfiFOUR$ and
$\sfiFOURl$. We have now $17$ preferred models, that is to say roughly
$9\%$ of the initial numbers of models. They correspond to only $9$
types of potential or scenarios. It is also interesting to notice that
$\kmiii$ is not in this set of preferred models since it has
$\Nuc_{\kmiii}\simeq 2.3$. While it cannot be concluded that the
models with the best Bayes factors and $0<\Nuc_{i}<1$ are the ``true''
models, they are the simplest and most effective inflationary
hypotheses that are compatible with the Planck 2013 CMB
data. Obviously, allowing for more unconstrained parameters increases
this list as displayed in figure~\ref{fig:evidcomp}.

Another interesting remark is that the $9$ potentials mentioned above
all belong to region $1$ in the Schwarz-Terrero-Escalante
classification (\ie there are all ``green''). This is of course
consistent with the findings of \Refc{Martin:2013tda} which has shown
that this region is the region favoured by the Planck data. This means
that the corresponding models all belong to ``plateau inflation'' for
which the potential does not necessarily grows to infinity when the
{\vev} of the field increases~\cite{Ijjas:2013vea}. This type of
potentials clearly appears to be the winners given the Planck data.

\section{Conclusions}
\label{sec:conclusion}

\begin{figure}
\begin{center}
\includegraphics[width=\wsingfig]{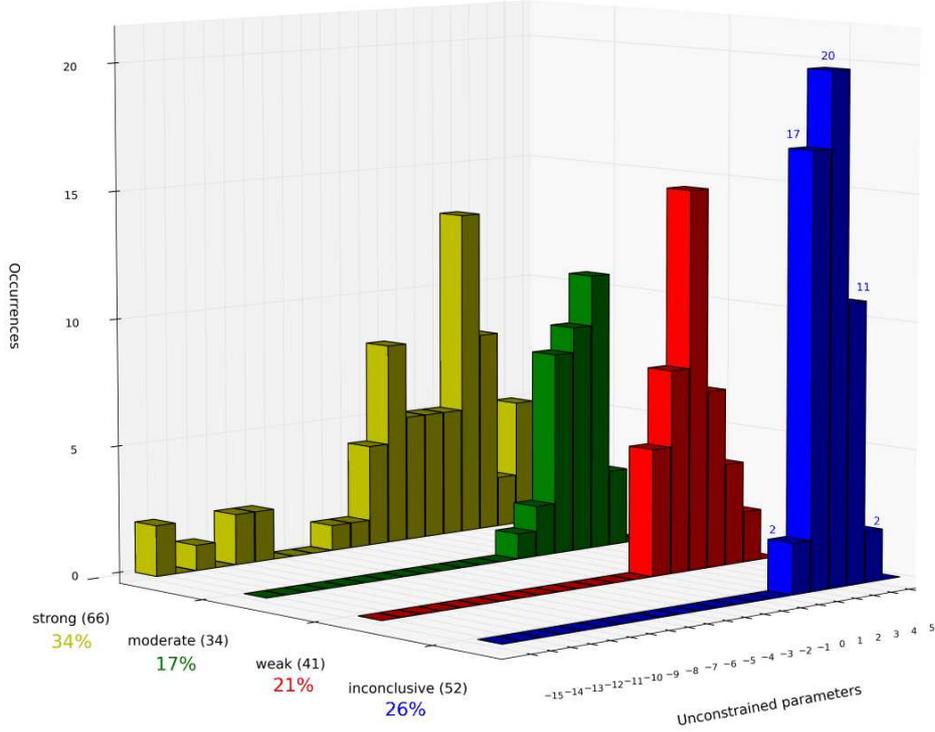}
\caption{Histogram of the {\EI} models within the four Jeffreys'
  categories (inconclusive: blue, weakly disfavoured: red, moderately
  disfavoured: green and strongly disfavoured: yellow) and for
  different number of unconstrained parameters. The number of
  preferred models is $17$, corresponding to $9$ different types of
  potential.}
\label{fig:histo}
\end{center}
\end{figure}

Let us now recap our main findings. Although this paper deals with
slow-roll single-field inflation only, we do not expect multifield
inflationary models to perform better than the optimal subset of
single-field models that have been delineated in this work. This is
because adding a field necessarily introduces extra-parameters
encoding the shape of the potential in this new direction. Therefore,
even if a multifield scenario would fit as well the Planck 2013 data
as the best slow-roll single-field models, such a model would be
penalised by its larger number of unconstrained parameters (in terms
of complexity). This conclusion could be modified if a multifields
model was able to fit the large scales glitches in the Planck data,
thus achieving a better evidence. However, those glitches are of
relatively weak statistical significance and cannot, currently,
greatly improve the overall fit. Furthermore, the fit improvement
would have to be sufficient to offset the extra Occam's factor penalty
implied by additional free parameters. Such a situation may however
change by considering additional and independent data sets which could
not be fitted by the class of slow-roll models discussed in this paper
such as, for instance, a small, but non-vanishing, level of
non-Gaussianities. The same remarks also apply for single-field
scenarios with non-minimal kinetic terms (or with features in the
potential). These models are not necessarily ruled out. However,
either they predict observable non-Gaussianities and the fact that
Planck sees a Gaussian sky implies that those models will be penalised
for this wasted parameter space. Or, they genuinely do not predict
non-Gaussianities but introduce additional parameters that increase
the model complexity (see for instance \Refc{Germani:2010gm,
  Germani:2010hd, Germani:2011ua}). Let us stress that, if we are not
considering the small gain that might be associated with fitting
Planck's glitches, the favoured models we have singled out in this
paper already saturate the maximal possible value for the
likelihood. As a result, even in the situation in which we would have
missed an extremely good fitting and simple model, its Bayesian
evidence would still be in the ``inconclusive region''.

Therefore, from a Bayesian point of view, it appears perfectly
legitimate to focus on single-field slow-roll inflation (with a
minimal kinetic term). These models have been studied and compared to
the recent Planck data in~\Refc{Martin:2013tda} which, therefore,
represents a complete cartography of the inflationary landscape
compatible with the most recent data. In the present article, we have
computed the Bayes factors and the Bayesian complexity for all these
\EI models. Our results are summarised in an histogram in
figure~\Fig{fig:histo}, which gives the number of models in each
Jeffreys' category (defined with respect to the best model) and for
each number of unconstrained parameters with $n<\Nuc<n+1$, where $n$
is an integer. This plot illustrates the power of the Planck data and
allows us to summarise our main results: from a large number of
models, one is able to single out a relatively small subset
corresponding to the ``best models''. We rule out $\simeq
\pourstrong\%$ of the models at a strong level of evidence and $\simeq
\pourincon\%$ of the models ($9\%$ if one includes the complexity) are
preferred. All the favoured scenarios belong to the category $1$ of the
the Schwarz--Terrero-Escalante classification and have a shape
consistent with ``plateau inflation''.

It is also worth pointing out that a few Bayesian evidences have been
calculated in~\Refc{Ade:2013uln}. The comparison is, however,
difficult to carry out since the priors on reheating assumed in that
paper greatly differ from those considered here\footnote{Let us also
  stress that the description made by~\Refcs{Ade:2013uln} of the work
  of~\Refc{Martin:2010kz} on reheating is incorrect. It is claimed
  that the study of~\Refc{Martin:2010kz} is restricted to equations of
  state of the form $\wrehbar=(p-2)/(p+2)$, which emerges in the case
  of a potential with the shape $\propto \phi^p$. This situation was
  indeed considered in~\Refc{Martin:2010kz} but only as a particular
  example. The completely generic case $-1/3<\wrehbar<1$ was in fact
  the main concern of \Refc{Martin:2010kz}.}. Indeed,
in~\Refc{Ade:2013uln}, a prior on $\Delta \Nstar$ is chosen while the
reheating energy density is arbitrarily fixed. There is no physical
motivations for picking up particular values of the reheating energy
density. Moreover, choosing a prior on $\Delta N_*$ is surprising
since this does not guarantee the validity of the physical prior,
namely $\rhonuc<\rhoreh<\rhoend$. Another side effect is that this
obviously modifies the calculation of the Bayesian evidences and, for
this reason, comparing the two approaches does not lead to interesting
insights.

To conclude this paper, let us present some speculations regarding
what we have learnt about the physics of inflation. Firstly, let us
stress that we have finally carried out one of the long standing task
of primordial cosmology, namely put constraints on the shape of the
inflationary potential. In some sense, this represents quite an
impressive achievement since we are able to say something about
physics at energy scales unreachable in accelerators. Indeed, with the
Large Hardron Collider (LHC), it would obviously be impossible to
establish the existence, at the Grand Unified Scale (GUT) scale, of a
scalar field with a potential having a plateau shape. This perfectly
illustrates the fact that cosmology can teach us something about high
energy physics. On the other hand, this conclusion should be toned
down: certainly, we have learnt a lot about the early Universe but,
clearly, this does not give us the Lagrangian of particle physics at
the GUT scales (\ie the field content, their interactions etc
\dots). As a consequence, our knowledge of physics at such a high
energy scale remains very limited. Hopefully, future analysis will
help us to learn more about these questions. In this respect,
constraining the reheating temperature of all the \EI models seems
promising since this can tell us something about the interaction of
the inflaton field with the rest of the world.

Finally, one cannot help making the connection between the results
obtained here and the recent works about ``conformal
inflation''~\cite{Futamase:1987ua, Einhorn:2009bh, Buchmuller:2012ex,
  Ellis:2013nxa, Kallosh:2013hoa, Kallosh:2013pby, Kallosh:2013lkr,
  Ferrara:2013rsa, Kallosh:2013daa, Kallosh:2013tua,
  Ellis:2013xoa}. It is well-known that it is difficult to control the
flatness of the inflaton potential that can easily be destroyed by
quantum corrections. However, if one starts with any shape of
$V(\phi)$, not necessarily very flat, and assumes a non-minimal
coupling (for instance, of the form $\xi \phi^2R$), then, in the
Einstein frame, the potential automatically flattens out and,
precisely takes the form of plateau inflation for some range of the
field. A striking example is provided in figure 4
of~\Refc{Kallosh:2013hoa}: far from the origin, the potential
automatically acquires the typical shape found in the present article
to be favoured by the Planck data (see in particular right bottom of
figure 4). Let us stress at this point that, although non-minimally
coupled to gravity, this class of models belong to the \ASPIC category
since, after a conformal transformation to the Einstein frame, these
models are in fact equivalent to single-field slow-roll inflation. In
this representation, the non-triviality of the non-minimal coupling
has been ``transferred'' to the complicated, non-minimal, interaction
of $\phi$ with the other degrees of freedom present in the early
Universe. In fact, Higgs inflation is the prototypical example of this
class of scenarios and the ingredients necessary to describe the
reheating phase in this case have been described in
\Refc{Martin:2013tda}. Therefore, we are in a situation where two
strong theoretical arguments (the flatness of the potential and the
presence of a non-minimal coupling to gravity --- recalling that,
according to the standard lore, a term that is not forbidden by a
symmetry must be present in the theory) point precisely to the models
that appear favoured by recent data. Whether this is just a
coincidence or whether we are starting to understand something deeper
about Nature will hopefully be answered in the near future when even
more accurate data become available.

\acknowledgments
This work is partially supported by the ESA Belgian Federal PRODEX
Grant No.~4000103071 and the Wallonia-Brussels Federation grant ARC
No.~11/15-040.

\appendix

\section{Choice of priors for inflationary models}
\label{sec:priors}

In this appendix we detail the priors used in this article, and report
the corresponding Bayesian evidences, complexities, number of
parameters and likelihoods at the best fit point of all \EI
scenarios. The priors are directly transcribed from considerations
presented in Ref.~\cite{Martin:2013tda}, which is assumed to be known
to the reader.

As discussed in section~\ref{subsec:priorsens}, there are cases where
it is difficult to numerically estimate the evidences. In particular,
this happens when one tries to extend the prior ranges in order to
study the impact of the prior choices on our physical
conclusions. However, most of the time, this prior sensitivity can be
trivially accounted for by means of simple analytical calculations
that we now briefly review. There are few instances in the following
where they are concretely used.

A common situation is when the support of the likelihood is included
in the prior range $[\tetamin,\tetamax]$, \ie $\calL(D\vert
\theta,\calM)\simeq 0$ for $\theta \notin [\tetamin,\tetamax]$. The
evidence of a model $\calM$ is given by
\begin{equation}
\calE(D\vert\calM)=\int_{\tetamin}^{\tetamax} \dd \theta 
\calL(D\vert \theta,\calM)\pi(\theta\vert \calM),
\end{equation}
where, for simplicity, we have assumed that there is only one
parameter, $\theta$ (the argument can be generalised to any
dimensions). For any proper (\ie, normalised) prior 
distribution $\pi(\theta|\calM)$, one has
\begin{equation}
  \pi(\theta\vert \calM)=\dfrac{\Pi(\theta)}
{\displaystyle \int_{\tetamin}^{\tetamax} \dd \theta \,
  \Pi(\theta)}\,, \quad \textrm{with} \quad\int_{\tetamin}^{\tetamax} \dd \theta
\,\pi(\theta\vert\calM)=1.
\end{equation}
Let us assume that we change the prior range for the parameter
$\theta$ and consider a new upper bound $\bartetamax$. The new prior
is now given by
\begin{equation}
  \pi(\theta\vert \calM)=\dfrac{\Pi(\theta)}
{\displaystyle \int_{\tetamin}^{\bartetamax} \dd \theta \, \Pi(\theta)}\,, 
\end{equation}
where, in accordance with the above discussion, the likelihood is
vanishing in $[\tetamax,\bartetamax]$. As a consequence, the value of
the evidence for the larger prior range is given by
\begin{equation}
\label{eq:rescaledevidence}
\bar{\calE}(D\vert\calM)=\int_{\tetamin}^{\bartetamax} \dd \theta
\calL(D\vert \theta,\calM)\pi(\theta\vert \calM)
=\calE(D\vert\calM)\dfrac{\displaystyle \int_{\tetamin}^{\tetamax} \dd
  \theta \, \Pi(\theta)} {\displaystyle\int_{\tetamin}^{\bartetamax} \dd \theta \,
  \Pi(\theta)}\,,
\end{equation}
and is obtained from the previous evidence value by simply rescaling
it by the ratio of the prior volumes.

If instead the likelihood is flat along the $\theta$ direction, \ie
the data do not constrain the parameter under consideration,
$\calL(D|\theta,\calM) = \calL_0$, then the evidence is unchanged by
modifying the prior bounds
\begin{equation}
\bar{\calE}(D\vert\calM) = \int_{\tetamin}^{\bartetamax} \dd \theta
\calL(D\vert \theta,\calM)\pi(\theta\vert \calM) = \calL_0
\int_{\tetamin}^{\bartetamax} \dd \theta \,\pi(\theta\vert\calM) = \calL_0 =
\calE(D\vert\calM),
\end{equation}
and one should evaluate the Bayesian complexity to distinguish between the
models.

Let us notice that the complexity may also be modified when the prior
range is extended to regions where the likelihood is known to be
negligible. However, contrary to the evidence, there is no simple
analytical treatment of how the complexity should be extrapolated in
this case. One can nevertheless make further simplifying assumptions
to roughly estimate how the complexity is sensitive to the choice of
priors.

Assuming that the prior and likelihood distributions are Gaussian, the
complexity is given by~\cite{Kunz:2006mc}
\begin{equation}
\label{eq:Comp:gaussian}
\Complexity = \sum_{i=1}^N {\frac{1}{1 +
    \left(\dfrac{\sigma_{\mathcal{L}}^i}{\sigma_\Pi^i}\right)^2}}
\simeq \frac{\Params}{1 +
  \left(\dfrac{\sigma_{\mathcal{L}}}{\sigma_\Pi} \right)^2} \,,
\end{equation}
where $\Params$ is the number of parameters, $\sigma_\Pi^i$ and
$\sigma_{\mathcal{L}}^i$ are the prior width and the standard
deviations of the likelihood covariance matrix along its
eigendirections $i$, respectively. The last approximation in the above
equation assumes that one can define the averaged values $\sigma_\Pi$
and $\sigma_{\mathcal{L}}$ over all the eigendirections. If the prior
is widened along $n$ directions (chosen among the $\Params$
parameters), its averaged volume $\sigma_\Pi^n$ gets multiplied by the
same ratio $\bar{\Evid}/\Evid$ as computed above, \ie
\begin{equation}
\sigma_{\bar{\Pi}} = \sigma_{\Pi}
\left(\dfrac{\Evid}{\bar{\Evid}}\right)^\frac{1}{n}\, .
\end{equation}
Plugging back this relation in \Eq{eq:Comp:gaussian}, one gets
\begin{equation}
\label{eq:complexity:Crescaling}
\bar{\Complexity}=\dfrac{\Params}{1 + \left(\dfrac{\bar{\Evid}}
  {\Evid}\right)^{2/n}\left(\dfrac{\Params}{\Complexity} - 1\right)}\,,
\end{equation}
where $\bar{\Evid}/\Evid$ is given by a volume ratio of the type
\Eq{eq:rescaledevidence}.

\color{black}

In the next subsections, we discuss, for each \EI scenarios, our
choice of priors. We also give the definition of all the acronyms used
in the paper, in particular in Fig.~\ref{fig:evid}.

\subsection{Higgs Inflation (HI)}

The Higgs inflation model the potential of which is given
by~\cite{Martin:2013tda}
\begin{equation}
V(\phi)=M^4\left(1-\ee^{-\sqrt{2/3}\phi/\Mp}\right)^2,
\end{equation}
which contains only one parameter: the mass scale $M$. However, as
discussed in section~\ref{subsec:computingEvid}, this one has been
traded for $\Pstar$ in our analysis and there are no other free
parameter in this potential. In total, including the reheating
parameter, one ends up with a two-parameters model. For this reason,
and besides the fact that it was actually the first model of inflation
ever proposed, we have chosen to take HI as the ``reference model''.

\begin{center}
\begin{baytabular}
$\hi$ & $-$ & $\Ehi$ & $\Chi$ &$\NPhi$ &$\BEhi$  \\
\end{baytabular}
\end{center}

\subsection{Radiatively Corrected Higgs Inflation (RCHI)}

This model is a one-parameter model. The shape of the potential
reads~\cite{Martin:2013tda}
\begin{equation}
V(\phi)=M^4\left(1-2\ee^{-2/\sqrt{6}\phi/\Mp}+\frac{\AI}{16 \pi^2}
\frac{\phi}{\sqrt{6}\Mp}\right).
\end{equation}
The parameter $\AI$ controls the amplitude of the radiative
corrections to the, tree level, HI potential. The one-loop expansion
is valid under the condition $\AI\ll 64\pi^2$, hence the physical
prior $\AI\in[-100,100]$. However, numerically, when $\AI<-65$, the
likelihood is so small that it cannot be calculated in a reliable
way. As a consequence, we choose the numerical prior to be
$\AI\in[-65,100]$. Anyhow, as already mentioned, the range
$\AI\in[-100,-65]$ does not contribute to the likelihood. On the other
hand, as discussed in \Refc{Martin:2013tda}, particle physics implies
$-48<\AI<-20$ and this defines a new model, the ``original'' one, that
we denote $\rchio$ in the following. We thus have two possible priors
for this scenario as indicated by the following table:

\begin{center}
\begin{baytabular}
$\rchi$ & $\AI\in[-65,100]$ & $\Erchi$ & $\Crchi$ & $\NPrchi$ &
  $\BErchi$ \\
$\rchio$ & $\AI \in [-48,-20]$ & $\Erchio$ & $\Crchio$ & $\NPrchio$ &
  $\BErchio$ \\
\end{baytabular}
\end{center}

\subsection{Large Field Inflation (LFI)}

Large field inflation is characterised by the following
potential~\cite{Martin:2013tda}
\begin{equation}
V(\phi)=M^4\left(\frac{\phi}{\Mp}\right)^p.
\end{equation}
This potential depends on a mass scale $M$ fixed by the CMB
normalisation and a free index $p$ of $\order{1}$ that can also take
specific integer or rational values. Hence, one may assume a general
prior on $p$ such that one can calculate the evidence of this class of
model. Here one takes $p\in[0.2,5]$ because, for $p>5$, one already
knows that the models are ruled out and $p>0.2$ instead of $p=0$ for
numerical reasons (in addition, the potential cannot be completely
flat since one needs to stop inflation). Another possibility is simply
to fix $p$ to some interesting values: $p=2/3$ corresponds to
monodromy inflation~\cite{Silverstein:2008sg} while $p=1, \cdots, 4$
represents interesting phenomenological scenarios.

\begin{center}
\begin{baytabular}
$\lfi$ & $p\in\left[0.2,5\right]$ & $\Elfi$ & $\Clfi$ & $\NPlfi$ & $\BElfi$ \\
$\lfiTWOTHREE$ & $p=2/3$ & $\ElfiTWOTHREE$ & $\ClfiTWOTHREE$ &
  $\NPlfiTWOTHREE$ & $\BElfiTWOTHREE$ \\
$\lfiONE$ & $p=1$ & $\ElfiONE$ & $\ClfiONE$ & $\NPlfiONE$ & $\BElfiONE$\\
$\lfiTWO$ & $p=2$ & $\ElfiTWO$ & $\ClfiTWO$ & $\NPlfiTWO$ &
  $\BElfiTWO$ \\
$\lfiTHREE$ & $p=3$ & $\ElfiTHREE$ & $\ClfiTHREE$ & $\NPlfiTHREE$ &
  $\BElfiTHREE$ \\
$\lfiFOUR$ & $p=4$ & $\ElfiFOUR$ & $\ClfiFOUR$ & $\NPlfiFOUR$ &
  $\BElfiFOUR$ \\
\end{baytabular}
\end{center}

\subsection{Mixed Large Field Inflation (MLFI)} 

This model possesses the following potential~\cite{Martin:2013tda} 
\begin{equation}
V(\phi)=M^4\left(\frac{\phi}{\Mp}\right)^2\left(1+\alpha 
\frac{\phi^2}{\Mp^2}\right).
\end{equation}
Beside the usual mass scale $M$ fixed by the CMB normalisation, MLFI
contains only one parameter, $\alpha$. Since the order of magnitude of
this parameter is a priori unknown, a Jeffreys prior on $\alpha $ is
assumed. In practice, when $\alpha<10^{-5}$, the likelihood is
numerically very close to that of LFI${}_2$ and when $\alpha>10$, the
likelihood is numerically very close to that of LFI${}_4$. As a
consequence, we take the prior given in the following table:

\begin{center}
\begin{baytabular}
$\gmlfiTWOTWO$ & $\log\left(\alpha\right)\in[-5,1]$ & 
$\EgmlfiTWOTWO$ & $\CgmlfiTWOTWO$ & $\NPgmlfiTWOTWO$ & $\BEgmlfiTWOTWO$ \\
\end{baytabular}
\end{center}

\subsection{Radiatively Corrected Massive Inflation (RCMI)}

The potential of this model is given by~\cite{Martin:2013tda}
\begin{equation}
V(\phi)=M^4\left(\frac{\phi}{\Mp}\right)^2\left[1-2\alpha 
\frac{\phi^2}{\Mp^2}\ln\left(\frac{\phi}{\Mp}\right)\right].
\end{equation}
It depends on one parameter, $\alpha$, which represents the amplitude
of the radiative corrections to the potential of the LFI${}_2$
scenario. Since the one-loop correction can vary over many orders of
magnitude, it is meaningful to choose a Jeffreys prior on
$\alpha$. Then, clearly one must require $\alpha\ll 1$ in order for
the perturbative expansion to be under control. On the other hand, the
shape of the potential has been derived under the assumption that
fermion loops dominate over self-interaction loops. This implies a
lower bound on $\alpha$, namely
$\alpha>10^{-15}$~\cite{Martin:2013tda}. However, when
$\alpha<10^{-7}$, the likelihood is numerically very close to that of
LFI${}_2$ and, therefore, it is not necessary to consider smaller
values of $\alpha$. There also exists an upper bound on $\alpha$
coming from the requirement of having a sufficient number of e-folds
during inflation, $\alpha <6\times 10^{-4}$. Moreover, when $\alpha >
10^{-3}$, the likelihood is so small that the evidence cannot be
properly computed. As a consequence, an upper bound on $\alpha$ of
$\simeq 10^{-3}$ seems to be an appropriate choice. Our choice is
summarised in the following table:

\begin{center}
\begin{baytabular}
$\rcmi$ & $\log\left(\alpha\right)\in[-7,-3]$ & 
$\Ercmi$ & $\Crcmi$ & $\NPrcmi$ & $\BErcmi$ \\
\end{baytabular}
\end{center}

\subsection{Radiatively Corrected Quartic Inflation (RCQI)} 

This model is a quartic large field model LFI${}_4$ plus radiative
corrections~\cite{Martin:2013tda}. The potential reads
\begin{equation}
V(\phi)=M^4\left(\frac{\phi}{\Mp}\right)^4\left[1-\alpha 
\ln\left(\frac{\phi}{\Mp}\right)\right].
\end{equation}
The amplitude of these corrections is controlled by the parameter
$\alpha$. As discussed in the previous subsection, the order of
magnitude of $\alpha$ is not known and, therefore, a Jeffreys prior
must be chosen. Moreover, the perturbative expansion making sense only
if the radiative correction is small, one must have $\alpha\ll 1$. The
physical prior is therefore $\log(\alpha)\in[-\infty,0]$. However, in
practice, when $\alpha<10^{-3}$, the likelihood is numerically very
close to that of LFI${}_4$ and when $\alpha > 10^{-0.1}$, the
likelihood is so small that it cannot be computed in a reliable
way. Hence, the prior that we choose is the one indicated in the
following table.

\begin{center}
\begin{baytabular}
$\rcqi$ & $\log\left(\alpha\right)\in[-3,-0.1]$ & 
$\Ercqi$ & $\Crcqi$ & $\NPrcqi$ & $\BErcqi$ \\
\end{baytabular}
\end{center}

\subsection{Natural Inflation (NI)} 

This is a one parameter model and the potential is given by
\cite{Martin:2013tda}
\begin{equation}
V(\phi)=M^4\left[1+\cos\left(\frac{\phi}{f}\right)\right].
\end{equation}
The order of magnitude of the free parameter $f$ is not known and,
therefore, a Jeffreys prior is chosen. Moreover, the model is
compatible with the CMB only if the mass scale $f$ is
super-Planckian. It is not clear whether this condition makes sense at
the fundamental level but, from the effective field point of view,
several mechanisms have been invented such that this condition can be
realised. In this situation $f$ can scale from a few $\Mp$ to $\sim
100 \Mp$, hence the prior $\log(f/\Mp)\in[0,2.5]$, see the following
table.

\begin{center}
\begin{baytabular}
$\nati$ & $\log\left(f/\Mp\right)\in[0,2.5]$ & 
$\Eni$ & $\Cni$ & $\NPni$ & $\BEni$ \\
\end{baytabular}
\end{center}

\subsection{Exponential SUSY Inflation (ESI)}

The potential of this model can be written as~\cite{Martin:2013tda}
\begin{equation}
V(\phi)=M^4\left(1-\ee ^{-q\phi/\Mp}\right),
\end{equation}
where $q$ is a free parameter. A priori, different priors on $q$ are
possible and this gives rise to different versions of this
scenario. If we view ESI as a phenomenological model, then one can
assume that the parameter $q$ is a free $\order{1}$ quantity. In that
case, a natural prior is $q\in[0.1,6]$. But one can also assume that
the order of magnitude of $q$ is not known (in the following, we
denote the corresponding version of the scenario by $\esil$). In this
situation, we must choose a Jeffreys prior, typically
$\logdec(q)\in[-3,3]$. However, when $q>1$, the model is numerically
difficult to track since it produces a too weak level of gravity
waves. Moreover, in this regime, the likelihood reaches a stationary
value. Therefore, as explained before, one can restrict ourselves to
the numerical prior $\logdec(q)\in[-3,1]$.

Another possible prior is based on the original derivation of the
$\esi$ scenario (we denote this version by $\esio$ in what
follows). Indeed, in that case, the model is based on supergravity and
one has $q=\sqrt{2/\beta}$, where $\beta$ is the coefficient which
appears in front of the K\"ahler potential of the model. Hence, it
seems reasonable to assume that this quantity is a coefficient of
order one. This justifies our choice for the ``original'' prior,
namely $\beta\in [1,4]$. Of course, specific values of $\beta $ are
also very relevant. In particular, $\beta=1$ or $\beta=3$ represents
the cases where the inflaton field is either a dilaton or a moduli
($\beta=3$ corresponds to the "no scale" structure). In the following,
we denote these versions of the $\esi$ scenario by $\esisqrtTWO$ and
$\esisqrtTWOTHREE$, respectively.

\begin{center}
\begin{baytabular}
$\esi$ & $q\in\left[0.1,6\right]$ & $\Eesi$ & $\Cesi$ & $\NPesi$ & $\BEesi$ \\
$\esil$ & $\logdec\left(q\right)\in[-3,1]$ & $\Eesil$ & $\Cesil$ &
  $\NPesil$ & $\BEesil$ \\
$\esio$ & $\beta=2/q^2\in[1,4]$ & $\Eesio$ & $\Cesio$ & $\NPesio$ &
  $\BEesio$ \\
$\esisqrtTWO$ & $q=\sqrt{2}$ & $\EesisqrtTWO$ & $\CesisqrtTWO$ &
  $\NPesisqrtTWO$ & $\BEesisqrtTWO$ \\
$\esisqrtTWOTHREE$ & $q=\sqrt{2/3}$ & $\EesisqrtTWOTHREE$ &
  $\CesisqrtTWOTHREE$ & $\NPesisqrtTWOTHREE$ & $\BEesisqrtTWOTHREE$ \\
\end{baytabular}
\end{center}

\subsection{Power Law Inflation (PLI)}

The potential of this class of models can be expressed as~\cite{Martin:2013tda}
\begin{equation}
V(\phi)=M^4\ee ^{-\alpha\phi/\Mp},
\end{equation}
where $\alpha$ is a positive coefficient. A priori, it is a small
quantity the order of magnitude of which is not known. As a
consequence, a Jeffreys prior seems to be the most natural choice and
we take $\logdec(\alpha)\in[-4,0]$. On a more phenomenological
viewpoint, inflation occurs when $\alpha<\sqrt{2}$ only and,
therefore, it makes also sense to choose a flat prior on $\alpha $,
namely $\alpha\in[0,\sqrt{2}]$ (in the following, we denote this
version of power law inflation by $\plip$). However, when
$\alpha>1.1$, the likelihood is so small that it cannot be properly
calculated. Hence, we will restrict ourselves to the prior
$\alpha\in[0,1.1]$.

\begin{center}
\begin{baytabular}
$\pli$ & $\logdec\left(\alpha\right)\in\left[-4,0\right]$ & $\Epli$ &
  $\Cpli$ & $\NPpli$ & $\BEpli$ 
\\
$\plip$ & $\alpha\in\left[0,1.1\right]$ & $\Eplip$ & $\Cplip$ &
$\NPplip$ & $\BEplip$ \\
\end{baytabular}
\end{center}

\subsection{K\"ahler Moduli Inflation (KMII)}

The potential of $\kmii$ inflation is given by~\cite{Martin:2013tda}
\begin{equation}
V(\phi)=M^4\left(1-\alpha\frac{\phi}{\Mp}\ee^{-\phi/\Mp}\right),
\end{equation}
where $\alpha $ is a free positive coefficient. As discussed in detail
in \Refc{Martin:2013tda}, in order for inflation to end by slow-roll
violation, one must have $\alpha \gtrsim 2.4095$. On the other hand,
the order of magnitude of this parameter is unspecified and this
suggests a Jeffreys prior on $\alpha $. Combining these two pieces of
information, we are led to the prior
$\logdec(\alpha)\in[\logdec(2.4095)\simeq 0.382,4]$.

On the other hand, one can also choose $\alpha $ such that the
potential is positive everywhere, as opposed to the previous situation
where, for some values of the field, the potential can be negative and
where one makes use of a finite portion of it only (the corresponding
version of the scenario is denoted by $\kmiivp$ in the following). In
that case, one has the extra condition $\alpha<e\simeq 2.7183$. Since
$e$ is close to $2.4095$, a Jeffreys prior no longer makes sense and a
linear prior now seems a sensible choice. Hence our second choice
$\alpha\in[2.4095,e\simeq 2.7183]$. Everything is summarised in the
following table:

\begin{center}
\begin{baytabular}
$\kmii$ & $\logdec\left(\alpha\right)\in[0.382,4]$ & $\Ekmii$ &
  $\Ckmii$ & $\NPkmii$ & $\BEkmii$
\\
$\kmiivp$ & $\alpha\in[2.4095,2.7183]$ & $\Ekmiivp$ & $\Ckmiivp$ &
$\NPkmiivp$ & $\BEkmiivp$ \\
\end{baytabular}
\end{center}

\subsection{Horizon Flow Inflation at first order (HF1I)}

The potential of $\hfONEi$ inflation reads~\cite{Martin:2013tda}
\begin{equation}
V(\phi)=M^4\left(1+A_1\frac{\phi}{\Mp}\right)^2\left[1-\frac23
\left(\frac{A_1}{1+A_1\phi/\Mp}\right)^2\right].
\end{equation}
This model is obtained by an integration of the horizon flow equations
truncated at a given order (here at second order). As such, this
scenario is in fact purely phenomenological. Moreover, it turns out
that the observational predictions are not very sensitive to the value
of the free parameter $A_1$. Therefore, since its order of magnitude
is not fixed, it makes sense to choose a Jeffrey prior on $A_1$ and we
take $\logdec(A_1)\in[-3,3]$ as indicated below.

\begin{center}
\begin{baytabular}
$\hfONEi$ & $\logdec\left(A_1\right)\in[-3,3]$ & $\EhfONEi$ &
  $\ChfONEi$ & $\NPhfONEi$ & $\BEhfONEi$
\\
\end{baytabular}
\end{center}

\subsection{Coleman Weinberg Inflation (CWI)} 

Coleman Weinberg inflation is based on the following
potential~\cite{Martin:2013tda}
\begin{equation}
V(\phi)=M^4\left[1+\alpha\left(\frac{\phi}{Q}\right)^4
\ln\left(\frac{\phi}{Q}\right)\right],
\end{equation}
with $\alpha =4e$ in order to have a vanishing minimum. The shape of
$V(\phi)$ is therefore characterised by only one parameter, $Q$. In
the original version of the scenario, $Q$ is fixed by the GUT scale,
$Q\sim 10^{14}-10^{15} \GeV$. Therefore, in this case, it is natural
to choose a flat prior on $Q$ (we denote this version of the scenario
by $\cwif$). On the other hand, if one considers a more general
situation, then there is a priori no criterion to fix the value (or
the order of magnitude) of $Q$ and, therefore, this justifies the
choice of a Jeffreys prior, namely $\logdec(Q/\Mp)\in[-5,-3]$ (we
denote the version of the scenario by $\cwil$).

\begin{center}
\begin{baytabular}
$\cwif$ & $Q/\Mp\in[5\times 10^{-5},5\times 10^{-4}]$ & $\Ecwif$ &
  $\Ccwif$ & $\NPcwif$ & $\BEcwif$
\\
$\cwil$ & $\logdec\left(Q/\Mp\right)\in[-5,-3]$ & $\Ecwil$ & $\Ccwil$
& $\NPcwil$ & $\BEcwil$
\\
\end{baytabular}
\end{center}

\subsection{Loop Inflation (LI)}

The potential of $\li$ inflation can be written
as~\cite{Martin:2013tda}
\begin{equation}
V(\phi)=M^4\left[1+\alpha\ln\left(\frac{\phi}{\Mp}\right)\right],
\end{equation}
where the parameter $\alpha$ controls the strength of the one loop
correction to the tree level $V(\phi)$ (here the constant term) and
must therefore be such that $\alpha\ll 1$. When $\alpha<0$, in order
to have a sufficient number of $e$-folds, one must require
$\alpha>\alphamin\simeq -0.3$~\cite{Martin:2013tda}. In principle, the
model makes sense only if inflation proceeds at sub-Planckian \vev's
which is, strictly speaking, not possible in this regime. If we allow
\vev's larger than the Planck mass, typically up to $\phi/\Mp\simeq
1000$, then this sets an additional condition, namely
$\alpha<-0.1$. When $\alpha >0$, there is no extra condition on
$\alpha $ except, as already signaled, that $\alpha $ must be small in
order for the perturbative expansion to make sense.

From the previous considerations, we assume a flat prior
$\alpha\in[-0.3,-0.1]$ in the case where $\alpha<0$ (we denote this
version of the scenario by $\lin$). We have seen that, when
$\alpha>0$, there exists no restrictions on this parameter. In
particular, its order of magnitude is not specified and, therefore, it
makes sense to choose a Jeffreys prior, namely
$\logdec\left(\alpha\right)\in[\logdec(0.003),\logdec(0.3)]$ (in the
following, this version of the scenario is denoted by
$\lip$). Finally, when the sign is left unspecified, we simply
consider a flat prior $\alpha\in [\alphamin,-0.1]\cup [0,0.3]$. These
priors are summarised in the following table.

\begin{center}
\begin{baytabular}
$\li$ & $\alpha\in[\alphamin,-0.1]\cup [0,0.3]$ & $\Eli$ & $\Cli$ &
  $\NPli$ & $\BEli$
\\
$\lip$ & $\logdec\left(\alpha\right)\in[\logdec(0.003),
\logdec(0.3)]$ & $\Elip$ & $\Clip$ & $\NPlip$ & $\BElip$
\\
$\lin$ & $\alpha\in[\alphamin,-0.1]$ & $\Elin$ & $\Clin$ & $\NPlin$ & $\BElin$
\\
\end{baytabular}
\end{center}

\subsection{$R+R^{2p}$ Inflation (RpI)}

The potential of $R+R^{2p}$ inflation can be expressed as
\begin{equation}
V(\phi)=M^4\ee^{-2 \sqrt{2/3}\phi/\Mp} \left|\ee^{\sqrt{2/3}\phi/\Mp} 
 - 1 \right|^{2p/(2p-1)},
\end{equation}
which depends on the parameter $p$. The case $p=1$ is peculiar and
corresponds to Higgs Inflation (HI). It has been shown in
\Refc{Martin:2013tda} that, if $p$ takes integer values different from
$p=1$, then the model is ruled out since it leads to values of $r$ and
$\nS$ that are not compatible with the Planck data. As a consequence,
$p$ must be sufficiently close to $1$, and therefore must be a real
number. When $p>1$, the potential possesses a maximum located at
\begin{equation}
\frac{\phimax}{\Mp}=\sqrt{\frac{3}{2}}\ln \left(\frac{2p-1}{p-1}\right).
\end{equation}
and two regimes of inflation exist (denoted by $\rpiONE$ and $\rpiTWO$
in what follows) depending on whether inflation takes place in $\phi
\in[0,\phimax]$ or in $\phi \in[\phimax,\infty]$. In the first case,
inflation stops by slow-roll violation and the model is therefore a
one parameter model. In the second case, however, inflation must stop
by instability at $\phiend$ and, hence, the corresponding model is in
fact a two parameters model, $p$ and $\phiend$. Since $p$ must be
close to one, we choose the flat prior $p\in[1,1.5]$. In the case of
$\rpiTWO$, the order of magnitude of $\phiend$ being unspecified, we
take the following Jeffreys prior on $\phiend$:
$\logdec(\phiend/\phimax)\in[0.5,2]$.

If $p<1$, then there is a single regime where inflation can
proceed. It is denoted by $\rpiTHREE$ in what follows. In that case,
inflation stops by violation of the slow-roll conditions and,
therefore, the model is a one parameter model. As a consequence, we
choose to consider the following flat prior on $p$: $p\in[0.8,1]$. 

\begin{center}
\begin{baytabular}
$\rpiONE $ & $p\in[1,1.5]$ & $\ErpiONE $ & $\CrpiONE$ & $\NPrpiONE$ &
  $\BErpiONE$ \\
\hline
\bicenter{$\rpiTWO$} & $p\in[1,1.5]$ & \bicenter{$\ErpiTWO$} 
& \bicenter{$\CrpiTWO$} & \bicenter{$\NPrpiTWO$} & \bicenter{$\BErpiTWO$} \\
& $\logdec(\phiend/\phimax)\in[0.8,1]$ & & & &\\
\hline
$\rpiTHREE $ & $p\in[0.8,1]$ & $\ErpiTHREE $ & $\CrpiTHREE $ &
$\NPrpiTHREE$ & $\BErpiTHREE$
\\
\end{baytabular}
\end{center}

\subsection{Double Well Inflation (DWI)}

Double Well inflation is a one parameter model characterised by the
following potential
\begin{equation}
V(\phi)=M^4\left[\left(\frac{\phi}{\phizero}\right)^2-1\right]^2.
\end{equation}
As shown in \Refc{Martin:2013tda}, slow-roll inflation takes place
only if $\phizero/\Mp>2\sqrt{2}$. On the other hand, \COBE normalising
the model allows us to express the mass scale $M$ in terms of the free
parameter $\phizero$. Then, the requirement $M/\Mp<1$ leads to to the
constraint $\phizero/\Mp\lesssim 10^{5}$. As a consequence, a Jeffreys
logarithmic prior on $\phizero$ is chosen, namely
$\logdec(\phizero/\Mp)\in[\logdec(2\sqrt{2})\simeq 0.45,5]$.

\begin{center}
\begin{baytabular}
$\dwi$ & $\logdec\left(\phi_0/\Mp\right)\in[\logdec(2\sqrt{2}),5]$ 
& $\Edwi$ & $\Cdwi$ & $\NPdwi$ & $\BEdwi$
\\
\end{baytabular}
\end{center}

\subsection{Mutated Hilltop Inflation (MHI)} 

The potential of Mutated Hilltop inflation is given by
\begin{equation}
V(\phi)=M^4\left[1-{\sech} \left(\frac{\phi}{\mu} \right)\right],
\end{equation}
and depends on one free parameter, $\mu$. This model is
phenomenological although it is supposed to emerge from supergravity
considerations. In this last case, only sub-Planckian values for $\mu$
probably make sense. This is the reason why it seems interesting to
consider different priors. Given that the order of magnitude of
$\mu/\Mp$ is not specified, we take three Jeffreys priors
corresponding to situations where $\mu$ is sub-Planckian (denoted by
$\mhil$), super-Planckian (denoted by $\mhis$) or not
specified. Those choices are summarised in the following table:

\begin{center}
\begin{baytabular}
$\mhi$ & $\logdec\left(\mu/\Mp\right)\in[-2,2]$ & $\Emhi$ & $\Cmhi$ &
  $\NPmhi$ & $\BEmhi$
\\
$\mhil$ & $\logdec\left(\mu/\Mp\right)\in[-2,0]$ & $\Emhil$ & $\Cmhil$
& $\NPmhil$ & $\BEmhil$
\\
$\mhis$ & $\logdec\left(\mu/\Mp\right)\in[0,2]$ & $\Emhis$ & $\Cmhis$
& $\NPmhis$ & $\BEmhis$
\\
\end{baytabular}
\end{center}

\subsection{Radion Gauge Inflation (RGI)} 

The potential of Radion Gauge inflation can be expressed as
\begin{equation}
V(\phi)=M^4\frac{\left(\phi/\Mp\right)^2}{\alpha+\left(\phi/\Mp\right)^2},
\end{equation}
where $\alpha $ is a dimensionless positive parameter. A priori,
smaller than unity values are preferred but, at the same time,
$\alpha>1$ is not forbidden. This is why it is interesting to study
how the Bayesian evidence of the model depends on the range of
variation of $\alpha$. Let us also notice that the order of magnitude
of this parameter is not specified. As a consequence, we choose three
Jeffreys priors, one such that $\logdec(\alpha)\in[-4,4]$, one
corresponding to a situation where $\alpha<1$, namely
$\logdec(\alpha)\in[-4,0]$ (and we denote this version of the model by
$\rgis$) and one corresponding to $\alpha>1$, namely
$\logdec(\alpha)\in[0,4]$ (this version being referred to as
$\rgil$). Finally, in \Refc{delaMacorra:1995qh}, the potential of
Radion Gauge inflation was also obtained in the context of S-dual
superstring models. In that case, the value of $\alpha $ is fixed and
given by $\alpha=1/16$ which leads to a fourth choice of
prior. Everything is summarised in the following table:

\begin{center}
\begin{baytabular}
$\rgi$ & $\logdec\left(\alpha\right)\in[-4,4]$ & $\Ergi$ & $\Crgi$ &
  $\NPrgi$ & $\BErgi$
\\
$\rgis$ & $\logdec\left(\alpha\right)\in[-4,0]$ & $\Ergis$ & $\Crgis$
& $\NPrgis$ & $\BErgis$
\\
$\rgil$ & $\logdec\left(\alpha\right)\in[0,4]$ & $\Ergil$ & $\Crgil$ &
$\NPrgil$ & $\BErgil$
\\
$\rgiONEONESIX $ & $\alpha=1/16$ & $\ErgiONEONESIX$ & $\CrgiONEONESIX$
& $\NPrgiONEONESIX$ & $\BErgiONEONESIX$
\\
\end{baytabular}
\end{center}

\subsection{MSSM Inflation (MSSMI)} 
\label{subsec:priormssmi}

In this scenario, inflation occurs along a flat direction of the MSSM
potential. This flat direction is usually lifted by higher order
non-renormalisable operators and SUSY soft terms. As a consequence,
one can show that the potential takes the form~\cite{Martin:2013tda}
\begin{equation}
V(\phi)=M^4\left[\left(\frac{\phi}{\phi_0}\right)^2-\frac{2}{3}
  \left(\frac{\phi}{\phi_0}\right)^6+\frac{1}{5}\left(
  \frac{\phi}{\phi_0}\right)^{10}\right],
\end{equation}
where $\phi_0$ is a free parameter which can be expressed as
\begin{equation}
\phi_0^8=\frac{\Mp^6m_\phi^2}{10\lambda_6^2}.
\end{equation}
The quantity $\lambda_6$ is a coupling constant that is taken to be of
order one while $m_\phi$ is a soft breaking mass and, thus, is chosen
to be around $\simeq 1\TeV$. As a consequence, one has $\phizero\simeq
10^{14}\GeV$. In this original form of the scenario (denoted in what
follows by $\mssmio$), it is therefore natural to take a flat prior on
$\phizero$ such that $\phizero/\Mp\in[2\times 10^{-5},2\times
10^{-4}]$.

This model can also be viewed as a phenomenological inflection point
potential (denoted by $\mssmip$) where the value of $\phizero$ is not
fixed by high energy physics considerations. In that case, a Jeffreys
prior on $\phizero$ is appropriate and, here, we take
$\logdec(\phizero/\Mp)\in[-3,3]$.

\begin{center}
\begin{baytabular}
$\mssmio$ & $\phi_0/\Mp\in[2\times 10^{-5},2\times 10^{-4}]$ &
  $\Emssmio$ & $\Cmssmio$ & $\NPmssmio$ & $\BEmssmio$
\\
$\mssmip$ & $\logdec\left(\phi_0/\Mp\right)\in[-3,3]$ & $\Emssmip$ &
$\Cmssmip$ & $\NPmssmip$ & $\BEmssmip$ 
\\
\end{baytabular}
\end{center}

\subsection{Renormalisable Inflection Point Inflation (RIPI)} 
\label{subsec:priorripi}

This model is derived in the same context as MSSM inflation except
that an additional term in the superpotential involving right handed
neutrinos is considered. The amplitude of this new term is controlled
by the dimensionless coupling constant $h\simeq 10^{-12}$. This gives
rise to a new flat direction parametrised by the inflaton field
$\phi$. This flat direction is lifted by the same mechanism discussed
previously in the context of MSSM inflation and leads to the following
potential~\cite{Martin:2013tda}
\begin{equation}
V(\phi)=M^4\left[\left(\frac{\phi}{\phizero}\right)^2 -
   \frac{4}{3} \left( \frac{\phi}{\phizero} \right)^3 + \frac{1}{2}
    \left( \frac{\phi}{\phizero} \right)^4\right],
\end{equation}
where 
\begin{equation}
\phizero=\sqrt{3}\frac{m_\phi}{h},
\end{equation}
$m_\phi$, as a soft breaking mass, being between $100\GeV$ and
$10\TeV$. As a consequence, one has $\phizero \sim 10^{14}\GeV$. For
this version of the model (denoted as the ``original version'',
$\ripio$), a flat prior on $\phizero$ represents the preferred choice,
$\phizero/\Mp\in[2\times 10^{-5},2\times 10^{-4}]$. As for MSSM
inflation, however, one can also see this scenario as a
phenomenological scenario where $\phizero $ is not specified (denoted
by $\ripip$ in what follows). In this case, a Jeffreys prior on
$\phizero$ is natural and we take
$\logdec(\phizero/\Mp)\in[-3,3]$. Finally, the above potential can
also arise in a supergravity framework with shift symmetry in the
K\"ahler potential (denoted by $\ripiS$) which allows for
super-Planckian \vev of the inflaton field $\phi$. For this reason, we
also consider the prior $\phizero/\Mp\in[10,50]$.

\begin{center}
\begin{baytabular}
$\ripio$ & $\phizero/\Mp\in[2\times 10^{-5},2\times
10^{-4}]$ & $\Eripio$ & $\Cripio$ & $\NPripio$ & $\BEripio$
\\
$\ripip$ & $\logdec\left(\phi_0/\Mp\right)\in[-3,3]$ & $\Eripip$ &
$\Cripip$ & $\NPripip$ & $\BEripip$
\\
$\ripiS$ & $\phizero/\Mp\sim [10,50]$ & $\EripiS$ & $\CripiS$ &
$\NPripiS$ & $\BEripiS$
\\
\end{baytabular}
\end{center}

\subsection{Arctan Inflation (AI)} 

The potential of $\ai$ can be expressed as
\begin{equation}
V(\phi)=M^4\left[1-\frac{2}{\pi}
    \arctan\left(\frac{\phi}{\mu}\right)\right],
\end{equation}
where $\mu$ is a free parameter. As shown in \Refc{Martin:2013tda},
inflation stops by slow-roll violation only if
$\mu/\Mp<0.512378$. This model is purely phenomenological and, as a
consequence, the scale $\mu$ is not fixed by any high energy physics
considerations. As a consequence, its order of magnitude is a priori
unspecified. Therefore, we choose a Jeffreys logarithmic prior on
$\mu$, namely
$\logdec(\mu/\Mp)\in[-3,\logdec(0.51\cdots)\simeq-0.29]$.

\begin{center}
\begin{baytabular}
$\ai$ & $\logdec\left(\mu/\Mp\right)\in[-3,-0.29]$  & $\Eai$ & $\Cai$
  & $\NPai$ & $\BEai$
\\
\end{baytabular}
\end{center}

\subsection{Constant ns A Inflation (CNAI)} 

The potential of $\cnai$ is given by the following expression
\begin{equation}
V(\phi)=M^4\left[3-\left(3+\alpha^2 \right) \tanh^2
    \left( \frac{\alpha}{\sqrt{2}} \frac{\phi}{\Mp} \right)\right],
\end{equation}
where $\alpha $ is a dimensionless free parameter. It was demonstrated
in \Refc{Martin:2013tda} that slow-roll inflation takes place provided
$\alpha<\alphamax\simeq 0.66$. This model is phenomenological and
is not based on high energy physics. It is in fact designed to produce
an exact power law spectrum of density perturbations. As a
consequence, the order of magnitude of $\alpha$ is not specified and
one chooses to work with a Jeffreys prior
$\logdec(\alpha)\in[-4,\logdec(\alphamax)\simeq -0.18]$.

\begin{center}
\begin{baytabular}
$\cnai$ & $\logdec\left(\alpha\right)\in[-4,-0.18]$  & $\Ecnai$ &
  $\Ccnai$ & $\NPcnai$ & $\BEcnai$
\\
\end{baytabular}
\end{center}

\subsection{Constant ns B Inflation (CNBI)} 

This model is very similar to $\cnai$ inflation. It is also a
phenomenological scenario designed to produce a constant spectral
index and also depends on one dimensionless parameter $\alpha$. The
corresponding potential can be expressed as
\begin{equation}
V(\phi)=M^4\left[\left(3-\alpha^2\right) \tan^2
    \left(\frac{\alpha}{\sqrt{2}}\frac{\phi}{\Mp} \right)-3\right].
\end{equation}
It was shown in \Refc{Martin:2013tda} that slow-roll inflation takes
place if $\alpha<\alphamax\simeq 0.2975$. If one CMB normalises the
model, then one can express the mass scale $M$ in terms of
$\alpha$. It follows that the requirement $M/\Mp<1$ implies
$\alpha\lesssim 10^{-9}$. As a consequence, we should take a Jeffreys
prior on $\alpha$, namely
$\logdec(\alpha)\in[-9,\logdec(\alphamax)\simeq-0.527]$. In practice,
however, when $\alpha>10^{-1.4}$, the likelihood is so small that it
cannot be properly calculated. Moreover, when $\alpha<10^{-5}$, the
value of the likelihood reaches a numerical stationary value and,
therefore, it is not necessary to numerically calculate it beyond that
point. As a consequence, we consider the following prior:
$\logdec(\alpha)\in[-5,-1.4]$.

\begin{center}
\begin{baytabular}
$\cnbi$ & $\logdec\left(\alpha\right)\in[-5,-1.4]$ & $\Ecnbi$ &
  $\Ccnbi$ & $\NPcnbi$ & $\BEcnbi$
\\
\end{baytabular}
\end{center}

\subsection{Open String Tachyonic Inflation (OSTI)}  

In this model, the inflaton field is a tachyon field on a D3-brane. In
principle, its kinetic term is non-minimal but when higher order terms
are neglected, it becomes a standard slow-roll model with a potential
given by the following expression
\begin{equation}
V(\phi)=-M^4\left(\frac{\phi}{\phizero}\right)^2
\ln\left[\left(\frac{\phi}{\phizero}\right)^2\right].
\end{equation}
In the original version of the model, $\phizero$ is set to the string
scale $\phizero\sim M_{\us}$. However, $\phizero$ can also be viewed
as a free sub-Planckian scale. In that case, a Jeffreys prior is
appropriate, for instance $\logdec(\phizero/\Mp)\in[0,4]$. However,
when $\phizero/\Mp<10$, the likelihood is so small that it cannot be
numerically calculated in a reliable way. As a consequence, in what
follows, we consider the prior $\logdec(\phizero/\Mp)\in[1,4]$ only.

\begin{center}
\begin{baytabular}
$\osti$ &  $\logdec\left(\phi_0/\Mp\right)\in[1,4]$ & $\Eosti$ &
  $\Costi$ & $\NPosti$ & $\BEosti$ 
\\
\end{baytabular}
\end{center}

\subsection{Witten-O'Raifeartaigh Inflation (WRI)}

The potential of WRI inflation can be expressed as
\begin{equation}
V(\phi)=M^4\ln\left(\frac{\phi}{\phizero}\right)^2.
\end{equation}
When the high energy justifications of the scenario are considered,
the condition $\phizero=\Mp$ holds. In what follows, we call this
version of the model the ``original WRI'' and we denote it as
$\wrio$. If this condition is relaxed (the corresponding version of
the model is then denoted by $\wrig$) and if the model is now viewed
as a more phenomenological one, then the order of magnitude and value
of $\phizero$ are unspecified and a Jeffreys prior is appropriate. We
choose $\logdec(\phizero/\Mp)\in[-3,3]$. These considerations are
summarised in the following table:

\begin{center}
\begin{baytabular}
$\wrio$ & $\phizero=\Mp$  & $\Ewrio$ & $\Cwrio$ & $\NPwrio$ & $\BEwrio$
\\
$\wrig$ & $\logdec\left(\phizero/\Mp\right)\in[-3,3]$ & $\Ewrig$ &
$\Cwrig$ & $\NPwrig$ & $\BEwrig$
\\
\end{baytabular}
\end{center}

\subsection{Small Field Inflation (SFI)}

Small field inflation is characterised by the following potential
\begin{equation}
V(\phi)=M^4\left[1-\left(\frac{\phi}{\mu}\right)^p\right],
\end{equation}
which depends on two parameters, the dimensionless index $p$ and the
mass scale $\mu$. In most of high energy physics implementations, only
the case $\mu<\Mp$ is sensible. It is, however, always possible to
take a more phenomenological point of view and also consider the case
$\mu>\Mp$. In what follows, for this reason, we will discuss a
``small'' version of the scenario for which
$\logdec(\mu/\Mp)\in[-1,0]$ and a ``large'' version for which
$\logdec(\mu/\Mp)\in[0,2]$. Two remarks are in order at this
point. Firstly, a Jeffreys prior is chosen on $\mu$ because, a priori,
its order of magnitude is unspecified. Secondly, in the small version
of the model, we only consider $\logdec(\mu/\Mp)\in[-1,0]$ (and not,
for instance, $\logdec(\mu/\Mp)\in[-2,0]$) because, when
$\mu/\Mp<0.1$, the likelihood is so small that it cannot be properly
numerically calculated.

The index $p$ is an $\order{1}$ parameter that can also take specific
integer values. We will treat the case where there is a flat prior on
$p$, namely $p\in[2,10]$, but also the case where $p$ has specific
values, $p=1$, $p=2$, $p=3$ and $p=4$. Let us also notice that for
$p=1$ and $p=2$, the small version of the SFI inflation does not exist
because slow-roll is violated in that case (for instance, for $p=2$,
one has $\epsilon_2 > 4$).

Our priors are summarised in the following table:

\begin{center}
\begin{baytabular}
\bicenter{$\sfi$} & $p\in\left[2,10\right]$  & 
\bicenter{$\Esfi$} & \bicenter{$\Csfi$} & \bicenter{$\NPsfi$} &
\bicenter{$\BEsfi$} 
\\
& $\logdec\left(\mu/\Mp\right)\in[-1,2]$ & & & &\\
\hline
\bicenter{$\sfis$} & $p\in\left[2,10\right]$ & 
\bicenter{$\Esfis$} & \bicenter{$\Csfis$} & \bicenter{$\NPsfis$} &
\bicenter{$\BEsfis$} 
\\
& $\logdec\left(\mu/\Mp\right)\in[-1,0]$ & & & &\\
\hline
\bicenter{$\sfil$} & $p\in\left[2,10\right]$ & 
\bicenter{$\Esfil$} & \bicenter{$\Csfil$} & \bicenter{$\NPsfil$} &
\bicenter{$\BEsfil$} 
\\
& $\logdec\left(\mu/\Mp\right)\in[0,2]$ & & & & \\
\hline
\bicenter{$\sfiONE$} & $p=1$ & 
\bicenter{$\EsfiONE$} & \bicenter{$\CsfiONE$} & \bicenter{$\NPsfiONE$}
& \bicenter{$\BEsfiONE$}
\\
& $\logdec\left(\mu/\Mp\right)\in[-1,2]$ & & & &\\
\hline
\bicenter{$\sfiTWO$} & $p=2$ & 
\bicenter{$\EsfiTWO$} & \bicenter{$\CsfiTWO$} & \bicenter{$\NPsfiTWO$}
& \bicenter{$\BEsfiTWO$}
\\
& $\logdec\left(\mu/\Mp\right)\in[-1,2]$ & & & & \\
\hline
\bicenter{$\sfiTWOl$} & $p=2$ & 
\bicenter{$\EsfiTWOl$} & \bicenter{$\CsfiTWOl$} &
\bicenter{$\NPsfiTWOl$} & \bicenter{$\BEsfiTWOl$}
\\
& $\logdec\left(\mu/\Mp\right)\in[0,2]$ & & & &\\
\hline
\bicenter{$\sfiTHREE$} & $p=3$ & 
\bicenter{$\EsfiTHREE$} & \bicenter{$\CsfiTHREE$} &
\bicenter{$\NPsfiTHREE$} & \bicenter{$\BEsfiTHREE$}
\\
& $\logdec\left(\mu/\Mp\right)\in[-1,2]$ & & & & \\
\hline
\bicenter{$\sfiTHREEs$} & $p=3$ & 
\bicenter{$\EsfiTHREEs$} & \bicenter{$\CsfiTHREEs$} &
\bicenter{$\NPsfiTHREEs$} & \bicenter{$\BEsfiTHREEs$}
\\
& $\logdec\left(\mu/\Mp\right)\in[-1,0]$ & & & & \\
\hline
\bicenter{$\sfiTHREEl$} & $p=3$ & 
\bicenter{$\EsfiTHREEl$} & \bicenter{$\CsfiTHREEl$} &
\bicenter{$\NPsfiTHREEl$} & \bicenter{$\BEsfiTHREEl$}
\\
& $\logdec\left(\mu/\Mp\right)\in[0,2]$ & & & &\\
\hline
\bicenter{$\sfiFOUR$} & $p=4$ & 
\bicenter{$\EsfiFOUR$} & \bicenter{$\CsfiFOUR$} &
\bicenter{$\NPsfiFOUR$} & \bicenter{$\BEsfiFOUR$}
\\
& $\logdec\left(\mu/\Mp\right)\in[-1,2]$ & & & &\\
\hline
\bicenter{$\sfiFOURs$} &$p=4$ & 
\bicenter{$\EsfiFOURs$} & \bicenter{$\CsfiFOURs$} &
\bicenter{$\NPsfiFOURs$} & \bicenter{$\BEsfiFOURs$}
\\
& $\logdec\left(\mu/\Mp\right)\in[-1,0]$ & & & &\\
\hline
\bicenter{$\sfiFOURl$} & $p=4$ & 
\bicenter{$\EsfiFOURl$} & \bicenter{$\CsfiFOURl$} &
\bicenter{$\NPsfiFOURl$} & \bicenter{$\BEsfiFOURl$}
\\
& $\logdec\left(\mu/\Mp\right)\in[0,2]$ & & & & \\
\end{baytabular}
\end{center}

\subsection{Intermediate Inflation (II)} 

Intermediate Inflation is a phenomenological model that can be defined
by demanding an equation of state during inflation of the form
\begin{equation}
\rho+p=\gamma \rho ^{\lambda},
\end{equation}
where $\gamma >0$ and $\lambda>1$ are dimensionless parameters, $\rho$
and $p$ being the energy density and pressure stored in the inflaton
field, respectively. This assumption is in fact equivalent to having a
scale factor given by $a(t)\propto \exp \left(At^f\right)$ where
\begin{equation}
f=\frac{2(1-\lambda)}{1-2\lambda}.
\end{equation} 
Given that $\lambda>1$, it follows that $0<f<1$. Finally, it is also
equivalent to postulate the following potential
\begin{equation}
V(\phi)=M^4\left[\left(\frac{\phi}{\Mp}\right)^{-\beta}
-\frac{\beta^2}{6}\left(\frac{\phi}{\Mp}\right)^{-\beta-2}\right],
\end{equation}
with 
\begin{equation}
\beta=4\left(\frac{1}{f}-1\right).
\end{equation}
In this scenario, inflation cannot stop by violation of the slow-roll
conditions and, hence, one needs to postulate an extra mechanism such
as tachyonic instability. This implies that the scenario depends on
another parameter, $\phiend$, the \vev at which inflation
ends. Intermediate inflation is therefore a two parameters models,
$\phiend$ and $\lambda$ (or $f$ or $\beta$).

Given the above considerations, one can choose to take a flat prior on
$\beta\in[0,10]$ (in the following, we denote the corresponding
version of the scenario by $\ii$). It makes also sense to work with a
flat prior on $f\in[0,1]$ (this version of the model is denoted
$\iif$). In fact, in order to avoid an infinite value of $\beta$, we
will consider the following prior $f\in[0.1,1]$. Finally, we also
investigate a Jeffreys prior on $\lambda$ (this version is denoted by
$\iilambda$), namely $\logdec(\lambda)\in[0.1,4]$, the lower bound
$\logdec(\lambda)>0.1$ being chosen to have finite values of $\beta$.

The prior on $\phiend$ also needs to be discussed. It was shown in
\Refc{Martin:2013tda} that the parameter $\xend=\phiend/\Mp$ must be
larger than some value $\xendmin$ in order to have a sufficient
number of $e$-folds during inflation. The parameter $\xend$ is only
known numerically and has been calculated in
\Refc{Martin:2013tda}. Moreover, the order of magnitude of $\xend$ is
not known and, therefore, this suggests a Jeffreys prior. As a
consequence, we take
$\logdec(\xend)\in[\logdec(\xendmin),4]$. Everything is
summarised in the following table.

\begin{center}
\begin{baytabular}
\bicenter{$\ii$} & $\beta\in\left[0,10\right]$ & 
\bicenter{$\Eii$} & \bicenter{$\Cii$} & \bicenter{$\NPii$} & \bicenter{$\BEii$}
\\
&  $\logdec\left(\xend\right)\in[\logdec\left(\xendmin\right),4]$ & &
& & \\
\hline
\bicenter{$\iif$} &  $f=1/\left(1+\beta/4\right)\in\left[0.1,1\right]$ & 
\bicenter{$\Eiif$} & \bicenter{$\Ciif$} & \bicenter{$\NPiif$} &
\bicenter{$\BEiif$}
\\
&  $\logdec\left(\xend\right)\in[\logdec\left(\xendmin\right),4]$ & &
& & \\
\hline
\bicenter{$\iilambda$} &  
$\logdec(\lambda)=\logdec\left(1+2/\beta\right)\in\left[0.1,4\right]$ & 
\bicenter{$\Eiilambda$} & \bicenter{$\Ciilambda$} &
\bicenter{$\NPiilambda$} & \bicenter{$\BEiilambda$}
\\
&  $\logdec\left(\xend\right)\in[\logdec\left(\xendmin\right),4]$ & &
& & \\
\end{baytabular}
\end{center}

\subsection{K\"ahler Moduli Inflation II (KMIII)} 

K\"ahler Moduli Inflation III is a stringy inspired scenario the
potential of which can be written as
\begin{equation}
V(\phi)=M^4\left[1-\alpha\left(\frac{\phi}{\Mp}\right)^{4/3}
\ee^{-\beta\left(\phi/\Mp\right)^{4/3}}\right].
\end{equation}
In this model, the inflaton field is a modulus field. The potential
depends on two parameters, $\alpha $ and $\beta $. As reviewed in
\Refc{Martin:2013tda}, the order of magnitude of the parameter $\beta
$ is in fact controlled by the compactification volume $\calV$. More
precisely, one can show that $\alpha=\order{\calV_\us^{5/3}}$
and $\beta=\order{\calV_\us^{2/3}}$ where $\calV_\us$ is
a dimensionless volume defined by
$\calV_\us=\calV/\ell_\us^6$, $\ell_\us$ being
the string length. Since typical values are usually chosen such that
$\calV_\us\sim 10^{6}$ and since the order of magnitude of
$\calV_\us$ is not precisely specified, we take a logarithmic
prior on $\calV_\us$, namely $\logdec(\calV)\in[5,7]$.

On the other hand, the ratio $\alpha/(\beta\calV)$ is a $\order{1}$
quantity, thanks to the scaling mentioned above. As a consequence, we
choose a flat prior $\alpha/(\beta\calV)\in[0.2,5]$. In practice, once
the number $\calV_\us$ is fixed, one calculate $\beta$ by means
of $\beta=\calV_\us^{2/3}$. Then, the ratio
$\alpha/(\beta\calV)$ is chosen and one deduces the value of $\alpha$.

\begin{center}
\begin{baytabular}
\bicenter{$\kmiii$} & $\logdec(\calV)\in[5,7]$ & 
\bicenter{$\Ekmiii$} & \bicenter{$\Ckmiii$} & \bicenter{$\NPkmii$} &
\bicenter{$\BEkmii$} 
\\
& $\alpha/(\beta\calV)\in[0.2,5]$  & & & &\\
\end{baytabular}
\end{center}

\subsection{Logamediate inflation (LMI)}

This model is a phenomenological model designed such that the scale
factor during inflation behaves as
\begin{equation}
\label{eq:scalefactorlmi}
a(t)=a_0\exp\left[A\left(\ln \frac{t}{t_0}\right)^{\lambda}\right], 
\end{equation}
where $A>0$ and $\lambda>1$ are two dimensionless parameters and $t_0$
is a third parameter the dimension of which is time. From this
expression of the scale factor, one can infer the shape of the
potential. Straightforward calculations~\cite{Martin:2013tda} lead to
\begin{equation}
\label{eq:potlmi}
V(\phi)=M^4\left(\frac{\phi}{\Mp}\right)^{4(1-\gamma)}
\exp\left[-\beta\left(\frac{\phi}{\Mp}\right)^\gamma\right]
\end{equation}
where the parameters $\gamma$ and $\beta $ can be expressed as
\begin{equation}
\label{eq:param}
\gamma=\frac{2}{\lambda+1}, \qquad
\beta=2\left(\frac{\lambda+1}{2\sqrt{2A\lambda}}\right)^{2/(\lambda+1)}.
\end{equation}
These relations, together with the conditions on $A$ and $\lambda$,
imply $0<\gamma\leq 1$ and $\beta >0$. The potential~(\ref{eq:potlmi})
has a maximum located at
\begin{equation}
x_\mathrm{max}\equiv \frac{\phi_\mathrm{max}}{\Mp}
=\left[\frac{4(1-\gamma)}{\beta \gamma}\right]^{1/\gamma}.
\end{equation}
This gives rise to two different versions of the
model~\cite{Martin:2013tda}: either inflation proceeds on the left
side of its maximum and the field \vev decreases during inflation (we
call this version LMI1 in the following) or it proceeds on the
right side of its maximum and the field \vev increases during
inflation (this version is denoted LMI2). In the case of
LMI1, inflation stops by slow-roll violation. The case of
LMI2 is more complicated but, in brief, one needs an extra
mechanism to end inflation and this introduces a new parameter in the
model, $\xend$, see \Refc{Martin:2013tda} for more details. LMI2
is therefore a three parameter model.

Regarding the priors, we essentially have two choices: either we
specify them on the parameters characterising the potential or we
specify them on the parameters controlling the behaviour of the scale
factor. In the following, we consider both cases.

Let us start with the case where we choose priors on the parameters of
the potential. In the following, we denote the two corresponding
versions of the scenario by $\lmiONEp$ and $\lmiTWOp$. For $\lmiONEp$,
it is natural to take a flat prior on $\gamma$, namely
$\gamma\in[0,1]$. In fact, $\gamma=0$ is numerically pathological and,
therefore, in practice, we consider $\gamma\in[0.1,1]$. For the
parameter $\beta $, one takes a flat prior
$\beta\in[0.01,\betamax(\gamma)]$, where
\begin{equation}
\betamax(\gamma)=2^{2-3\gamma/2}\left(0.1\right)^{\gamma/2}
\frac{(1-\gamma)^{1-\gamma/2}}{\gamma^{1+\gamma/2}}.
\end{equation}
As discussed in \Refc{Martin:2013tda}, the condition $\beta
<\betamax(\gamma)$ is mandatory in order for the slow-roll
conditions to be valid.

Let us now turn to $\lmiTWOp$. For this case, we also consider a flat
prior on $\gamma$, $\gamma\in[0.1,0.99]$. For this model, there is no
condition on $\beta $ in order to satisfy the slow-roll and,
therefore, one takes a flat prior on this parameter, namely
$\beta\in[0.01,10]$. Finally, the order of magnitude of $\xend$ is not
specified and this suggests a Jeffreys prior. Notice also that one
must have $\xend>\xendmin\left(\gamma,\beta, \Delta \Nmin\right)$ in
order to have at least $\Delta \Nmin$ e-folds during inflation
(typically $\Delta N _{\umin}\simeq 50$). Combining these two pieces
of information leads us to the following prior
$\logdec\left(\xend\right)\in[\logdec\left(\xendmin\right),
\logdec\left(100\,\xendmin\right)]$.

Let us now treat the case where the priors are chosen from
considerations based on the form of the scale
factor~(\ref{eq:scalefactorlmi}). We denote these versions $\lmiONEo$
and $\lmiTWOo$. This means that we first choose $A$ and $\lambda$ and
then infer $\gamma $ and $\beta $ from Eqs.~(\ref{eq:param}). For the
$\lmiONEo$ model, since $\lambda$ is a $\order{1}$ parameter, one
takes a flat prior on this parameter, namely $\lambda\in[1,6]$. For
the parameter $A$, one needs to take into account the fact that there
is a maximum value of $\beta $, see the above discussion. In fact, it
is possible to invert Eqs.~(\ref{eq:param}) and to express $A$ in
terms of $\beta $ and $\gamma$. One finds
\begin{equation}
A=\left(\frac{2}{\beta}\right)^{2/\gamma}
\left(\frac{2}{\gamma}\right)^2\frac{1}{8(2/\gamma-1)}.
\end{equation}
Given that $2/\gamma>1$, the presence of a $\betamax$ implies a
$A_\umin$ which can be expressed as
\begin{equation}
A_\umin=\left(\frac{2}{\betamax}\right)^{2/\gamma}
\left(\frac{2}{\gamma}\right)^2\frac{1}{8(2/\gamma-1)}.
\end{equation}
In addition, since the order of magnitude of $A$ is a priori not
fixed, one chooses to work with a Jeffreys prior. We therefore take
$\logdec(A)\in[A_\umin(\lambda),2]$.
 
Let us finally examine the $\lmiTWOo$ version. We take the same prior
on $\lambda$ and $\xend$ as before. Since there is no maximum value of
$\beta $ anymore, there is no minimal value of $A$. As a consequence,
we work with the following prior on $A$: $\logdec(A)\in[-2,2]$. 

All the above considerations are summarised in the following table:

\begin{center}
\begin{baytabular}
\bicenter{$\lmiONEp$} & $\gamma\in\left[0.1,1\right]$ & 
\bicenter{$\ElmiONEp$} & \bicenter{$\ClmiONEp$}  &
\bicenter{$\NPlmiONEp$} & \bicenter{$\BElmiONEp$}
\\
& $\beta\in[0.01,\betamax(\gamma)]$  & & & &\\
\hline
\bicenter{$\lmiONEo$} & $\lambda \in\left[1,6\right]$  & 
\bicenter{$\ElmiONEo$} & \bicenter{$\ClmiONEo$} &
\bicenter{$\NPlmiONEo$} & \bicenter{$\BElmiONEo$}
\\
& $\logdec\left(A\right) \in[A_\umin(\lambda),2]$ & & & & \\
\hline
& $\gamma\in\left[0.1,0.99\right]$ & & & & \\
$\lmiTWOp$ & $\beta\in[0.01,10]$  & $\ElmiTWOp$ & $\ClmiTWOp$ &
$\NPlmiTWOp$ & $\BElmiTWOp$ \\
& $\logdec\left(\xend\right)\in[\logdec\left(\xendmin\right),
\logdec\left(100\,\xendmin\right)]$ & & & &\\
\hline
& $\lambda \in\left[1.1,6\right]$ & & & & \\
$\lmiTWOo$ & $\logdec\left(A\right)\in[-2,2]$ & $\ElmiTWOo$ &
$\ClmiTWOo$ & $\NPlmiTWOo$ & $\BElmiTWOo$ \\
& $\logdec\left(\xend\right)\in[\logdec\left(\xendmin\right),
\logdec\left(100\,\xendmin\right)]$ & & & &\\
\end{baytabular}
\end{center}

\subsection{Twisted inflation (TWI)} 

The potential of Twisted Inflation (TWI) is given by the following
expression
\begin{equation}
V(\phi)=M^4\left[1-A\left(\frac{\phi}{\phizero}\right)^2
\ee^{-\phi/\phizero}\right],
\end{equation}
where the two parameters $M$ and $\phizero$ can be expressed as
\begin{equation}
M^4=\frac{8\calN }{A\pi^2(2\pi R)^4}, \quad 
\frac{\phizero}{\Mp}=\frac{1}{2\pi R\Mp}, 
\end{equation}
the constant $A$ being defined by $A=32/[93\zeta(5)]\simeq 0.33$. This
model is based on higher dimensional supersymmetric gauge theories,
more precisely $\mbox{U}(\calN)$ Yang-Mills theory, and $R$ represents
the radius of compactification. The above potential is valid provided
$R\Mp\gg 1$, that is to say $\phizero/\Mp\ll 1$. In fact, the model
makes sense if $\phi\ll \Mp$ for any \vev and not only $\phizero$, see
\Refc{Martin:2013tda} for more detail. Inflation cannot stop by
violation of the slow-roll conditions and, as a consequence, one needs
to introduce another mechanism which is characterised by a new
parameter, $\phiend$. TWI inflation is therefore a two parameter
model.

Let us now discuss the priors. We have just seen that $\phizero$ must
be sub-Planckian. Since its order of magnitude is a priori unknown, it
seems natural to take a Jeffreys prior, namely
$\logdec(\phizero/\Mp)\in[-4,-1]$. Concerning the $\vev$ at which
inflation ends, we know that $\phiend/\phizero>2$ because the minimum
of the potential is located at $\phi/\phizero=2$. Otherwise, as
already discussed, the only other constraint is $\phiend\ll
\Mp$. However, in practice, for values of $\phiend $ approaching the
Planck mass, the potential is so flat that this regime is already
strongly disfavoured (because $\nS\simeq 1$). Therefore, it is better
to choose an upper bound for $\logdec(\phiend/\phizero)$ supplemented
with the hard prior $\phiend<\Mp$. Then, one can study if the evidence
is changed if we modify the upper bound. Since the order of magnitude
of this parameter is a priori not specified, we must also take a
Jeffreys prior on $\phizero$. To summarise, we consider the two
following priors
$\logdec(\phiend/\phizero)\in[\logdec(2),\logdec(20)]$ and
$\logdec(\phiend/\phizero)\in[\logdec(2),\logdec(40)]$ and check that,
indeed, the final result is not sensitive to the upper bound. In the
following, we denote these priors by $\twiAONE$ and $\twiATWO$.

At the fundamental level, Twisted Inflation is in fact characterised
by $\calN$ and not by $\phizero$. If we CMB normalise the model, one
can express the latter in terms of the former, namely
$\phizero/\Mp\simeq 10^{-5}/\sqrt{\mathcal{N}}$. In this version of
the model, denoted by $\twiBONE$ and $\twiBTWO$ in what follows, the
prior choices are now fixed on $\calN$ (and the value of $\phizero$ is
calculated using the above equation). Since $\calN$ is a priori a
number of order one, it makes sense to take a flat prior and we choose
$\mathcal{N}\in[1,100]$. Concerning $\phiend$, we just take the same
priors as before.

\begin{center}
\begin{baytabular}
& $\logdec\left(\phizero/\Mp\right)\in[-4,-1]$ & & & & \\
$\twiAONE $ & $\logdec\left(\phiend/\phizero\right)\in[\logdec(2),
    \logdec(20)]$  
& $\EtwiAONE $ & $\CtwiAONE$ & $\NPtwiAONE$ & $\BEtwiAONE$ \\
& $\phiend<\Mp$ & & & &\\
\hline
& $\logdec\left(\phizero/\Mp\right)\in[-4,-1]$ & & & & \\
$\twiATWO $ & $\logdec\left(\phiend/\phizero\right)\in[\logdec(2), \logdec(40)]$ 
& $\EtwiATWO $ & $\CtwiATWO$ & $\NPtwiATWO$ & $\BEtwiATWO$ \\
& $\phiend<\Mp$ & & & &\\
\hline
&  $\mathcal{N}=10^{-10}(\phizero/\Mp)^{-2}\in[1,100]$ & & & & \\
$\twiBONE $ & $\logdec\left(\phiend/\phizero\right)\in[\logdec(2), \logdec(20)]$ 
& $\EtwiBONE $ & $\CtwiBONE$ & $\NPtwiBONE$ & $\BEtwiBONE$ \\ 
& $\phiend<\Mp$ & & & &\\
\hline
& $\mathcal{N}=10^{-10}(\phizero/\Mp)^{-2}\in[1,100]$ & & & & \\
$\twiBONE $ & $\logdec\left(\phiend/\phizero\right)\in[\logdec(2),
  \logdec(40)]$  
& $\EtwiBTWO $ & $\CtwiBTWO$ & $\NPtwiBTWO$ & $\BEtwiBTWO$ \\ 
& $\phiend<\Mp$ & & & &\\
\end{baytabular}
\end{center}

\subsection{Generalised MSSM Inflation (GMSSMI)} 
\label{subsec:priorgmssmi}

This model is a generalisation of MSSMI studied in
section~\ref{subsec:priormssmi}. The potential can be expressed
as~\cite{Martin:2013tda}
\begin{equation}
V(\phi)=M^4\left[\left(\frac{\phi}{\phi_0}\right)^2-\frac{2}{3}\alpha
  \left(\frac{\phi}{\phi_0}\right)^6+\frac{\alpha}{5}\left(
  \frac{\phi}{\phi_0}\right)^{10}\right].
\end{equation}
This is a two-parameters model, $\phizero$ and $\alpha$, and the
potential of MSSMI is recovered for $\alpha=1$. As already discussed
in section~\ref{subsec:priormssmi} and in \Refc{Martin:2013tda}, the
typical value for the \vev $\phizero$ is $\phizero\simeq
10^{14}\GeV$. The model can also be viewed as a phenomenological one,
that is to say as a representative of the class of the so-called
inflection point inflationary scenario.

Let us now discuss the priors. Viewed as a phenomenological model
(denoted by $\gmssmi$ in what follows), the model is such that the
scale of $\phizero$ is unspecified and, therefore, a Jeffreys prior is
appropriate. We choose to work with
$\logdec(\phizero/\Mp)\in[-5,5]$. On the other hand, the parameter
$\alpha$ is of order one and, as a consequence, we take a flat prior:
$\alpha\in[0.9,1.1]$. Finally, a hard prior has been implemented to
reject all non slow-roll cases (defined to have $|\epsilon_2|>0.2$).

If we now want to calculate the evidence of the model motivated by
particle physics, we must include in the analysis the fact that the
$\vev $ $\phizero$ is around $10^{14}\GeV$. For this reason, we choose
a flat prior such that $\phizero/\Mp\in[2\times 10^{-5},2\times
10^{-4}]$. One also knows that, if $\alpha $ is not precisely tuned
around $\alpha=1$, then the model can not support slow-roll inflation
and is, therefore, ruled out. Moreover, requiring at least $\Delta N
\simeq 60$ \efold during inflation leads to the constraint
\begin{equation}
\label{eq:condgmssmi}
\vert \alpha -1\vert <\frac{\phizero^4}{\Mp^4}
\frac{\pi^2}{900 \Delta N^2},
\end{equation}
see \Refc{Martin:2013tda}. This formula tells us that, if $\vert
\alpha -1\vert \gtrsim 10^{-20}$, then the model is ruled out. This
illustrates the extreme fine-tuning needed for this model to be
compatible with the Planck data. When $\alpha >1$, we implement this
fine-tuning through two different choices of priors satisfying the
above condition, namely $\logdec( 1-\alpha)\in[-28,-23]$ and
$\log(1-\alpha) \in [-28,-20]$, corresponding to the $\gmssmiomA$ and
$\gmssmiomB$ versions of the model, $\gmssmiomB$ being on the validity
threshold. If $\alpha>1$, we define two other models denoted
$\gmssmiopA$ and $\gmssmiopB$ such that
$\logdec(\alpha-1)\in[-28,-23]$ and $\log(\alpha-1) \in
[-28,-21.75]$. Our choices for the priors are summarised in the
following table:

\begin{center}
\begin{baytabular}
\bicenter{$\gmssmi$} & $\logdec\left(\phizero/\Mp\right)\in[-5,5]$ & 
\bicenter{$\Egmssmi$} & \bicenter{$\Cgmssmi$} & \bicenter{$\NPgmssmi$}
& \bicenter{$\BEgmssmi$} \\ 
& $\alpha \in[0.9,1.1]$  & & & & \\
\hline
& $\phizero/\Mp \in [2\times 10^{-5},2\times 10^{-4}]$ & & & & \\
$\gmssmiopA $ &  $\log \left(\alpha-1\right)\in\left[-28,-23\right]$
& $\EgmssmiopA $ & $\CgmssmiopA$ & $\NPgmssmiopA$ & $\BEgmssmiopA$ \\
& $\ln R \in \left[-46, 0\right]$, $\Delta N > 60$ & & & &\\
\hline
& $\phizero/\Mp \in [2\times 10^{-5},2\times 10^{-4}]$ & & & & \\
$\gmssmiopB $ &  $\log \left(\alpha-1\right)\in\left[-28,-21.75\right]$
& $\EgmssmiopB $ & $\CgmssmiopB$ & $\NPgmssmiopB$ & $\BEgmssmiopB$ \\
& $\ln R \in \left[-46, 0\right]$, $\Delta N > 60$ & & & & \\
\hline
& $\phizero/\Mp \in [2\times 10^{-5},2\times 10^{-4}]$ & & & & \\
$\gmssmiomA $ &  $\log \left(1 - \alpha\right)\in\left[-28,-23\right]$
& $\EgmssmiomA $ & $\CgmssmiomA$ & $\NPgmssmiomA$ & $\BEgmssmiomA$ \\
& $\ln R \in \left[-46, 0\right]$, $\Delta N > 60$ & & & &\\
\hline
&  $\phizero/\Mp \in [2\times 10^{-5},2\times 10^{-4}]$ & & & & \\
$\gmssmiomB $ &  $\log \left(1 -\alpha\right)\in\left[-28,-20\right]$
& $\EgmssmiomB $ & $\CgmssmiomB$ & $\NPgmssmiomB$ & $\BEgmssmiomB$ \\
& $\ln R \in \left[-46, 0\right]$, $\Delta N > 60$ & & & &\\
\end{baytabular}
\end{center}

\begin{figure}
\begin{center}
\includegraphics[width=\wsingfig]{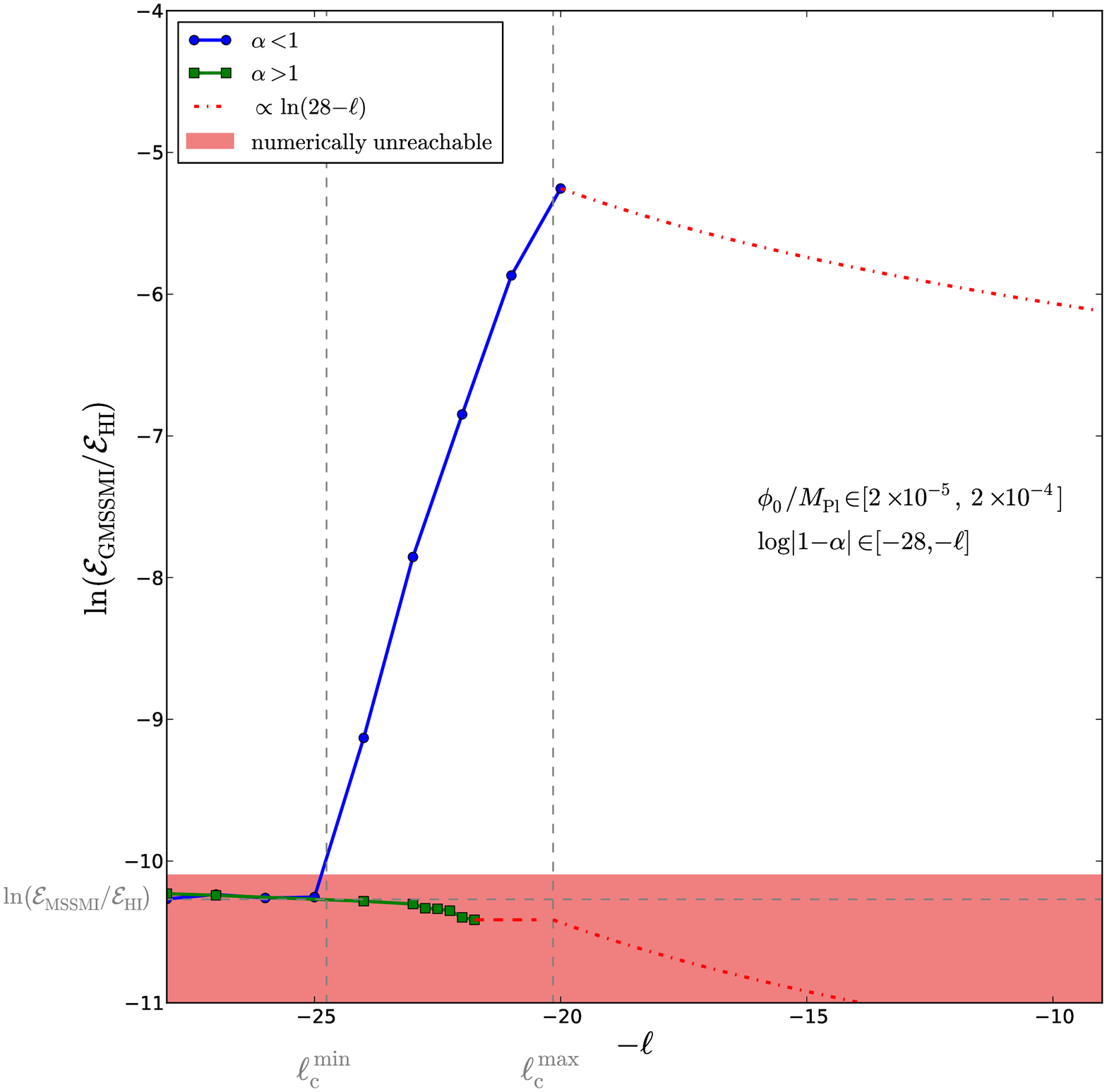}
\caption{Evolution of the GMSSMI Bayes factor versus the upper bound
  $-\ell$ of the prior range on $\alpha $ for $\alpha>1$ and
  $\alpha<1$. The green squares and blue circles represent numerical
  values of the evidence. The dotted red curves represent the
  analytical laws giving the behaviour of the Bayes factor versus
  $-\ell$ for $-\ell \gtrsim \ell_{\mathrm{c}}^\umax$ according to
  Eqs.~(\ref{eq:gmssipcalib}) and~(\ref{eq:evidgmssmmell}). These
  equations predict how the Bayes factor behaves with $-\ell $ and,
  therefore, can be used to extrapolate in regimes where $\alpha$
  becomes of order one.}
\label{fig:finetuning:gmssmi}
\end{center}
\end{figure}

One may also wonder how the evidence would be changed if one
penetrates the regime where \Eq{eq:condgmssmi} is not satisfied (and
where the slow-roll approximation is not satisfied). In that case,
since all non slow-roll models are incompatible with the Planck data,
the evidence should only be rescaled by the ratio of the prior
volumes. Therefore, in the following, we study the more general
situation where $\log\vert \alpha-1\vert \in [-28,-\ell]$, where $\ell
$ is the variable with respect to which we want to study the behaviour
of the Bayesian evidence. In the prior plane
$[\phizero/\Mp,\log\vert1-\alpha\vert]$, \Eq{eq:condgmssmi} defines a
line above which the likelihood vanishes (since, in that case and as
already mentioned, the model becomes incompatible with the data). This
curve approximately goes from $(2\times 10^{-5},-24\equiv
\ell_{\mathrm{c}}^\umin)$ to $(2\times 10^{-4},-20\equiv
\ell_{\mathrm{c}}^\umax)$ and, therefore, defines three different
regions according to whether $-\ell <\ell_{\mathrm{c}}^\umin$, $-\ell
\in [\ell_{\mathrm{c}}^\umin, \ell_{\mathrm{c}}^\umax]$ or $-\ell
>\ell_{\mathrm{c}}^\umax$.

Let us first assume that $\alpha>1$ and $\log(\alpha-1) \in
[-28,-\ell]$.  If $-\ell \lesssim -24 $, then $\alpha -1$ is so small
that one expects the model to be equivalent to MSSMI. If $\ell \in
[\ell_{\mathrm{c}}^\umin, \ell_{\mathrm{c}}^\umax]$ (denoted the
``transition region'' in what follows), then only numerical
calculations can track the behaviour of the evidence. Notice that
$\gmssmiopA$ and $\gmssmiopB$ belongs to this region. Finally, for
$-\ell >\ell_{\mathrm{c}}^\umax$, one expects the evidence to scale
with the ratio of the prior volumes. These expectations are confirmed
in Fig.~\ref{fig:finetuning:gmssmi} (solid green line). However, for
numerical reasons, we are in fact unable to follow the evidence beyond
the point $-\ell \simeq -21.75$ ($\gmssmiopB$ model) which is still in
the transition region. One can nevertheless assume that the evidence
does not change much between that point and the edge of the transition
region (hence the small horizontal dashed red segment inside the
transition region in Fig.~\ref{fig:finetuning:gmssmi}). In that case,
in the regime $-\ell >\ell_{\mathrm{c}}^\umax$, one can write
\begin{equation}
\label{eq:gmssipcalib}
\ln\left[\frac{\EvidName{{\log(\alpha -1)\in
        [-28,-\ell]}}}{\EvidRefName}
\right] \simeq \ln\left(\frac{\EvidName{\gmssmiopB}}{\EvidRefName}
\right)+\ln \left(28-\ell_{\mathrm{c}}^\umax\right)
-\ln \left(28-\ell\right).
\end{equation}
This rough approximation can be considered as reasonable because it
gives an upper bound on the value of the evidence (since the evidence
can only decrease in the transition region) which is, anyhow, in a
regime where the model is strongly disfavoured. Moreover, one should
also keep in mind that we are close to a regime where the numerical
calculations cannot really be trusted (light red shaded region).

The case $\alpha <1$ is very similar and in
Fig.~\ref{fig:finetuning:gmssmi}, we have represented different
numerical values of the Bayes factor versus $-\ell$ (blue solid
line). The interpretation is very similar and one notices that, this
time, one can track the evidence until the end of the transition
regime, \ie until the $\gmssmiomB$ model. Then, one can extrapolate it
using again the ratio of the prior volumes and this leads to the
following expression
\begin{equation}
\label{eq:evidgmssmmell}
\ln\left(\frac{\EvidName{\log (1-\alpha) \in
    [-28,-\ell]}}{\EvidRefName} \right)=
\ln\left(\frac{\EvidName{{\gmssmiomB}}}{\EvidRefName}
\right)+\ln \left(28-\ell_{\mathrm{c}}^\umax\right)
-\ln \left(28-\ell\right).
\end{equation}
This expression is plotted as the (upper) dotted red line in
Fig.~\ref{fig:finetuning:gmssmi} and allows us to extrapolate, in a
reliable way, the Bayes factor for values $-\ell\gtrsim -20$. Let us
also notice that, in this case, the calculation is performed in a
regime where numerical calculations are trustful.

In Fig.~\ref{fig:finetuning:gmssmi}, one also notices that, for
$\alpha >1$, the evidence decreases in the transition region while,
for $\alpha <1$, it grows. This is because the spectral index of
GMSSMI decreases with the value of $\alpha$ starting from the MSSMI
value $\nS \simeq 0.9$ when $\vert \alpha -1\vert \simeq 0$. As a
consequence, when $\alpha <1$, if $-\ell $ is increased then $\ns $
grows and, therefore, crosses the Planck best fit region. For this
reason, the blue curve in Fig.~\ref{fig:finetuning:gmssmi} increases
in the transition region. In the case $\alpha>1$, one observes the
opposite behaviour since, in that situation, the model moves away from
the Planck best fit region.

The previous considerations allow us to extrapolate the evidence
analytically to the theoretical prior in which $\alpha$ varies up to
unity. Those two extrapolated models have been named $\gmssmiep$ and
$\gmssmiem$ in the next table and their evidence have been estimated
using the two equations derived above, namely \Eqs{eq:gmssipcalib} and
(\ref{eq:evidgmssmmell}).
\begin{center}
\begin{baytabular}
&  $\phizero/\Mp \in [2\times 10^{-5},2\times 10^{-4}]$ & & & & \\
$\gmssmiep $ &  $\log \left(\alpha-1\right)\in\left[-28,-0\right]$
& $\Egmssmiep $ & $\Cgmssmiep$ & $\NPgmssmiep$ & $\BEgmssmiep$ \\
& $\ln R \in \left[-46, 0\right]$ & & & &\\
\hline
&  $\phizero/\Mp \in [2\times 10^{-5},2\times 10^{-4}]$ & & & & \\
$\gmssmiem $ &  $\log \left(1 -\alpha\right)\in\left[-28,-0\right]$
& $\Egmssmiem $ & $\Cgmssmiem$ & $\NPgmssmiem$ & $\BEgmssmiem$ \\
& $\ln R \in \left[-46, 0\right]$ & & & &\\
\end{baytabular}
\end{center}

In the above table, the complexities have been rescaled following the
rough estimate given by \Eq{eq:complexity:Crescaling}. 

\subsection{Generalised Renormalisable Point Inflation (GRIPI)}

In the very same way as GMSSMI is a generalisation of MSSMI, see
section~\ref{subsec:priorgmssmi}, the GRIPI potential is a
generalisation of the RIPI one, see
section~\ref{subsec:priorripi}. This potential can be written as
\begin{equation} 
V(\phi)=M^4\left[\left(\frac{\phi}{\phizero}\right)^2 -
   \frac{4\alpha}{3} \left( \frac{\phi}{\phizero} \right)^3 + \frac{\alpha}{2}
    \left( \frac{\phi}{\phizero} \right)^4\right],
\end{equation}
and depends on two parameters, $\phizero$ and $\alpha$. The case
$\alpha=1$ corresponds to the RIPI potential. As discussed in
\Refc{Martin:2013tda}, the typical value of the \vev $\phizero$ is
given by $\phizero\simeq 10^{14}\GeV$ and/or $\phizero\simeq
10^{17}\GeV$. In fact, in the case $\phizero\simeq 10^{14}\GeV$, the
amount of fine-tuning is similar to the GMSSMI case. For this reason,
it is not so interesting to replicate the discussion of the previous
section and, here, one focuses on the case
$\phizero\simeq 10^{17}\GeV$ where one can expect the fine tuning
problem to be sligthly less severe.

Let us now discuss the priors. The GRIPI potential can always be
viewed as a phenomenological model (simply denoted $\gripi$ in what
follows). In that case, the order of magnitude of the parameter
$\phizero$ is not specified and, therefore, one chooses a Jeffreys
prior, namely $\logdec(\phizero/\Mp)\in[-5,5]$. Regarding the
parameter $\alpha$, since it is of order one, we simply take
$\alpha\in[0.9,1.1]$. As for GMSSMI, we have also added a hard prior
boundary, enforcing $|\epsilon_2|<0.2$, as otherwise some regions of the
parameter space would predict non-slow-roll inflation.

Returning to the original version of the model and considering the
fact that, in this case, the \vev $\phizero$ is specified, we choose
the prior $\phizero/\Mp\in[2\times 10^{-2},2\times 10^{-1}]$. As for
GMSSMI, if $\alpha $ is not tuned around $\alpha=1$, the model becomes
inconsistent. Requiring at least $\Delta N \simeq 60$ \efold during
inflation leads to the condition
\begin{equation}
  \label{eq:condgripi}
\vert \alpha -1\vert <\frac{\phizero^4}{\Mp^4}
\frac{\pi^2}{576 \Delta N^2}\,,
\end{equation}
see \Refc{Martin:2013tda} and, therefore, if $\vert \alpha -1\vert
\gtrsim 10^{-8}$, then the model is a priori ruled out. As a
consequence, when $\alpha >1$, we consider two cases satisfying the
above constraint namely $\logdec( 1-\alpha)\in[-15,-10]$ and
$\log(1-\alpha) \in [-28,-8]$, thus defining the $\gripiomA$ and
$\gripiomB$ models. If $\alpha>1$, we define two other models denoted
$\gripiopA$ and $\gripiopB$ such that $\logdec(\alpha-1)\in[-28,-10]$
and $\log(\alpha-1) \in [-28,-8]$.

Finally, the GRIPI potential can also arise in a supergravity
framework (we denote this version of the scenario $\gripiS$). In that
case, there is usually a shift symmetry which allows us to consider
super-Planckian \vev of the field. For this reason, we also
investigate the prior $\phizero/\Mp\in[10,50]$. The prior on $\alpha$
is still taken to be with $\alpha\in[0.9,1.1]$ in agreement with the
previous discussion. All the considerations presented in this section
are summarised in the table below.

\begin{center}
\begin{baytabular}
\bicenter{$\gripi$} & $\logdec\left(\phizero/\Mp\right)\in[-5,5]$ & 
\bicenter{$\Egripi$} & \bicenter{$\Cgripi$} & \bicenter{$\NPgripi$} &
  \bicenter{$\BEgripi$} \\ 
& $\alpha \in[0.9,1.1]$  & & & & \\
\hline
&  $\phizero/\Mp \in [2\times 10^{-2},2\times
        10^{-1}]$& & & & \\
$\gripiopA$ &  $\log \left(\alpha-1\right)\in\left[-15,-10\right]$  &
$\EgripiopA$ & $\CgripiopA$ &
$\NPgripiopA$ & $\BEgripiopA$ \\ 
& $\ln R \in \left[-46, 0\right]$, $\Delta N > 60$ & & & &\\
\hline
&  $\phizero/\Mp \in [2\times 10^{-2},2\times
        10^{-1}]$& & & & \\ 
$\gripiopB$ & $\log
      \left(\alpha-1\right)\in\left[-15,-8\right]$ & 
$\EgripiopB$ & $\CgripiopB$ &
$\NPgripiopB$ & $\BEgripiopB$\\ 
& $\ln R \in \left[-46, 0\right]$, $\Delta N > 60$ & & & &\\
\hline
&  $\phizero/\Mp \in [2\times 10^{-2},2\times 10^{-1}]$& & & & \\
$\gripiomA$ & $\log \left(1 -
      \alpha\right)\in\left[-15,-10\right]$ & 
$\EgripiomA$ & $\CgripiomA$ &
$\NPgripiomA$ & $\BEgripiomA$ \\ 
& $\ln R \in \left[-46, 0\right]$, $\Delta N > 60$ & & & &\\
\hline
& $\phizero/\Mp \in [2\times 10^{-2},2\times 10^{-1}]$& & & & \\
$\gripiomB$ & $\log \left(1 -
      \alpha\right)\in\left[-15,-8\right]$ & 
$\EgripiomB$ & $\CgripiomB$ &
$\NPgripiomB$ & $\BEgripiomB$  \\ 
& $\ln R \in \left[-46, 0\right]$, $\Delta N > 60$ & & & &\\
\hline
\bicenter{$\gripiS$} &  $\phizero/\Mp \in [10,50]$& 
\bicenter{$\EgripiS$} & \bicenter{$\CgripiS$} & \bicenter{$\NPgripiS$}
& \bicenter{$\BEgripiS$} \\ 
& $\alpha\in\left[0.9,1.1\right]$ & & & &\\
\end{baytabular}
\end{center}

\begin{figure}
\begin{center}
\includegraphics[width=\wsingfig]{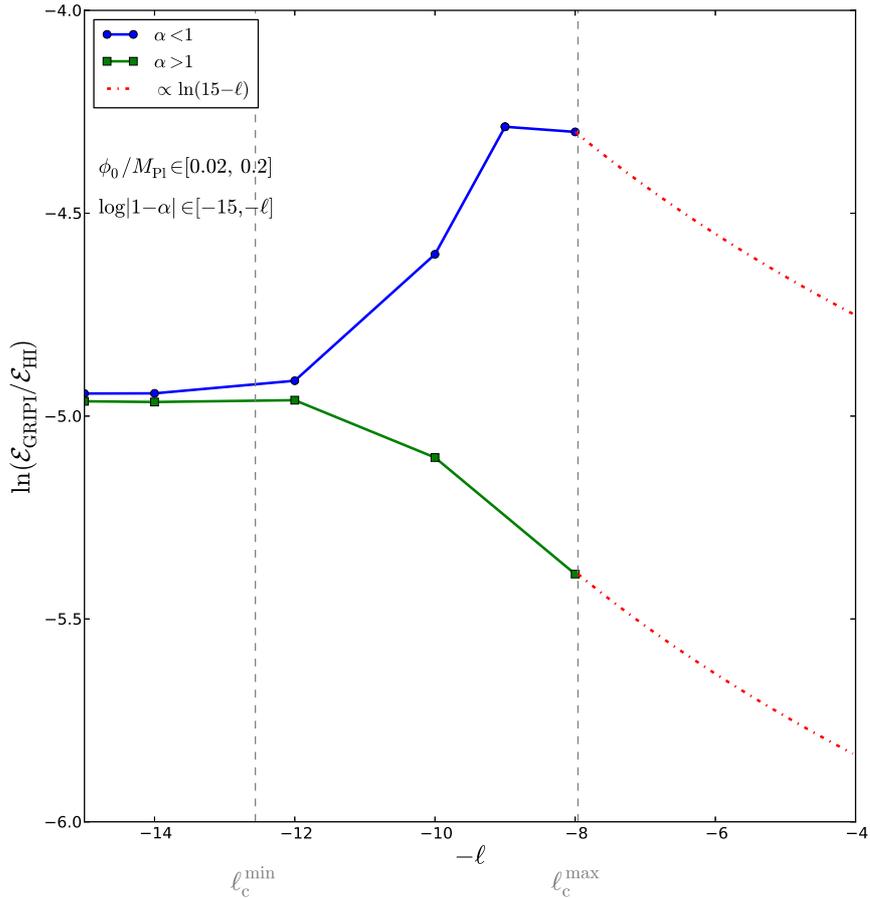}
\caption{Evolution of the GRIPI Bayes factor versus the upper bound
  $-\ell$ of the prior range on $\alpha $ for $\alpha>1$ and
  $\alpha<1$. The green squares and blue circles represent numerical
  values of the evidence. The dotted red curves represent the
  analytical laws giving the behaviour of the Bayes factor versus
  $-\ell$ for $-\ell \gtrsim \ell_{\mathrm{c}}^\umax$ according to
  Eqs.~(\ref{eq:evidgripipell}) and~(\ref{eq:evidgripimell}). These
  equations predict how the Bayes factor behaves with $-\ell $ and,
  therefore, can be used to extrapolate in regimes where $\alpha$
  becomes of order one. The behaviour of the evidences is very similar
  to what was found in the GMSSMI case, see
  Fig.~\ref{fig:finetuning:gmssmi}. However, a difference with GMSSMI
  is that, in the case $\alpha >1$, one is now able to track the Bayes
  factors through the entire transition region.}
\label{fig:finetuning:gripi}
\end{center}
\end{figure}

As was done in the case of GMSSM inflation in the previous section,
one can also study how the choice of the prior on $\alpha$ affects the
determination of the Bayesian evidence. For this reason, we consider
the following priors: $\log(\alpha-1)\in [-15,-\ell]$ for $\alpha>1$
and $\log(1-\alpha)\in [-15,-\ell]$ for $\alpha<1$. The dependence of
the evidence with respect to $\ell$ can be derived as in the previous
section. In the prior plane $[\phizero/\Mp,\log\vert1-\alpha\vert]$,
\Eq{eq:condgripi} defines a line above which the likelihood is tiny
and can be considered to be vanishing. This line divides the prior
space into two parts and goes from $(2\times 10^{-2},-12\equiv
\ell_{\mathrm{c}}^\umin)$ to $(2\times 10^{-1},-8\equiv
\ell_{\mathrm{c}}^\umax)$ and, therefore, defines three different
regions according to whether $-\ell <\ell_{\mathrm{c}}^\umin$, $-\ell
\in [\ell_{\mathrm{c}}^\umin, \ell_{\mathrm{c}}^\umax]$ or $-\ell
>\ell_{\mathrm{c}}^\umax$.

Let us first assume that $\alpha>1$.  If $-\ell \lesssim -15 $, then
$\alpha -1$ is tiny and one expects GRIPI to be equivalent to RIPI
(with the same value of $\phizero$). If $\ell \in
[\ell_{\mathrm{c}}^\umin, \ell_{\mathrm{c}}^\umax]$, then only
numerical calculations can track the behaviour of the evidence. Notice
that $\gripiopA$ and $\gripiopB$ belongs to this region. Finally, for
$-\ell >\ell_{\mathrm{c}}^\umax$, one expects the evidence to scale
with the ratio of the prior volumes. In that case, one can write
\begin{equation}
\label{eq:evidgripipell}
\ln\left(\frac{\EvidName{\log (\alpha-1) \in [-15,-\ell]}}{\EvidRefName}
\right)=\ln\left(\frac{\EvidName{\gripiopB}}{\EvidRefName}
\right)+\ln (15-\ell_{\mathrm{c}}^\umax)-\ln (15-\ell).
\end{equation}
Here, we have taken $\gripiopB$ as the calibration model, a natural
choice considering that this model lies at the frontier of the
transition region. The corresponding results are represented in
Fig.~\ref{fig:finetuning:gripi} (solid green line for the numerical
results and dashed red line for the extrapolated evidences).

For $\alpha<1$, taking $\gripiomB$ as a calibration model, exactly the
same discussion applies and one is led to (again, see
Fig.~\ref{fig:finetuning:gripi})
\begin{equation}
\label{eq:evidgripimell}
\ln\left(\frac{\EvidName{\log (1-\alpha) \in [-15,-\ell]}}{\EvidRefName}
                   \right)=\ln\left(\frac{\EvidName{\gripiomB}}{\EvidRefName}
                   \right)+\ln (15-\ell_{\mathrm{c}}^\umax)-\ln (15-\ell).
\end{equation}
We can now use these formulae to rescale the evidence if $\alpha$
varies up to unity. Naming the two corresponding models $\gripiep$ and
$\gripiem$, their evidences have been reported below.
\begin{center}
\begin{baytabular}
\bicenter{$\gripiep$} &  $\phizero/\Mp \in [2\times 10^{-2},2\times
        10^{-1}]$& 
\bicenter{$\Egripiep$} & \bicenter{$\Cgripiep$} &
\bicenter{$\NPgripiep$} & \bicenter{$\BEgripiep$} \\ 
& $\log \left(\alpha-1\right)\in\left[-15,0\right]$  & & & & \\
\hline
\bicenter{$\gripiem$} &  $\phizero/\Mp \in [2\times 10^{-2},2\times
        10^{-1}]$& 
\bicenter{$\Egripiem$} & \bicenter{$\Cgripiem$} &
\bicenter{$\NPgripiem$} & \bicenter{$\BEgripiem$} \\ 
& $\log \left(1-\alpha\right)\in\left[-15,0\right]$  & & & & \\
\end{baytabular}
\end{center}
In this table, complexities have also been rescaled following the
rough estimate given by \Eq{eq:complexity:Crescaling}.

\subsection{Brane SUSY Breaking Inflation (BSUSYBI)}

The potential is a sum of two exponential and reads
\begin{equation}
V(\phi) = M^4 \left(\ee^{\sqrt{6} \frac{\phi}{\Mp}} + \ee^{\sqrt{6}
    \gamma \frac{\phi}{\Mp}} \right).
\end{equation}
In addition to the parameter $\gamma$, the field value $\xend =
\phiend/\Mp$ at which inflation ends has to be specified. Within the
superstring scenario from which this model is inspired,
$0<\gamma<1/\sqrt{3}$~\cite{Dudas:2012vv}. However, the upper limit
would already implies to $\epsilon_1(x) > 3 \gamma^2 \simeq 1$ and
slow-roll is violated everywhere. We have therefore limited the prior
on $\gamma$ to slightly lower values $\gamma<0.3$ considering either a
flat prior or a Jeffreys prior. Concerning, $\xend$, one notices that
inflation proceeds at decreasing field values and is confined in a
region $x<\xepsoneOne$, $\xepsoneOne$ being the solution of
$\epsilon_1(x)=1$. As a result, there is a maximal bound $\xendmax$
which has been defined such that inflation last more than $120$
{\efolds}. The quantity is only known numerically and is obtained by
integrating the field trajectory from $\xini = \xepsoneOne$ during
$120$ {\efolds}. On the contrary, there is no lower limit on the
allowed values of $\xend$ and the limit $\xend \rightarrow -\infty$,
would correspond to $\nS=1$ and $r=0$. Therefore, for $\xend$ negative
enough, the likelihood, and therefore the evidence, becomes
independent on the lower bound on $\xend$. We have therefore
considered the following priors:
\begin{center}
\begin{baytabular}
\bicenter{$\bsusybif$} & $\gamma\in[0,0.3]$ &
\bicenter{$\Cbsusybif$} & \bicenter{$\Cbsusybif$} &
\bicenter{$\NPbsusybif$} & \bicenter{$\BEbsusybif$} \\
& $\xend \in [-200,\xendmax]$ & & & &\\
\hline
\bicenter{$\bsusybil$} & $\log(\gamma)\in[-3,-1]$ &
\bicenter{$\Cbsusybil$} & \bicenter{$\Cbsusybif$} &
\bicenter{$\NPbsusybil$} & \bicenter{$\BEbsusybil$} \\
& $\xend \in [-200,\xendmax]$ & & & & \\
\end{baytabular}
\end{center}

\subsection{Tip Inflation (TI)}

This string inspired potential has two parameters, a dimensionless
coupling $\alpha$ and a typical {\vev} $\mu$:
\begin{equation}
V(\phi) = M^4 \left[1 + \cos\left(\dfrac{\phi}{\mu}\right) + \alpha
  \sin^2 \left(\dfrac{\phi}{\mu} \right) \right].
\end{equation}
As made explicit in Ref.~\cite{Martin:2013tda}, these parameters
encode combinations of geometrical quantities related to the relative
position of branes within a conifold geometry. This potential supports
inflation at its top provided $\alpha \simeq 1/2$, which amounts to
some level of fine-tuning. When this condition is satisfied, $\mu$
actually gives the volume of the extra-dimensions
\begin{equation}
\dfrac{\mu}{\Mp} \simeq2 \times 10^8 \sigma_0^{9/4},
\end{equation}
where $\sigma_0$ is the stabilised value of the volume modulus in the
absence of uplifting terms~\cite{Pajer:2008uy}. A typical value for
$\sigma_0$ is $\sigma_0 \simeq 10^2$ which translates into $\mu/\Mp
\simeq 10^{-4}$, up to a few orders of magnitude. Following these
considerations, we have examined various priors designed to measure
how important is the fine-tuning over $\alpha$. In particular, the
three sub-classes $\alpha \gtrsim 1/2$, $\alpha=1/2$ and $\alpha
\lesssim 1/2$ yield different observable predictions and have been
treated as separated models. They are summarised below.
\begin{center}
\begin{baytabular}
\bicenter{$\tiBONE$} & $\alpha\in]0.5,0.5 + 10^{-7}]$ &
\bicenter{$\EtiBONE$} & \bicenter{$\CtiBONE$} & \bicenter{$\NPtiBONE$}
& \bicenter{$\BEtiBONE$} \\
& $\log(\mu/\Mp) \in [-5,-3]$ & & & & \\
\hline
\bicenter{$\tiBTWO$} & $\alpha\in]0.5,0.5 + 2\times 10^{-7}]$ &
\bicenter{$\EtiBTWO$} & \bicenter{$\CtiBTWO$} & \bicenter{$\NPtiBTWO$}
& \bicenter{$\BEtiBTWO$} \\
& $\log(\mu/\Mp) \in [-5,-3]$ & & & &\\
\hline
\bicenter{$\tiBTHREE$} & $\alpha\in]0.5,0.5 + 10^{-6}]$ &
\bicenter{$\EtiBTHREE$} & \bicenter{$\CtiBTHREE$} &
\bicenter{$\NPtiBTHREE$} & \bicenter{$\BEtiBTHREE$} \\
& $\log(\mu/\Mp) \in [-5,-3]$ & & & &\\
\hline
\bicenter{$\tiONETWO$} & $\alpha=1/2$ &
\bicenter{$\EtiONETWO$} & \bicenter{$\CtiONETWO$} &
\bicenter{$\NPtiONETWO$} & \bicenter{$\BEtiONETWO$} \\
& $\log(\mu/\Mp) \in [-5,-3]$ & & & &\\
\hline
\bicenter{$\tiDONE$} & $\alpha\in[0.5-10^{-7},0.5[$ &
\bicenter{$\EtiDONE$} & \bicenter{$\CtiDONE$} & \bicenter{$\NPtiDONE$}
& \bicenter{$\BEtiDONE$} \\
& $\log(\mu/\Mp) \in [-5,-3]$ & & & &\\
\hline
\bicenter{$\tiDTWO$} & $\alpha\in[0.5-10^{-6},0.5[$ &
\bicenter{$\EtiDTWO$} & \bicenter{$\CtiDTWO$} & \bicenter{$\NPtiDTWO$}
& \bicenter{$\BEtiDTWO$} \\
& $\log(\mu/\Mp) \in [-5,-3]$ & & & &\\
\hline
\bicenter{$\tiDTHREE$} & $\alpha\in[0.5-10^{-5},0.5[$ &
\bicenter{$\EtiDTHREE$} & \bicenter{$\CtiDTHREE$} &
\bicenter{$\NPtiDTHREE$} & \bicenter{$\BEtiDTHREE$} \\
& $\log(\mu/\Mp) \in [-5,-3]$ & & & &\\
\end{baytabular}
\label{table:tialpha}
\end{center}
For completeness, we have also considered models in which $\alpha
\simeq 1/2$ without any prior prejudice on the sign of
$\alpha-1/2$. One gets the following evidences:
\begin{center}
\begin{baytabular}
\bicenter{$\tiFONE$} & $\alpha\in[0.5-10^{-7},0.5 + 10^{-7}]$ &
\bicenter{$\EtiFONE$} & \bicenter{$\CtiFONE$} & \bicenter{$\NPtiFONE$}
& \bicenter{$\BEtiFONE$} \\
& $\log(\mu/\Mp) \in [-5,-3]$ & & & & \\
\hline
\bicenter{$\tiFTWO$} & $\alpha\in[0.5-10^{-6},0.5 + 10^{-6}]$ &
\bicenter{$\EtiFTWO$} & \bicenter{$\CtiFTWO$} &\bicenter{$\NPtiFTWO$}
& \bicenter{$\BEtiFTWO$}  \\
& $\log(\mu/\Mp) \in [-5,-3]$ & & & & \\
\hline
\bicenter{$\tiFTHREE$} & $\alpha\in[0.5-10^{-5},0.5 + 10^{-5}]$ &
\bicenter{$\EtiFTHREE$} & \bicenter{$\CtiFTHREE$} &
\bicenter{$\NPtiFTHREE$} & \bicenter{$\BEtiFTHREE$} \\
& $\log(\mu/\Mp) \in [-5,-3]$ & & & & \\
\end{baytabular}
\end{center}

One can also wonder what happens if one detunes the prior on $\alpha $
since, after all, this fine-tuning is not theoretically motivated. Let
us first consider the case where $\alpha >1/2$. We want to calculate
the evidence if the prior on $\alpha $ is chosen such that $\alpha \in
]0.5,0.5+a]$. We assume that, for $a>10^{-6}$, the likelihood vanishes
(which is, according to the results presented in the tables, a
realistic hypothesis). Applying the considerations presented earlier
[see Eq.~(\ref{eq:rescaledevidence})] and taking as a calibration
model $\tiBTHREE$, one obtains that for $a>10^{-6}$
\begin{equation}
\calE_{\alpha \in ]0.5,0.5+a]}=\calE_{\alpha \in
]0.5,0.5+10^{-6}]}\frac{10^{-6}}{a}\,,
\end{equation}
or 
\begin{equation}
\ln\left(\frac{\EvidName{\alpha \in ]0.5,0.5+a]}}{\EvidRefName}
                   \right)=\ln\left(\frac{\EvidName{\tiBTHREE}}{\EvidRefName}
                   \right)-6\ln(10)-\ln(a) \simeq \EtiBTHREE-6\ln(10)-\ln(a).
\end{equation}
If we now assume $\alpha < 1/2$, the same considerations lead to a
similar formula, namely
\begin{equation}
\ln\left(\frac{\EvidName{\alpha \in ]0.5-a,0.5]}}{\EvidRefName}
                   \right)=\ln\left(\frac{\EvidName{\tiDTHREE}}{\EvidRefName}
                   \right)-5\ln(10) -\ln(a) \simeq \EtiDTHREE-5\ln(10)-\ln(a),
\end{equation}
for $a>10^{-5}$.
Finally, the case where the sign of $\alpha-1/2$ is not specified
yields
\begin{equation}
\ln\left(\frac{\EvidName{\alpha \in ]0.5-a,0.5+a]}}{\EvidRefName}
                   \right)=\ln\left(\frac{\EvidName{\tiFTHREE}}{\EvidRefName}
                   \right)-5\ln(10)-\ln(a) \simeq \EtiFTHREE-5\ln(10)-\ln(a),
\end{equation}
for $a>10^{-5}$. As expected, we see on these last three formulae
that, if one increases the range of the prior in a region where the
likelihood vanishes, then the corresponding models get penalised for
the wasted parameter space. Therefore, the above calculations
concretely illustrate the Occam's razor effect.

Applying these formulae allows us to extrapolate the evidence for a
natural prior choice having $a=10^{-1}$, \ie assuming only
$\vert\alpha-0.5\vert<0.1$.
\begin{center}
\begin{baytabular}
\bicenter{$\tiem$} & $\alpha\in[0.4,0.5]$ &
\bicenter{$\Etiem$} & \bicenter{$\Ctiem$} & \bicenter{$\NPtiem$}
& \bicenter{$\BEtiem$} \\
& $\log(\mu/\Mp) \in [-5,-3]$ & & & & \\
\hline
\bicenter{$\tiep$} & $\alpha\in[0.5,0.6]$ &
\bicenter{$\Etiep$} & \bicenter{$\Ctiep$} &\bicenter{$\NPtiep$}
& \bicenter{$\BEtiep$}  \\
& $\log(\mu/\Mp) \in [-5,-3]$ & & & & \\
\hline
\bicenter{$\tie$} & $\alpha\in[0.4,0.6]$ &
\bicenter{$\Etie$} & \bicenter{$\Ctie$} &
\bicenter{$\NPtie$} & \bicenter{$\BEtie$} \\
& $\log(\mu/\Mp) \in [-5,-3]$ & & & & \\
\end{baytabular}
\end{center}
In this table, complexities have also been rescaled following the
rough estimate given by \Eq{eq:complexity:Crescaling}. Here the volume
ratio is so big that the rescaled complexities end up being very close
to the number of parameters, which is certainly overestimated due to
our crude assumptions in deriving \Eq{eq:complexity:Crescaling}.

\subsection{Beta Exponential Inflation (BEI)}

This model is an extension of PLI to the generalised exponential
function $\expbeta$ defined by $\expbeta(x)=(1+\beta x)^{1/\beta}$ for
$1 + \beta x>0$ and $\expbeta(x)=0$ otherwise. The potential therefore
reads
\begin{equation}
V(\phi) = M^4 \expbeta\left(\-\lambda \dfrac{\phi}{\Mp} \right),
\end{equation}
where $\lambda>0$ is a dimensionless parameter. As detailed in
Ref.~\cite{Martin:2013tda}, inflation ends naturally only for
$\beta>0$, which will be our prior. The model being phenomenological,
there is no natural value for $\lambda$ and we have chosen a Jeffreys
prior. Moreover, one can show that the slow-roll observable
predictions does not depend on $\lambda$, and thus the prior
boundaries do not affect the evidence. In the limit $\beta \rightarrow
0$, the model becomes strongly disfavoured such that, changing the
lower limit of the $\beta$-prior accordingly decreases the evidence of
the model. This is summarised in the following table:
\begin{center}
\begin{baytabular}
\bicenter{$\bei$} & $\log(\lambda)\in[-3,3]$ &
\bicenter{$\Ebei$} & \bicenter{$\Cbei$} & \bicenter{$\NPbei$} &
\bicenter{$\BEbei$}  \\
& $\log(\beta) \in [-1.5,3]$ & & & & \\
\end{baytabular}
\end{center}

\subsection{Pseudo Natural Inflation (PSNI)}

The potential of PSNI reads
\begin{equation}
V(\phi) = M^4 \left[ 1+ \alpha \ln\left(\cos\dfrac{\phi}{f}\right) \right],
\end{equation}
where $\alpha$ is a dimensionless coupling and $f$ is an energy
scale. In order for the model to be consistent, one should have
$f<\mpl=\sqrt{8\pi}\Mp$ and $\alpha \ll
1$~\cite{ArkaniHamed:2003mz}. As discussed in
Ref.~\cite{Martin:2013tda}, the above potential has $\epsilon_2 >
\epstwomin=2 \alpha \Mp^2/f^2$ and slow-roll inflation can occur only
if $\alpha \Mp^2/f^2$ is constrained to be small.

A first phenomenological choice of priors therefore consists in
adopting prior boundaries for uniform priors in the quantities
$\log(\alpha\Mp^2/f^2)$ and $\log(f/\Mp)$. Taking
$\log(f/\Mp)\in[-2,1]$ and $\log(\alpha\Mp^2/f^2)<-1$, different lower
bounds on $\alpha\Mp^2/f^2$ have been studied, corresponding to
different levels of fine-tuning of this parameter.
\begin{center}
\begin{baytabular}
\bicenter{$\psniftONE$} & $\log(\alpha \Mp^2/f^2) \in[-5,-1]$ &
\bicenter{$\EpsniftONE$} & \bicenter{$\CpsniftONE$} &
\bicenter{$\NPpsniftONE$} & \bicenter{$\BEpsniftONE$} \\
& $\log(f/\Mp)\in[-2,1]$ & & & &  \\
\hline
\bicenter{$\psniftTWO$} & $\log(\alpha \Mp^2/f^2) \in[-3,-1]$ &
\bicenter{$\EpsniftTWO$} & \bicenter{$\CpsniftTWO$} &
\bicenter{$\NPpsniftTWO$} & \bicenter{$\BEpsniftTWO$} \\
& $\log(f/\Mp)\in[-2,1]$ & & & & \\
\hline
\bicenter{$\psniftTHREE$} & $\log(\alpha \Mp^2/f^2) \in[-2,-1]$ &
\bicenter{$\EpsniftTHREE$} & \bicenter{$\CpsniftTHREE$} &
\bicenter{$\NPpsniftTHREE$} & \bicenter{$\BEpsniftTHREE$} \\
& $\log(f/\Mp)\in[-2,1]$ & & & & \\
\hline
\bicenter{$\psniftFOUR$} & $\log(\alpha \Mp^2/f^2) \in[-1.5,-1]$ &
\bicenter{$\EpsniftFOUR$} & \bicenter{$\CpsniftFOUR$} &
\bicenter{$\NPpsniftFOUR$} & \bicenter{$\BEpsniftFOUR$} \\
& $\log(f/\Mp)\in[-2,1]$ & & & & \\
\end{baytabular}
\end{center}
One can see that the evidence increases as the lower bound on
$\log(\alpha \Mp^2/f^2)$ decreases because the likelihood is better in
the region where $\log(\alpha \Mp^2/f^2)$ is small.

\begin{figure}
\begin{center}
\includegraphics[width=\wdblefig]{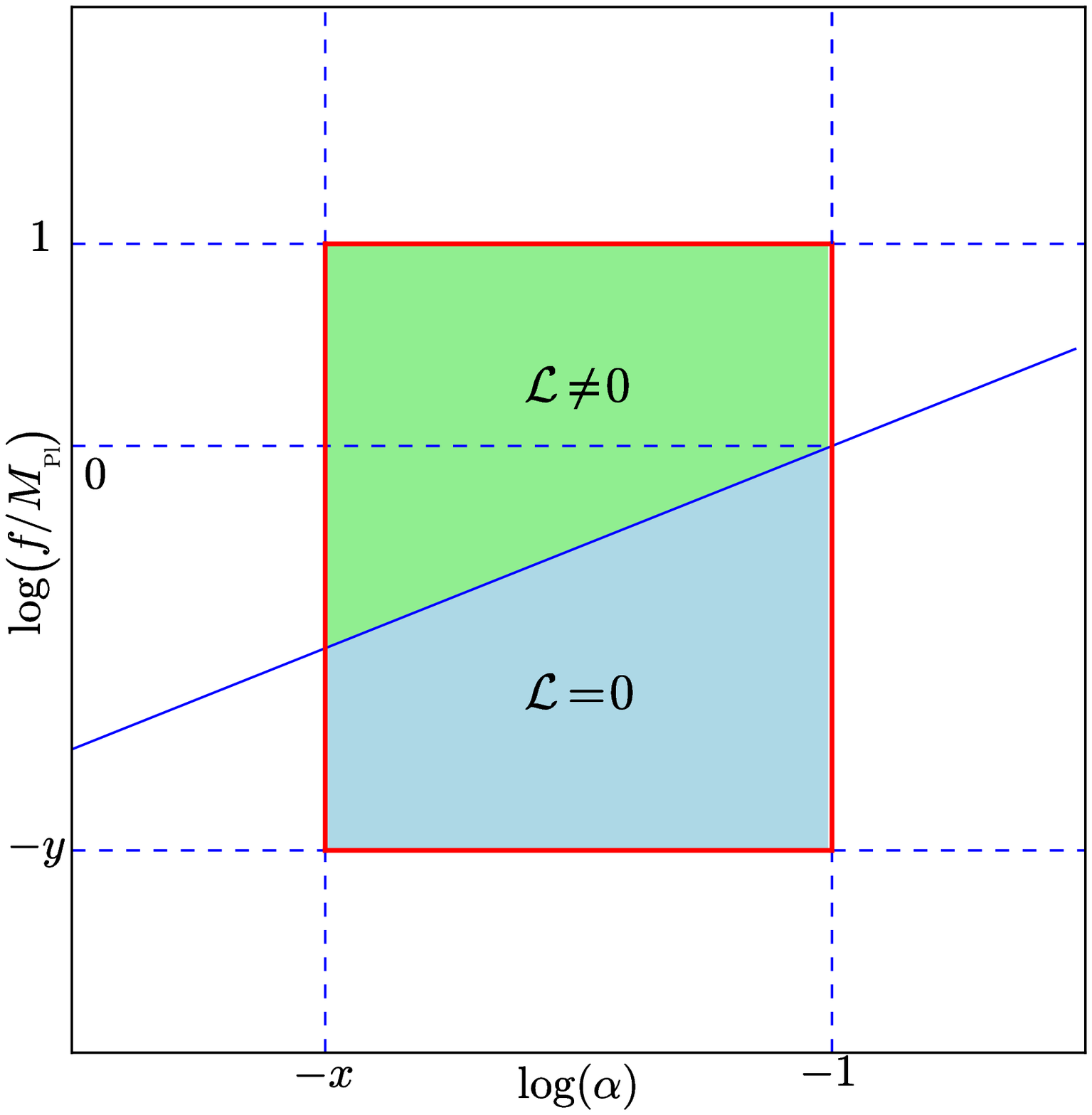}
\includegraphics[width=\wdblefig]{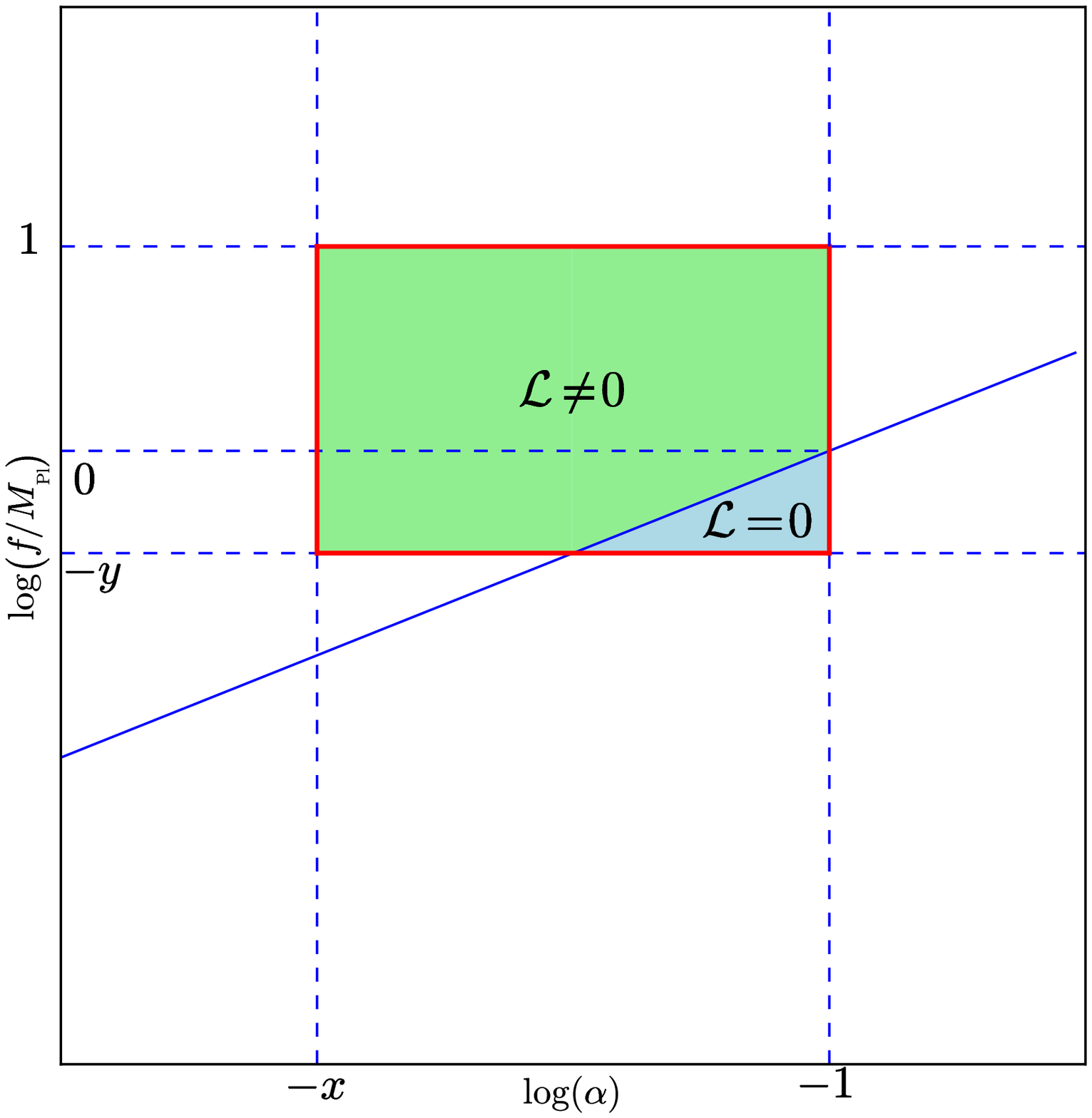}
\caption{Priors for PSNI inflation in the plane $\left[\log \alpha,
    \log(f/\Mp)\right]$. The red rectangle represents the theoretical
  motivated prior. The blue line represents the condition of validity
  of the slow-roll approximation, namely $\alpha f^2/\Mp^2 <
  0.1$. Above this line (green region), slow-roll is satisfied and
  below (blue region) slow-roll is strongly violated. As a
  consequence, the likelihood vanishes in the blue region and is
  different from zero in the green one. Numerically, one can only
  determine the Bayesian evidence with a prior corresponding to the
  green region. The evidence corresponding to the red rectangle can be
  derived from analytical considerations (see text). The left panel
  corresponds to the situation where $y>(x-1)/2$ while the right panel
  is for $y<(x-1)/2$.}
\label{fig:psniprior}
\end{center}
\end{figure}

Another sensible choice of priors, based on the previous
considerations, is to take uniform priors on the theoretical
motivated parameter $\log(\alpha) \in[-x,-1]$ and $\log(f/\Mp) \in
[-y,1]$, where $x$ and $y$ are positive numbers left unspecified for
the moment. In the plane $[\log (f/\Mp),\log \alpha]$ the
two-dimensional prior range is represented in \Fig{fig:psniprior} by
the red rectangle. As discussed previously, $\alpha \Mp^2/f^2$ must be
a small quantity for slow-roll to be satisfied. In the numerical
calculations, we have assumed that it is smaller than $0.1$, \ie , we
use the same hard prior boundary as before, $\epsilon_2 < 0.2$. Such a hard
prior cuts the theoretical prior domain of figure~\ref{fig:psniprior}
along the curve
\begin{equation}
\label{eq:conditionpsni}
\log \left(\frac{f}{\Mp}\right)>\frac{1}{2}\log \alpha+\frac{1}{2}\,,
\end{equation}
which is represented by a blue line in \Fig{fig:psniprior}. As for
GMSSMI and GRIPI, we can estimate analytically how the evidence would
be rescaled by removing this hard prior, but first we need to estimate
how much prior volume is affected.

When $y\ge(x-1)/2$ (left panel of \Fig{fig:psniprior}), the hard prior boundary line
intersects the right vertical edge of the red rectangle at the point
$(-1,0)$ and the left vertical edge at $(-x,1/2-x/2)$. In the case
where $y\le (x-1)/2$ (right panel of \Fig{fig:psniprior}), this line
still intersects the right vertical edge of the red rectangle at the
point $(-1,0)$ but now meets the bottom horizontal edge at
$(-1-2y,-y)$. The condition~(\ref{eq:conditionpsni}) corresponds to
the green region in \Fig{fig:psniprior}, where slow-roll is valid and
the likelihood non-vanishing. The complementary domain has been
represented in blue on the same figure. In this domain, slow-roll is
violated and the predictions cannot be in agreement with the
observations. As a consequence, the likelihood function $\calL$ is
very small and for the purpose of our analytical extrapolation it will
be assumed to vanish. In the following table, we have numerically
computed the evidences in the green domain, \ie in the region where our
computations can be trusted, for various prior choices.
\begin{center}
\begin{baytabular}
&  $\log(\alpha) \in[-7,-1]$  & & & &\\
$\psnioA$ & $\log(f/\Mp)\in[-2,1]$ &
$\EpsnioA$ & $\CpsnioA$ &
$\NPpsnioA$ & $\BEpsnioA$  \\
& $\alpha\Mp^2/f^2<10^{-1}$ & & & &  \\
\hline
&  $\log(\alpha) \in[-5,-1]$  & & & &\\
$\psnioB$ & $\log(f/\Mp)\in[-2,1]$ &
$\EpsnioB$ & $\CpsnioB$ &
$\NPpsnioB$ & $\BEpsnioB$  \\
& $\alpha\Mp^2/f^2<10^{-1}$ & & & &  \\
\hline
&  $\log(\alpha) \in[-3,-1]$  & & & &\\
$\psnioC$ & $\log(f/\Mp)\in[-2,1]$ &
$\EpsnioC$ & $\CpsnioC$ &
$\NPpsnioC$ & $\BEpsnioC$  \\
& $\alpha\Mp^2/f^2<10^{-1}$ & & & &  \\
\end{baytabular}
\end{center}

We can now use these evidences calculated with the green domain prior
and rescale them appropriately to obtain the evidences over the full
domain, including the slow-roll violating regions, that is to say in
the red rectangle. From Eq.~(\ref{eq:rescaledevidence}), generalised
to a two-dimensional prior, one gets
\begin{equation}
\calE_{\mathrm{red}}=\dfrac{\int_{\mathrm{green}} \dd\log(\alpha) \, \dd
  \log(f/\Mp)}
{\int _{\mathrm{red}} \dd \log(\alpha) \,  \dd \log(f/\Mp)}\calE_{\mathrm{green}},
\end{equation}
\ie the evidence is rescaled according to the ratio of the prior
volumes between the green and red domains. Explicitly, one gets
\begin{equation}
\int _{\mathrm{red}} \dd \log(\alpha) \,  \dd \log(f/\Mp) = (x-1)(y+1),
\end{equation}
and 
\begin{equation}
\int _{\mathrm{green}}  \dd \log(\alpha) \,  \dd \log(f/\Mp)
=
\left\lbrace
\begin{aligned}
\dfrac{(x-1)(x+3)}{4}
& \quad \textrm{if} \quad  y\ge\frac{x-1}{2}\,,\\
(y+1)(x-1)-y^2
& \quad \textrm{if} \quad y\le\frac{x-1}{2}\,,\\
\end{aligned}
\right.
\end{equation}
such that
\begin{equation}
\ln\left(\dfrac{\calE_{\mathrm{red}}}{\EvidRefName}\right) =
\ln \left( \dfrac{\calE_{\mathrm{green}}}{\EvidRefName} \right) +  \left\lbrace
\begin{aligned}
\ln\left[\frac{x+3}{4(y+1)}\right]
& \quad \textrm{if} \quad y\ge\frac{x-1}{2}\,,\\
\ln \left[1-\frac{y^2}{(y+1)(x-1)} \right]
& \quad \textrm{if} \quad y\le\frac{x-1}{2}\,.\\
\end{aligned}
\right.
\end{equation}
Therefore, the rescaled evidence (for the red domain) can be obtained
from the one in the green region by using the correction factor given
by the above formula. The results are summarised below.

\begin{center}
\begin{baytabular}
\bicenter{$\psniepA$} & $\log(\alpha) \in[-7,-1]$ &
\bicenter{$\EpsniepA$} & \bicenter{$\CpsniepA$} &
\bicenter{$\NPpsniepA$} & \bicenter{$\BEpsniepA$} \\
& $\log(f/\Mp)\in[-2,1]$ & & & &  \\
\hline
\bicenter{$\psniepB$} & $\log(\alpha) \in[-5,-1]$ &
\bicenter{$\EpsniepB$} & \bicenter{$\CpsniepB$} &
\bicenter{$\NPpsniepB$} & \bicenter{$\BEpsniepB$} \\
& $\log(f/\Mp)\in[-2,1]$ & & & &  \\
\hline
\bicenter{$\psniepC$} & $\log(\alpha) \in[-3,-1]$ &
\bicenter{$\EpsniepC$} & \bicenter{$\CpsniepC$} &
\bicenter{$\NPpsniepC$} & \bicenter{$\BEpsniepC$} \\
& $\log(f/\Mp)\in[-2,1]$ & & & &  \\
\end{baytabular}
\end{center}
In this table, complexities have also been rescaled following the
rough estimate given by \Eq{eq:complexity:Crescaling}.

\subsection{Non Canonical K\"ahker Inflation (NCKI)}

The model has two dimensionless parameters $\alpha$ and $\beta$ and
its potential reads
\begin{equation}
V(\phi) = M^4 \left[1 + \alpha \ln \left(\dfrac{\phi}{\Mp} \right) +
  \beta \left(\dfrac{\phi}{\Mp} \right)^2 \right].
\end{equation}
The logarithmic term encodes loop corrections to the monomial part of
the potential~\cite{Boubekeur:2005zm} and, therefore, natural values
of $\alpha$ are such $0<\alpha \ll 1$ whereas $\beta = \order{1}$. As
discussed in Ref.~\cite{Martin:2013tda}, for $\beta>0$, the first
Hubble flow function has a maximum $\epsonemax \simeq \beta/2$
($\alpha \ll 1$) at $\xepsoneMax \simeq 1/\sqrt{\beta}$, with $x
\equiv \phi/\Mp$. Therefore, we require $\beta$ to small enough to
have $\epsonemax \ll 1$ to ensure slow-roll inflation. If this
condition is not satisfied, inflation could still process in the large
field limit, but would be equivalent to the LFI models. A similar
requirement exists for $\beta<0$ by noticing that the second Hubble
flow function verifies $\epstwomin > -8 \beta$ ($\alpha \ll 1$), which
should be less than unity. The lower limit of $|\beta|$ is arbitrary
but cannot not be too small in order to maintain the hierarchy between
the loop corrections and the monomial term. On purely phenomenological
grounds, taking the limit $\beta \rightarrow 0$ gives back the LI
potential. We have accordingly chosen the following priors:

\begin{center}
\begin{baytabular}
\bicenter{$\nckip$} & $\log(\alpha)\in[-4,-1]$ &
\bicenter{$\Enckip$} & \bicenter{$\Cnckip$} & \bicenter{$\NPnckip$} &
\bicenter{$\BEnckip$}  \\
& $\beta\in[0.02,0.2]$ & & & & \\
\hline
\bicenter{$\nckim$} & $\log(\alpha)\in[-4,-1]$ &
\bicenter{$\Enckim$} & \bicenter{$\Cnckim$} & \bicenter{$\NPnckim$} &
\bicenter{$\BEnckim$}  \\
& $\beta\in[-0.1,-0.02]$ & & & & \\
\end{baytabular}
\end{center}

\subsection{Constant Spectrum Inflation (CSI)}

This potential is designed to produce a scale invariant power spectrum
$\nS\simeq 1$ and reads
\begin{equation}
V(\phi) = \dfrac{M^4}{\left(1 - \alpha \dfrac{\phi}{\Mp} \right)^2}\,,
\end{equation}
where $\alpha$ is supposed to be small. This potential requires the
field value $\xend=\phiend/\Mp$ at which inflation stops to be
specified. In the branch $x=\phi/\Mp<1/\alpha$, inflation proceeds at
decreasing field value while it cannot start at too large initial
field value $\xini$. Indeed, one has $\epsilon_1(x)<1$ only for
$x<\xepsoneOne$ such that $\xini$ is bounded from above $\xini <
\xepsoneOne$. This implies that there is a maximal bound for $\xend$,
that is numerically determined by requesting inflation to last, at
least, $120$ {\efolds} from $\xini=\xepsoneOne$ to $\xend=\xendmax$. A
priori, there is no lower limit for $\xend$. However, for $\xendmin
\rightarrow -\infty$, all the slow-roll functions vanish, $\nS
\rightarrow 1$, $r \rightarrow 0$. Therefore, the likelihood values
become independent on $\xendmin$, as is the evidence. For convenience,
we have chosen $\xendmin(\alpha)$ such that $\epsilon_1(\xendmin) >
10^{-16}$, the machine precision limit. Everything is summarised in
the following table.

\begin{center}
\begin{baytabular}
\bicenter{$\csi$} & $\log(\alpha) \in [-5,-1]$ &
\bicenter{$\Ecsi$} & \bicenter{$\Ccsi$} & \bicenter{$\NPcsi$} &
\bicenter{$\BEcsi$}  \\
& $\xend \in [\xendmin,\xendmax]$ & & & &\\
\end{baytabular}
\end{center}

\subsection{Orientifold Inflation (OI)}

The potential of these models has two parameters, a coupling $\alpha$
and a {\vev} $\phizero$, and reads
\begin{equation}
V(\phi) = M^4 \left(\dfrac{\phi}{\phizero}\right)^4 \left[
  \ln^2\left(\dfrac{\phi}{\phizero} \right) - \alpha \right].
\end{equation}
As the model is motivated by super Yang-Mills orientifold theories,
the {\vev} $\phizero$ should be related to the Grand Unified energy
scale and the coupling $\alpha$ should be small since
$\alpha=\order{1/N_\uc}$, $N_\uc \gg 1$ being the number of
colours~\cite{Channuie:2012bv}. Therefore, we have chosen the following
priors.

\begin{center}
\begin{baytabular}
\bicenter{$\oi$} & $\log(\alpha) \in [-3,-1]$ &
\bicenter{$\Eoi$} & \bicenter{$\Coi$} & \bicenter{$\NPoi$} &
\bicenter{$\BEoi$} \\
& $\log(\phizero/\Mp) \in [-3,-1]$ & & & & \\
\end{baytabular}
\end{center}

\subsection{Constant $\nS$ C Inflation (CNCI)}

This is the class ``C'' of potentials, according to the classification
of Ref.~\cite{Martin:2013tda}, which produces a constant spectral
index. The potential is parametrised by one parameter $\alpha$ and
reads
\begin{equation}
V(\phi) = M^4 \left[\left(3 + \alpha^2 \right) \coth^2
  \left(\dfrac{\alpha}{\sqrt{2}} \dfrac{\phi}{\Mp} \right) - 3 \right].
\end{equation}
In addition to $\alpha$, the model requires the field value $\xend
\equiv \phiend/\Mp$ at which inflation ends to be specified. These
scenarios are phenomenological and motivated for $\alpha$
small. Moreover, inflation proceeds at increasing field values and
there is a region at small $x=\phi/\Mp$ in which $\epsilon_1(x)>1$. As
the result, $\xini>\xepsoneOne$, with $\xepsoneOne$ the solution of
$\epsilon_1=1$. Requesting inflation to support at least $120$
{\efolds} from $\xepsoneOne$ implies the existence of minimal value for
$\xend>\xendmin(\alpha)$. These considerations lead us to the
following priors:

\begin{center}
\begin{baytabular}
\bicenter{$\cnci$} & $\log(\alpha)\in[-5,-1]$ &
\bicenter{$\Ecnci$} & \bicenter{$\Ccnci$} & \bicenter{$\NPcnci$} &
\bicenter{$\BEcnci$}  \\
& $\xend/\xendmin\in [1,10]$ & & & & \\
\end{baytabular}
\end{center}

\subsection{Supergravity Brane Inflation (SBI)}

The potential depends on two dimensionless parameters $\alpha$ and
$\beta$ and reads
\begin{equation}
V(\phi) = M^4 \left\{1 + \left[-\alpha + \beta
  \ln\left(\dfrac{\phi}{\Mp} \right)
  \right] \left(\dfrac{\phi}{\Mp}\right)^4 \right\}.
\end{equation}
As discussed in Ref.~\cite{Martin:2013tda}, the logarithmic term comes
from loop corrections that should not dominate the field dynamics. As
such, the potential supports inflation in the small field region in
which it is convex. For inflation to end, one requires $\alpha \ge
\alphamin(\beta)$ where $\alphamin = (\beta/4)[1-\ln(\beta/4)]$, and
$\beta$ should be a small parameter. For $\alpha>\alphamin$, inflation
is well defined but, at larger field values, the potential exhibits a
negative minimum showing that it cannot be extended to those
regions. On the other hand, for $\alpha=\alphamin$, the potential has
a vanishing minimum and is well defined everywhere. We have therefore
considered these two cases.

\begin{center}
\begin{baytabular}
\bicenter{$\sbi$} & $\log(\alpha) \in [-5,-2]$ &
\bicenter{$\Esbi$} & \bicenter{$\Csbi$} & \bicenter{$\NPsbi$} &
\bicenter{$\BEsbi$} \\
& $\log(\beta) \in [-4,-1]$ & & & &\\
\hline
\bicenter{$\sbialphamin$} & $\alpha = \alphamin$ &
\bicenter{$\Esbialphamin$} & \bicenter{$\Csbialphamin$} &
\bicenter{$\NPsbialphamin$} & \bicenter{$\BEsbialphamin$}  \\
& $\log(\beta) \in [-4,-1]$ & & & &\\
\end{baytabular}
\end{center}

\subsection{Spontaneous Symmetry Breaking Inflation (SSBI)}

The SSBI models are described by potentials of the form
\begin{equation}
V(\phi) = M^4 \left[1 + \alpha \left(\dfrac{\phi}{\Mp}\right)^2 +
  \beta \left(\dfrac{\phi}{\Mp} \right)^4 \right],
\end{equation}
where $\alpha$ and $\beta$ are the two dimensionless parameters. As
discussed in Ref.~\cite{Martin:2013tda}, this potential supports six
different inflationary regimes according the relative signs of
$\alpha$ and $\beta$. They are $\ssbiONE$ for $\alpha>0$, $\beta>0$;
$\ssbiTWO$ for $\alpha<0$, $\beta<0$; $\ssbiTHREE$ and $\ssbiFOUR$ for
$\alpha>0$, $\beta<0$; $\ssbiFIVE$ and $\ssbiSIX$ for $\alpha<0$,
$\beta>0$. A priori the parameters $\alpha$ and $\beta$ may take very
small values, or not, depending on the underlying theoretical
motivations (see Ref.~\cite{Martin:2013tda}). As a result, we have
both considered a Jeffreys and flat prior for those two
parameters. There are however some additional restrictions. For
$\ssbiONE$, inflation ends only for $\alpha> \alphamin(\beta)$ which
fixes an absolute lower limit for the $\alpha$-prior. Moreover, even
when this condition is satisfied, $\ssbiONE$ is strongly disfavoured
when $\alpha$ becomes small and we have only considered $\alpha >
\max(10^{-3},\alphamin)$. For $\ssbiTHREE$ and $\ssbiFOUR$, inflation
proceeds from the top of the potential, either at increasing field
values or at decreasing field values. The shape of the SSBI potential
is such that this may occur in a non slow-rolling way with
$\epsilon_2$ large. These situations violates the slow-roll
approximation, and are strongly disfavoured. Therefore, we have added
a hard prior rejecting all model parameter values yielding
$\epsilon_2(\xVtop)>0.2$, $\xVtop$ being the field value at the top of
the potential. Finally, for $\ssbiFIVE$ and $\ssbiSIX$, there is
another value $\alphamax(\beta)$ above which inflation never ends. As
a result, for those scenarios, the prior on $\alpha$ verifies
$\alpha<\alphamax(\beta)$. The following table summarises all the SSBI
models considered with the Jeffreys prior choices:

\begin{center}
\begin{baytabular}
\bicenter{$\ssbiONE$} & $\max[-3,\log(\alphamin)] <\log(\alpha)<1$ &
\bicenter{$\EssbiONE$} & \bicenter{$\CssbiONE$} &
\bicenter{$\NPssbiONE$} & \bicenter{$\BEssbiONE$} \\
& $\log(\beta) \in [-5,1]$ & & & &\\
\hline
\bicenter{$\ssbiTWO$} & $\log(-\alpha) \in [-5,1]$ &
\bicenter{$\EssbiTWO$} & \bicenter{$\CssbiTWO$} &
\bicenter{$\NPssbiTWO$} & \bicenter{$\BEssbiTWO$} \\
& $\log(-\beta) \in [-5,1]$ & & & &\\
\hline
& $\log(\alpha) \in [\log(\alphamin),1]$ & & & &\\
$\ssbiTHREE$ & $\log(-\beta) \in [-5,1]$ & $\EssbiTHREE$ &
$\CssbiTHREE$ & $\NPssbiTHREE$ & $\BEssbiTHREE$ \\
& $\epsilon_2(\xVtop) < 0.2$ & & & &\\
\hline
& $\log(\alpha) \in [\log(\alphamin),1]$ & & & &\\
$\ssbiFOUR$ & $\log(-\beta) \in [-5,1]$ & $\EssbiFOUR$ &
$\CssbiFOUR$ & $\NPssbiFOUR$ & $\BEssbiFOUR$ \\
& $\epsilon_2(\xVtop) < 0.2$ & & & &\\
\hline
\bicenter{$\ssbiFIVE$} & $\log(-\alpha) \in [\log(-\alphamax),1]$ &
\bicenter{$\EssbiFIVE$} & \bicenter{$\CssbiFIVE$} &
\bicenter{$\NPssbiFIVE$} & \bicenter{$\BEssbiFIVE$} \\
& $\log(\beta) \in [-5,1]$ & & & &\\
\hline
\bicenter{$\ssbiSIX$} & $\log(-\alpha) \in [\log(-\alphamax),1]$ &
\bicenter{$\EssbiSIX$} & \bicenter{$\CssbiSIX$} &
\bicenter{$\NPssbiSIX$} & \bicenter{$\BEssbiSIX$} \\
& $\log(\beta) \in [-5,1]$ & & & &\\
\end{baytabular}
\end{center}

We have also considered the same models but when the natural values of
$\alpha$ and $\beta$ are considered as being $\order{1}$, \ie with
flat priors rather than Jeffreys priors. They are listed in the table
below.

\begin{center}
\begin{baytabular}
\bicenter{$\ssbiONEf$} & $\alpha \in [\max(10^{-3},\alphamin),10] $ &
\bicenter{$\EssbiONEf$} & \bicenter{$\CssbiONEf$} &
\bicenter{$\NPssbiONEf$} & \bicenter{$\BEssbiONEf$}  \\
& $\beta \in [10^{-5},10]$ & & & & \\
\hline
\bicenter{$\ssbiTWOf$} & $\alpha \in [-10,-10^{-5}]$ &
\bicenter{$\EssbiTWOf$} & \bicenter{$\CssbiTWOf$} &
\bicenter{$\NPssbiTWOf$} & \bicenter{$\BEssbiTWOf$} \\
& $\log(-\beta) \in [-5,1]$ & & & &\\
\hline
& $\alpha \in [\alphamin,10]$ & & & & \\
$\ssbiTHREEf$ & $\beta \in [-10,-10^{-5}]$ & $\EssbiTHREEf$ &
$\CssbiTHREEf$ & $\NPssbiTHREEf$ & $\BEssbiTHREEf$  \\
& $\epsilon_2(\xVtop) < 0.2$ & & & &\\
\hline
& $\alpha \in [\alphamin,10]$ & & & & \\
$\ssbiFOURf$ & $\beta \in [-10,-10^{-5}]$ & $\EssbiFOURf$ &
$\CssbiFOURf$ & $\NPssbiFOURf$ & $\BEssbiFOURf$ \\
& $\epsilon_2(\xVtop) < 0.2$ & & & &\\
\hline
\bicenter{$\ssbiFIVEf$} & $\alpha \in [-10,\alphamax]$ &
\bicenter{$\EssbiFIVEf$} & \bicenter{$\CssbiFIVEf$} &
\bicenter{$\NPssbiFIVEf$} & \bicenter{$\BEssbiFIVEf$} \\
& $\beta \in [10^{-5},10]$ & & & &\\
\hline
\bicenter{$\ssbiSIXf$} & $\alpha \in [-10,\alphamax]$ &
\bicenter{$\EssbiSIXf$} & \bicenter{$\CssbiSIXf$} &
\bicenter{$\NPssbiSIXf$} & \bicenter{$\BEssbiSIXf$} \\
& $\beta \in [10^{-5},10]$ & & & &\\
\end{baytabular}
\end{center}

\subsection{Inverse Monomial Inflation (IMI)}

The potential is a extension of the large field inflation potential to
negative power indices and read
\begin{equation}
V(\phi) = M^4 \left(\dfrac{\phi}{\Mp} \right)^{-p},
\end{equation}
with $p>0$. Inflation proceeds at increasing field values and ends at
the field value $\xend = \phiend/\Mp$, which is an additional model
parameter. There is however a region, at small field values, which
does not support inflation as $\epsilon_1(x)>1$. Denoting
$\xepsoneOne$ the solution of $\epsilon_1=1$, this implies that
$\xini>\xepsoneOne$ and there is a minimal acceptable value for
$\xend$ such that inflation lasts more than $120$ {\efolds}. As for
the other models, this value $\xendmin$ is numerically determined by
solving the field trajectory starting at $\xini = \xepsoneOne$ for the
specified amount of {\efolds}. In the absence of definite constraints
on $\xend$, we have chosen a flat prior for $\xend/\xendmin \in
[1,100]$ as well as various fixed values of $p=\order{1}$. They are
summarised below.

\begin{center}
\begin{baytabular}
\bicenter{$\imi$} & $p\in [1,6]$ & \bicenter{$\Eimi$} &
\bicenter{$\Cimi$} & \bicenter{$\NPimi$} & \bicenter{$\BEimi$} \\
& $\xend/\xendmin\in[1,100]$ & & & &\\
\hline
\bicenter{$\imiONE$} & $p=1$ & \bicenter{$\EimiONE$} &
\bicenter{$\CimiONE$} & \bicenter{$\NPimiONE$} &
\bicenter{$\BEimiONE$} \\
& $\xend/\xendmin\in[1,100]$ & & & & \\
\hline
\bicenter{$\imiTWO$} & $p=2$ & \bicenter{$\EimiTWO$} &
\bicenter{$\CimiTWO$} & \bicenter{$\NPimiTWO$} &
\bicenter{$\BEimiTWO$} \\
& $\xend/\xendmin\in[1,100]$ & & & & \\
\hline
\bicenter{$\imiTHREE$} & $p=3$ & \bicenter{$\EimiTHREE$} &
\bicenter{$\CimiTHREE$} & \bicenter{$\NPimiTHREE$} &
\bicenter{$\BEimiTHREE$}  \\
& $\xend/\xendmin\in[1,100]$ & & & &\\
\hline
\bicenter{$\imiFOUR$} & $p=4$ & \bicenter{$\EimiFOUR$} &
\bicenter{$\CimiFOUR$} & \bicenter{$\NPimiFOUR$} & \bicenter{$\BEimiFOUR$} \\
& $\xend/\xendmin\in[1,100]$ & & & &\\
\hline
\bicenter{$\imiFIVE$} & $p=5$ & \bicenter{$\EimiFIVE$} &
\bicenter{$\CimiFIVE$} & \bicenter{$\NPimiFIVE$} &
\bicenter{$\BEimiFIVE$} \\
& $\xend/\xendmin\in[1,100]$ & & & & \\
\hline
\bicenter{$\imiSIX$} & $p=6$ & \bicenter{$\EimiSIX$} &
\bicenter{$\CimiSIX$} & \bicenter{$\NPimiSIX$} &
\bicenter{$\BEimiSIX$}  \\
& $\xend/\xendmin\in[1,100]$ & & & &\\
\end{baytabular}
\end{center}

\subsection{Brane Inflation (BI)}

The potential of brane inflation reads
\begin{equation}
V_{\sss{\bi}}(\phi) = M^4 \left[1 - \left(\dfrac{\phi}{\mu}\right)^{-p} \right],
\label{eq:bi}
\end{equation}
and depends explicitly on two parameters $\mu$ and $p$. This is an
approximated expression derived from KKLMMT-like inflationary
scenarios, in which $p=4$ and $\mu \ll \Mp$~\cite{Kachru:2003aw,
  Kachru:2003sx}. In the following, we define $x \equiv \phi/\mu$ and
inflation proceeds at decreasing $x$. It is induced by the motion of a
brane inside the throat of some compactified extra-dimensions, $\phi$
referring to the position of this brane. Brane inflation can either
ends naturally, \ie when the acceleration of the universe stops, or
before if a tachyonic preheating is triggered by brane
annihilation. The model has therefore an additional parameter,
$\xstg$, which is the field value at which brane annihilation
occurs. Denoting by $\xepsoneOne$ the solution of $\epsilon_1(x)=1$,
inflation actually ends at the field value $\xend =
\max(\xstg,\xepsoneOne)$. The parameter $\xstg$ is related to various
hidden string parameters such as the flux conserved quantum numbers
and the volume of the throat. As shown in Ref.~\cite{Lorenz:2007ze},
the internal consistency of the model implies that $\xstg > 1$, its
order of magnitude remaining unknown. Moreover, there is a maximal
field value, $\phiUV$, which corresponds to the brane position at the
edge of the throat. As the model only describes brane interactions
within the throat, one should impose $\phi<\phiUV$. As discussed in
Ref.~\cite{Lorenz:2007ze}, the internal consistency of the model
imposes that $\phiUV < 2 \Mp$.

Following these considerations, we have first considered strict priors
associated with the string scenario ($\bistg$), namely $p=4$,
$\log(\mu/\Mp)\in[-6,\log(2)]$, $\log(\xstg) \in[0,3]$ and
$\log(\phiUV/\Mp)\in[-2,\log(2)]$. For the sake of generality, we have
also considered the non-approximated potential associated with the
KKLMMT model, namely
\begin{equation}
V_{\sss{\kklti}}(\phi) = \dfrac{M^4}{1 + \left(\dfrac{\phi}{\mu}\right)^{-p}}\,,
\label{eq:kklt}
\end{equation}
under the label $\kkltistg$, and with the same priors as $\bistg$. As
one can check in the following table, there is no difference between
the two models under those priors.
\begin{center}
\begin{baytabular}
\qdcenter{$\bistg$} & $p=4$ & \qdcenter{$\Ebistg$} &
\qdcenter{$\Cbistg$} & \qdcenter{$\NPbistg$} & \qdcenter{$\BEbistg$} \\
& $\log(\mu/\Mp)\in[-6,\log(2)]$ &  & & & \\
& $\log(\xstg) \in [0,3]$ & & & & \\
& $\log(\phiUV/\Mp) \in [-2,\log(2)]$ & & & &\\
\hline
\qdcenter{$\kkltistg$} & $p=4$ & \qdcenter{$\Ekkltistg$} &
\qdcenter{$\Ckkltistg$} & \qdcenter{$\NPkkltistg$} &
\qdcenter{$\BEkkltistg$} \\
& $\log(\mu/\Mp)\in[-6,\log(2)]$ & & & &  \\
& $\log(\xstg) \in [0,3]$ & & & & \\
& $\log(\phiUV/\Mp) \in [-2,\log(2)]$ & & & & \\
\end{baytabular}
\end{center}

Then, we have allowed for other phenomenological scenarios that would
be based on the same potentials by relaxing $p$ and allowing $\mu$ to
become super-Planckian. Out of the string framework, there is no
motivation to keep $\xstg$ and $\phiUV$ as extra model parameters and
we have instead assumed that inflation ends at
$\xend=\xepsoneOne$. However, for $\mu > \Mp$, inflation within the
potential~\eqref{eq:bi} or \eqref{eq:kklt} may yield different
observable predictions. As a result, we have separated the models in
which $\mu < \Mp$ from those in which $\mu$ can take any values. The
phenomenological models considered, and their priors, are enumerated
in the following table.

\begin{center}
\begin{baytabular}
\bicenter{$\kklti$} & $p\in[2,10]$ & \bicenter{$\Ekklti$}&
\bicenter{$\Ckklti$} & \bicenter{$\NPkklti$} & \bicenter{$\BEkklti$} \\
& $\log(\mu/\Mp)\in[-3,3]$ & & & &\\
\hline
\bicenter{$\kkltis$} & $p\in[2,10]$ & \bicenter{$\Ekkltis$}&
\bicenter{$\Ckkltis$} & \bicenter{$\NPkkltis$} &
\bicenter{$\BEkkltis$} \\
& $\log(\mu/\Mp)\in[-3,0]$ & & & &\\
\hline
\bicenter{$\bi$} & $p\in[2,10]$ & \bicenter{$\Ebi$}&
\bicenter{$\Cbi$} & \bicenter{$\NPbi$} & \bicenter{$\BEbi$}  \\
& $\log(\mu/\Mp)\in[-3,3]$ & & & & \\
\hline
\bicenter{$\bis$} & $p\in[2,10]$ & \bicenter{$\Ebis$}&
\bicenter{$\Cbis$} & \bicenter{$\NPbis$} & \bicenter{$\BEbis$} \\
& $\log(\mu/\Mp)\in[-3,0]$ & & & &\\
\hline
\bicenter{$\biONEs$} & $p=1$ & \bicenter{$\EbiONEs$}&
\bicenter{$\CbiONEs$} & \bicenter{$\NPbiONEs$} & \bicenter{$\BEbiONEs$} \\
& $\log(\mu/\Mp)\in[-3,0]$ & & & &\\
\hline
\bicenter{$\biTWOs$} & $p=2$ & \bicenter{$\EbiTWOs$}&
\bicenter{$\CbiTWOs$} & \bicenter{$\NPbiTWOs$} &
\bicenter{$\BEbiTWOs$}  \\
& $\log(\mu/\Mp)\in[-3,0]$ & & & & \\
\hline
\bicenter{$\biTHREEs$} & $p=3$ & \bicenter{$\EbiTHREEs$}&
\bicenter{$\CbiTHREEs$} & \bicenter{$\NPbiTHREEs$} &
\bicenter{$\BEbiTHREEs$}  \\
& $\log(\mu/\Mp)\in[-3,0]$ & & & & \\
\hline
\bicenter{$\biFOURs$} & $p=4$ & \bicenter{$\EbiFOURs$}&
\bicenter{$\CbiFOURs$} & \bicenter{$\NPbiFOURs$} &
\bicenter{$\BEbiFOURs$}  \\
& $\log(\mu/\Mp)\in[-3,0]$ & & & & \\
\hline
\bicenter{$\biFIVEs$} & $p=5$ & \bicenter{$\EbiFIVEs$}&
\bicenter{$\CbiFIVEs$} & \bicenter{$\NPbiFIVEs$} &
\bicenter{$\BEbiFIVEs$}  \\
& $\log(\mu/\Mp)\in[-3,0]$ & & & & \\
\hline
\bicenter{$\biSIXs$} & $p=6$ & \bicenter{$\EbiSIXs$}&
\bicenter{$\CbiSIXs$} & \bicenter{$\NPbiSIXs$} &
\bicenter{$\BEbiSIXs$} \\
& $\log(\mu/\Mp)\in[-3,0]$ & & & & \\
\end{baytabular}
\end{center}

\subsection{Running-mass Inflation (RMI)}

The running-mass inflationary models, denoted RMI, have a potential of
the form
\begin{equation}
V(\phi) = M^4\left[1-\frac{c}{2}\left(-\frac{1}{2} +\ln
\frac{\phi }{\phizero}\right)\frac{\phi ^2}{\Mp^2}\right],
\end{equation}
which supports four different inflationary regimes, namely RMI1
($\phi<\phizero$, $c>0$), RMI2 ($\phi>\phizero$, $c>0$), RMI3
($\phi<\phizero$, $c<0$) and RMI4 ($\phi>\phizero$, $c<0$), see
Ref.~\cite{Martin:2013tda}. In addition to the constant $c$ and the
{\vev} $\phizero$, the field value $\phiend$ at which inflation ends
has to be specified making RMI a three-parameters model. The model
describing loop corrections over a polynomial expansion, the constant
$c$ cannot be too small and the {\vev} $\phizero$ must be
sub-Planckian. The order of magnitude of $\phizero$ being unspecified,
we have chosen a Jeffreys prior in $\log(\phizero/\Mp)\in[-2,0]$. For
RMI1, RMI2 and RMI3, the likelihood has a flat direction along the
parameter $\phizero$ such that the evidence is independent of the
lower bound on $\phizero$. For RMI4, the likelihood is vanishing when
$\phizero$ becomes small, and the evidence is accordingly decreased if
the prior lower bound on $\log(\phizero/\Mp)$ is pushed to smaller
values.

As discussed in Ref.~\cite{Martin:2013tda}, within supersymmetry,
natural values of $c\simeq 10^{-2}$ to $10^{-1}$ for soft masses
values matching the energy scale of inflation. This suggest to take a
flat prior for $c$ encompassing those values. For other type of
couplings, $c$ may take smaller values and we therefore consider
another motivated prior in which the order of magnitude of $c$ is
unknown, e.g. $\log(c)\in[-3,-1]$.

Finally, the field value $\phiend$ is constrained to be in the
inflationary region of interest. The shape of the potential therefore
gives the natural prior bounds for $\xend\equiv \phiend/\phizero$,
i.e. $\xend \in [1/\ee,1]$ for RMI1 and RMI3, $\xend \in [1,\ee]$ for
RMI2. For RMI4, one still has $\xend>1$ but choosing the prior upper
limit requires some precaution. Indeed, the potential is an increasing
function of $\phi$, which approaches large field inflation
asymptotically, and inflation proceeds at decreasing field values,
bounded from below by $\xend$. Since the large field regime is not
acceptable for RMI4, one has to require the \emph{initial} field value
$\xini< \xepsoneMax$. Here $\xepsoneMax$ is the field value at which
the first Hubble flow function is maximal, which is the frontier
between the vacuum dominated regime and the large field one. As for
the other models, adding the ``hard prior'' that inflation lasts
longer than $120$ {\efolds} ensures the existence of a maximal value
$\xendmax$, which is obtained by integrating the field trajectory from
$\xini=\xepsoneMax$. This is a complicated functions of the other
parameters which is only known numerically. The parameter space of
RMI4 is therefore sampled with a flat prior for $\xend \in
[1,\xendmax]$.

Finally, for RMI1 and RMI2, we have added another ``hard prior'' to
avoid an infinite number of {\efolds} to occur at the top of the
potential by requiring $\epsilon_1(\xini)$ to be larger than the
numerical machine precision. This has no observable consequences as
the parameter space volume cut remains extremely small and those cases
would correspond otherwise to $\nS=1$ and are disfavoured. All these
considerations are summarised in the following table.

\begin{center}
\begin{baytabular}
& $c\in[0.01,0.2]$ & & & & \\
$\rmiONE$ &  $\log(\phizero/\Mp)
      \in[-2,0]$ & $\ErmiONE$ & $\CrmiONE$ & $\NPrmiONE$ & $\BErmiONE$
      \\
& $\xend\in[1/\ee,1]$ & & & &\\
\hline
& $\log(c)\in[-3,-1]$ & & & & \\
$\rmiONEl$ &  $\log(\phizero/\Mp)
      \in[-2,0]$ & $\ErmiONEl$ & $\CrmiONEl$ & $\NPrmiONEl$ &
      $\BErmiONEl$ \\
& $\xend\in[1/\ee,1]$ & & & &\\
\hline
& $c\in[0.01,0.2]$ & & & & \\
$\rmiTWO$ &  $\log(\phizero/\Mp)
      \in[-2,0]$ & $\ErmiTWO$ & $\CrmiTWO$ & $\NPrmiTWO$ & $\BErmiTWO$
      \\
& $\xend\in[1,\ee]$ & & & & \\
\hline
& $\log(c)\in[-3,-1]$ & & & & \\
$\rmiTWOl$ &  $\log(\phizero/\Mp)
      \in[-2,0]$ & $\ErmiTWOl$ & $\CrmiTWOl$ & $\NPrmiTWOl$ &
      $\BErmiTWOl$ \\
& $\xend\in[1,\ee]$ & & & &\\
\hline
& $c\in[-0.2,-0.01]$ & & & & \\
$\rmiTHREE$ &  $\log(\phizero/\Mp)
      \in[-2,0]$ & $\ErmiTHREE$ & $\CrmiTHREE$ & $\NPrmiTHREE$ &
      $\BErmiTHREE$ \\
& $\xend\in[1/\ee,1]$ & & & &\\
\hline
& $\log(-c)\in[-3,-1]$ & & & & \\
$\rmiTHREEl$ &  $\log(\phizero/\Mp)
      \in[-2,0]$ & $\ErmiTHREEl$ & $\CrmiTHREEl$ & $\NPrmiTHREEl$ &
      $\BErmiTHREEl$ \\
& $\xend\in[1/\ee,1]$ & & & &\\
\hline
& $c\in[-0.2,-0.01]$ & & & & \\
$\rmiFOUR$ &  $\log(\phizero/\Mp)
      \in[-2,0]$ & $\ErmiFOUR$ & $\CrmiFOUR$ & $\NPrmiFOUR$ &
      $\BErmiFOUR$ \\
& $\xend\in[1,\xendmax]$ & & & & \\
\hline
& $\log(-c)\in[-3,-1]$ & & & & \\
$\rmiFOURl$ &  $\log(\phizero/\Mp)
      \in[-2,0]$ & $\ErmiFOURl$ & $\CrmiFOURl$ & $\NPrmiFOURl$ &
      $\BErmiFOURl$  \\
& $\xend\in[1,\xendmax]$ & & & & \\
\end{baytabular}
\end{center}

\subsection{Valley Hybrid Inflation (VHI)}

The potential is parametrised by two parameters $p$ and $\mu$ and reads
\begin{equation}
V(\phi)=M^4\left[1 +\left(\frac{\phi}{\mu} \right)^{p} \right],
\end{equation}
$p>0$ being the power index and $\mu$ is a typical {\vev}. Because
this expression only describes inflation along the valley of the
genuine two-field hybrid inflationary scenario, the {\vev} $\mu$ is
forced to be super-Planckian. As discussed in
Refs.~\cite{Clesse:2008pf, Clesse:2009ur, Clesse:2010iz,
  Clesse:2012dw, Clesse:2013jra}, this condition is required to get
enough {\efolds} of inflation occurring in the valley. Another
implicit prior is to assume that the parameters associated with the
other field are such that the regime of waterfall inflation does not
take place. As discussed in Ref.~\cite{Martin:2013tda}, the dynamics
of VHI is significantly different if $p>1$ or $p<1$ and the two
classes are considered. In addition to $\mu$ and $p$, hybrid inflation
ends by tachyonic instability and the field value $\xend \equiv
\phiend/\mu$ at which this occurs is an extra model parameter. As for
RMI, our prior is to restrain VHI to the vacuum dominated regime only,
i.e. $\xini < \xepsoneMax$ where $\xepsoneMax$ is the frontier between
the vacuum dominated regime and the large field one. From this limit,
requiring inflation to support at least $120$ {\efolds} gives a
numerical upper bound $\xend < \xendmax$. The quantity $\xendmax$ is
determined numerically using the {\ASPIC} code by integrating the
field trajectory starting at $\xepsoneMax$. For the cases $p \le 1$,
the VHI potential does not support inflation around $x=0$ as
$\epsilon_1$ diverges in this limit. For those, we therefore consider
a prior $\xend>\xendmin$ where $\xendmin=\xepsoneOneMinus$ is the
solution of $\epsilon_1=1$ in the vacuum dominated region. For $p>1$,
the tachyonic instability can take place at arbitrarily small field
values and $\xendmin=0$ (up to machine precision limitations). Notice
that the upper bounds of the $p$ and $\mu$ priors have been fixed to
arbitrary values. All the models considered for the VHI scenarios are
listed below and are all ruled out, independently of their priors.

\begin{center}
\begin{baytabular}
& $p \in ]1,6]$ & & & &  \\
$\vhi$ & $\log(\mu/\Mp) \in [0,3]$ & $\Evhi$ & $\Cvhi$ & $\NPvhi$ &
$\BEvhi$ \\
& $\xend \in [\xendmin,\xendmax]$ & & & &\\
\hline
& $p \in [0,0.9]$ & & & & \\
$\vhilONE$ & $\log(\mu/\Mp) \in [0,3]$ & $\EvhilONE$ & $\CvhilONE$ &
$\NPvhilONE$ & $\BEvhilONE$ \\
& $\xend \in [\xendmin,\xendmax]$ & & & & \\
\hline
& $p=1/2$ & & & & \\
$\vhiONETWO$ & $\log(\mu/\Mp) \in [0,3]$ & $\EvhiONETWO$ &
$\CvhiONETWO$ & $\NPvhiONETWO$ & $\BEvhiONETWO$ \\
& $\xend \in [\xendmin,\xendmax]$ & & & & \\
\hline
& $p=1$ & & & &  \\
$\vhiONE$ & $\log(\mu/\Mp) \in [0,3]$ & $\EvhiONE$ & $\CvhiONE$ &
$\NPvhiONE$ & $\BEvhiONE$ \\
& $\xend \in [\xendmin,\xendmax]$ & & & & \\
\hline
& $p=2$ & & & &  \\
$\vhiTWO$ & $\log(\mu/\Mp) \in [0,3]$ & $\EvhiTWO$ & $\CvhiTWO$ &
$\NPvhiTWO$ & $\BEvhiTWO$ \\
& $\xend \in [\xendmin,\xendmax]$ & & & & \\
\hline
& $p=3$ & & & &  \\
$\vhiTHREE$ & $\log(\mu/\Mp) \in [0,3]$ & $\EvhiTHREE$ & 
$\CvhiTHREE$ & $\NPvhiTHREE$ & $\BEvhiTHREE$ \\
& $\xend \in [\xendmin,\xendmax]$ & & & &  \\
\hline
& $p=4$ & & & & \\
$\vhiFOUR$ & $\log(\mu/\Mp) \in [0,3]$ & $\EvhiFOUR$ & $\CvhiFOUR$ &
$\NPvhiFOUR$ & $\BEvhiFOUR$ \\
& $\xend \in [\xendmin,\xendmax]$ & & & & \\
\end{baytabular}
\end{center}

\subsection{Dynamical Supersymmetric Inflation (DSI)}

The potential is an extension of the VHI one to negative power index
and reads
\begin{equation}
V(\phi) = M^4\left[ 1+\left(\frac{\phi}{\mu} \right)^{-p} \right],
\end{equation}
while this class of model naturally appears in supersymmetric theories
(see Ref.~\cite{Martin:2013tda}). As such, the {\vev} $\mu$ should be
always sub-Planckian. For the priors, we have either considered the
typical values of Refs.~\cite{Kinney:1997hm,Kinney:1998dv}, i.e. a
flat prior for $\mu$ around $10^{-7}$ (model $\dsio$), which is also
relaxed to allow for any other phenomenological models of the same
kind ($\dsi$). Inflation takes place at increasing field value and the
end of inflation $\xend = \phiend/\mu$ is an additional
parameter. Moreover, as discussed in Ref.~\cite{Martin:2013tda},
inflation can only take place in the region $x>\xepsoneOne$, where
$\xepsoneOne$ is the solution of $\epsilon_1=1$. This provides a lower
bound for $\xini$, and therefore, complemented with our hard prior
that inflation lasts more than $120$ {\efolds}, this gives $\xend>
\xendmin$. As for VHI, the quantity $\xendmin$ has to be numerically
evaluated by integrating the field trajectory over $120$ {\efolds}
starting at $\xini=\xepsoneOne$. Moreover, within the supersymmetric
framework in which this potential is derived, there are additional
terms lifting $V(\phi)$ at large field values which can be ignored
provided $\phi$ is not too large. This gives a natural upper bound for
the prior on $\xend$. More specifically, these terms are of the form
$\Delta V = \phi^{q+4}/\Mp^q$ such that the corrected potential
exhibits a global minimum at a field value $\xVmin$. Requiring $\xend
\ll \xVmin$ gives the absolute upper bound
\begin{equation}
\xend \ll \xendmax \equiv \left[43200\pi^2\frac{p^3}{q+4} \Pstar 
\left(\frac{\Mp}{\mu}\right)^{q+6} \right]^{1/(3p+q+6)}.
\end{equation}
As a motivated case, we have chosen $q=8$ and a Jeffreys' prior on
$\xend$ in $[\xendmin,\xendmax]$. The case $p=2$ has been considered
as an independent model as it corresponds to the so-called inverse
mutated scenarios. In summary, the following models and priors have
been considered:

\begin{center}
\begin{baytabular}
& $ p \in ]1,6]$ & & & & \\
$\dsi$ & $\log(\mu/\Mp) \in [-5,0]$ & $\Edsi$ & $\Cdsi$ & $\NPdsi$ &
$\BEdsi$ \\
& $\log \left( \dfrac{\xend-\xendmin}{\xendmax-\xendmin} \right) \in [-5,-0.7]$ & & & &\\
\hline
& $p\in ]1,6]$ & & & &\\
$\dsio$ & $\mu/\Mp \in \left[10^{-9},10^{-6}\right]$ & $\Edsio$ & $\Cdsio$ &
$\NPdsio$ & $\BEdsio$  \\
&  $\log \left( \dfrac{\xend-\xendmin}{\xendmax-\xendmin} \right)\in [-5,-0.7]$ & & & & \\
\hline
& $p=2$ & & & & \\
$\dsiTWO$ & $\mu/\Mp \in \left[10^{-9},10^{-6}\right]$ & $\EdsiTWO$ & $\CdsiTWO$
& $\NPdsiTWO$ & $\BEdsiTWO$ \\
&  $\log \left( \dfrac{\xend-\xendmin}{\xendmax-\xendmin} \right) \in [-5,-0.7]$ & & & &\\
\end{baytabular}
\end{center}

\subsection{Generalised Mixed Large Field Inflation (GMLFI)}

The potential mixes two large field monomials and reads
\begin{equation}
V(\phi) = M^4 \left(\dfrac{\phi}{\Mp} \right)^p \left[1 + \alpha
  \left(\dfrac{\phi}{\Mp} \right)^q \right],
\end{equation}
where $p$ and $q$ are power indices and $\alpha$ a constant. The model
has three parameters and their priors have been chosen on
phenomenological grounds. In particular, because GMLFI allows to
discuss the effects stemming from combining together two LFI models,
it motivates to fix $p$ and $q$ to all the possible theoretically
motivated combination of pure LFI models. One can also view GMLFI as a
new class of models and let $p$, $q$ and $\alpha$ freely varying. A
priori, the parameter $\alpha$ can be very small such that it should
be sampled along a Jeffreys prior. We have considered the following
cases:

\begin{center}
\begin{baytabular}
& $p\in[1,6]$ & & & & \\
$\gmlfi$ & $ \log(\alpha) \in [-5,1]$ & $\Egmlfi$ & $\Cgmlfi$ &
  $\NPgmlfi$ & $\BEgmlfi$  \\
& $q\in[1,6]$ & & & &\\
\hline
& $p=2/3$ & & & &\\
$\gmlfiTWOTHREEONETHREE$ & $\log(\alpha) \in [-5,1]$ & $\EgmlfiTWOTHREEONETHREE$ &
  $\CgmlfiTWOTHREEONETHREE$ & $\NPgmlfiTWOTHREEONETHREE$ &
$\BEgmlfiTWOTHREEONETHREE$ \\
& $q=1/3$ & & & &\\
\hline
& $p=2/3$ & & & & \\
$\gmlfiTWOTHREEFOURTHREE$ & $\log(\alpha) \in [-5,1]$ &
$\EgmlfiTWOTHREEFOURTHREE$ & $\CgmlfiTWOTHREEFOURTHREE$ &
$\NPgmlfiTWOTHREEFOURTHREE$ & $\BEgmlfiTWOTHREEFOURTHREE$ \\
& $q=4/3$ & & & & \\
\hline
& $p=1$ & & & & \\
$\gmlfiONEONE$ & $\log(\alpha) \in [-5,1]$ & $\EgmlfiONEONE$ &
  $\CgmlfiONEONE$ & $\NPgmlfiONEONE$ & $\BEgmlfiONEONE$ \\
& $q=1$ & & & & \\
\hline
& $p=1$ & & & & \\
$\gmlfiONETWO$ & $\log(\alpha) \in [-5,1]$ & $\EgmlfiONETWO$ &
  $\CgmlfiONETWO$ & $\NPgmlfiONETWO$ & $\BEgmlfiONETWO$ \\
& $q=2$ & & & & \\
\hline
& $p=1$ & & & & \\
$\gmlfiONETHREE$ & $\log(\alpha) \in [-5,1]$ & $\EgmlfiONETHREE$ &
  $\CgmlfiONETHREE$ & $\NPgmlfiONETHREE$ & $\BEgmlfiONETHREE$ \\
& $q=3$ & & & & \\
\hline
& $p=2$ & & & & \\
$\gmlfiTWOONE$ & $\log(\alpha) \in [-5,1]$ & $\EgmlfiTWOONE$ &
  $\CgmlfiTWOONE$ & $\NPgmlfiTWOONE$ & $\BEgmlfiTWOONE$ \\
& $q=1$ & & & & \\
\hline
& $p=2$ & & & & \\
$\gmlfiTWOTWO$ & $\log(\alpha) \in [-5,1]$ & $\EgmlfiTWOTWO$ &
  $\CgmlfiTWOTWO$ & $\NPgmlfiTWOTWO$ & $\BEgmlfiTWOTWO$ \\
& $q=2$ & & & & \\
\hline
& $p=2$ & & & & \\
$\gmlfiTWOTHREE$ & $\log(\alpha) \in [-5,1]$ & $\EgmlfiTWOTHREE$ &
  $\CgmlfiTWOTHREE$ & $\NPgmlfiTWOTHREE$ & $\BEgmlfiTWOTHREE$ \\
& $q=3$ & & & & \\
\hline
& $p=3$ & & & & \\
$\gmlfiTHREEONE$ & $\log(\alpha) \in [-5,1]$ & $\EgmlfiTHREEONE$ &
  $\CgmlfiTHREEONE$ & $\NPgmlfiTHREEONE $ & $\BEgmlfiTHREEONE$ \\
& $q=1$ & & & & \\
\hline
& $p=3$ & & & & \\
$\gmlfiTHREETWO$ & $\log(\alpha) \in [-5,1]$ & $\EgmlfiTHREETWO$ &
  $\CgmlfiTHREETWO$ & $\NPgmlfiTHREETWO$ & $\BEgmlfiTHREETWO$ \\
& $q=2$ & & & & \\
\hline
& $p=3$ & & & & \\
$\gmlfiTHREETHREE$ & $\log(\alpha) \in [-5,1]$ & $\EgmlfiTHREETHREE$ &
  $\CgmlfiTHREETHREE$ & $\NPgmlfiTHREETHREE$ & $\BEgmlfiTHREETHREE$ \\
& $q=3$ & & & &  \\
\end{baytabular}
\end{center}
Notice that the case $p=2$, $q=2$ is also referred to as $\gmlfiTWOTWO$
in Ref.~\cite{Martin:2013tda}.

\subsection{Logarithmic Potential Inflation (LPI)}

These scenarios are parametrised by a potential of the form
\begin{equation}
V(\phi) = M^4 \left(\dfrac{\phi}{\phizero} \right)^p \left( \ln
\dfrac{\phi}{\phizero} \right)^q.
\end{equation}
Some specific combinations of $p$ and $q$ match various Yang-Mills
composite models $\lpiONEFOURONE$ ($p=4$, $q=1$), $\lpiONEFOURTWO$
($p=1$, $q=2$) and $\lpiONEFOURTHREE$ ($p=4$,
$q=3$)~\cite{Bezrukov:2011mv,Channuie:2012bv}. Others combinations are
phenomenological~\cite{Barrow:1995xb}. Because the potential admits a
local maximum at $x=\xVmax$, with $x=\phi/\phizero$, inflation can
take place in three domains: LPI1 for $x>1$, LPI2 for $\xVmax <x <1$
and LPI3 for $x<\xVmax$. Let us notice that for both LPI2 and LPI3,
the potential is well-defined only if $q$ is an even integer. For
LPI1, both $p$ and $q$ can take arbitrary real values. The {\vev}
$\phizero$ is not constrained for LPI1, and we have chosen a Jeffreys
prior encompassing both sub-Planckian and super-Planckian values. On
the contrary, in the LPI1 and LPI2 domains, $\phizero$ must be deeply
super-Planckian to allow for slow-rolling inflation. The models and
priors considered are listed below.

\begin{center}
\begin{baytabular}
& $p \in [1,6]$ & & & & \\
$\lpiONE$ & $\log(\phizero/\Mp) \in [-3,3]$ & $\ElpiONE$ & $\ClpiONE$
  & $\NPlpiONE$ & $\BElpiONE$ 
  \\
& $q\in[1,6]$ & & & & \\
\hline
& $p=4$ & & & & \\
$\lpiONEFOURONE$ & $\log(\phizero/\Mp) \in [-3,3]$ & $\ElpiONEFOURONE$
& $\ClpiONEFOURONE$ & $\NPlpiONEFOURONE$ & $\BElpiONEFOURONE$ 
  \\
& $q=1$ & & & & \\
\hline
& $p=4$ & & & & \\
$\lpiONEFOURTWO$ & $\log(\phizero/\Mp) \in [-3,3]$ & $\ElpiONEFOURTWO$
& $\ClpiONEFOURTWO$ & $\NPlpiONEFOURTWO$ & $\BElpiONEFOURTWO$ 
  \\
& $q=2$ & & & & \\
\hline
& $p=4$ & & & & \\
$\lpiONEFOURTHREE$ & $\log(\phizero/\Mp) \in [-3,3]$ & $\ElpiONEFOURTHREE$
& $\ClpiONEFOURTHREE$ & $\NPlpiONEFOURTHREE$ & $\BElpiONEFOURTHREE$
  \\
& $q=3$ & & & &  \\
\hline
& $p\in[1,6]$ & & & & \\
$\lpiTWOTWO$ & $\log(\phizero/\Mp) \in [2,5]$ & $\ElpiTWOTWO$
& $\ClpiTWOTWO$ & $\NPlpiTWOTWO$ & $\BElpiTWOTWO$ 
  \\
& $q=2$ & & & &  \\
\hline
& $p\in[1,6]$ & & & &  \\
$\lpiTWOFOUR$ & $\log(\phizero/\Mp) \in [2,5]$ & $\ElpiTWOFOUR$
& $\ClpiTWOFOUR$ & $\NPlpiTWOFOUR$ & $\BElpiTWOFOUR$ 
  \\
& $q=4$ & & & &\\
\hline
& $p\in[1,6]$ & & & &  \\
$\lpiTWOSIX$ & $\log(\phizero/\Mp) \in [2,5]$ & $\ElpiTWOSIX$
& $\ClpiTWOSIX$ & $\NPlpiTWOSIX$ & $\BElpiTWOSIX$
  \\
& $q=6$ & & & &\\
\hline
& $p\in[1,6]$ & & & &  \\
$\lpiTHREETWO$ & $\log(\phizero/\Mp) \in [2,5]$ & $\ElpiTHREETWO$
& $\ClpiTHREETWO$ & $\NPlpiTHREETWO$ & $\BElpiTHREETWO$ \\
& $q=2$ & & & &\\
\hline
& $p\in[1,6]$ & & & & \\
$\lpiTHREEFOUR$ & $\log(\phizero/\Mp) \in [2,5]$ & $\ElpiTHREEFOUR$
& $\ClpiTHREEFOUR$ & $\NPlpiTHREEFOUR$ & $\BElpiTHREEFOUR$ 
  \\
& $q=4$ & & & &\\
\hline
& $p\in[1,6]$ & & & &  \\
$\lpiTHREESIX$ & $\log(\phizero/\Mp) \in [2,5]$ & $\ElpiTHREESIX$
& $\ClpiTHREESIX$ & $\NPlpiTHREESIX$ & $\BElpiTHREESIX$
  \\
& $q=6$ & & & & \\
\end{baytabular}
\end{center}

\subsection{Constant $\nS$ D Inflation (CNDI)}

The potential has two parameters $\alpha$, $\beta$ and reads
\begin{equation}
V(\phi) = \dfrac{M^4}{\left[1 + \beta \cos\left(\alpha \dfrac{\phi}{\Mp}
\right) \right]^{2}}\,.
\end{equation}
As discussed in Ref.~\cite{Martin:2013tda}, the only regime of
cosmological interest has $x=\phi/\Mp$ small and $\beta>1$. In that
situation, the field value at which inflation ends should be
specified, namely $\xend=\phiend/\Mp$. Moreover, if $x$ becomes too
large, inflation cannot even start because there exists a
``forbidden'' range of field values in which $\epsilon_1(x)>1$. As a
result, there is a maximal value for $\xini = \xepsoneOneMinus$,
$\xepsoneOneMinus$ being the smallest root of the equation
$\epsilon_1=1$. As for the other models, by imposing to get at least
$120$ {\efolds} of inflation, the maximal values of $\xini$ translates
into a maximal value $\xendmax$ thereby constituting the upper bound
of the $\xend$'s prior. Concerning the parameter $\alpha$, the genuine
CNDI model is designed to produce a constant spectral index and, as
discussed in Ref.~\cite{Martin:2013tda}, this occurs for not too
small, neither not too large values of $\alpha$. The priors chosen are
summarised in the following table.
\begin{center}
\begin{baytabular}
& $\beta\in[1.1,6]$ & & & & \\
$\cndi$ & $\log(\alpha)\in[-2,-1]$ & $\Ecndi$ & $\Ccndi$ & $\NPcndi$ &
  $\BEcndi$ \\
& $\xend \in [0,\xendmax]$ & & & & \\
\end{baytabular}
\end{center}

\bibliographystyle{JHEP}
\bibliography{biblio}

\providecommand{\href}[2]{#2}\begingroup\raggedright\begin{thebibliography}{10}

\bibitem{Ade:2013zuv}
{\bf Planck Collaboration} Collaboration, P.~Ade et~al., {\it {Planck 2013
  results. XVI. Cosmological parameters}},
  \href{http://xxx.lanl.gov/abs/1303.5076}{{\tt arXiv:1303.5076}}.

\bibitem{Trotta:2007hy}
R.~Trotta, {\it {Forecasting the Bayes factor of a future observation}},  {\em
  Mon.Not.Roy.Astron.Soc.} {\bf 378} (2007) 819--824,
  [\href{http://xxx.lanl.gov/abs/astro-ph/0703063}{{\tt astro-ph/0703063}}].

\bibitem{Ade:2013uln}
{\bf Planck Collaboration} Collaboration, P.~Ade et~al., {\it {Planck 2013
  results. XXII. Constraints on inflation}},
  \href{http://xxx.lanl.gov/abs/1303.5082}{{\tt arXiv:1303.5082}}.

\bibitem{Ade:2013ydc}
{\bf Planck Collaboration} Collaboration, P.~Ade et~al., {\it {Planck 2013
  Results. XXIV. Constraints on primordial non-Gaussianity}},
  \href{http://xxx.lanl.gov/abs/1303.5084}{{\tt arXiv:1303.5084}}.

\bibitem{Martin:2013tda}
J.~Martin, C.~Ringeval, and V.~Vennin, {\it {Encyclopaedia Inflationaris}},
  \href{http://xxx.lanl.gov/abs/1303.3787}{{\tt arXiv:1303.3787}}.

\bibitem{Ringeval:2013lea}
C.~Ringeval, {\it {Fast Bayesian inference for slow-roll inflation}},
  \href{http://xxx.lanl.gov/abs/1312.2347}{{\tt arXiv:1312.2347}}.

\bibitem{Cox:1946}
R.~T. {Cox}, {\it {Probability, Frequency and Reasonable Expectation}},  {\em
  American Journal of Physics} {\bf 14} (Jan., 1946) 1--13.

\bibitem{Jeffreys:1961}
H.~{Jeffreys}, {\it {Theory of probability}},  {\em Oxford University Press,
  Oxford Classics series (reprinted 1998)} (1961).

\bibitem{deFinetti:1974}
B.~\protect{de Finetti}, {\em Theory of probability}.
\newblock John Wiley \& Sons, Chichester, UK, 1974.
\newblock reprinted 1995.

\bibitem{Bernardo:1994}
J.~M. Bernardo and A.~F.~M. Smith, {\em Bayesian Theory}.
\newblock John Wiley \& Sons, Chicester, UK, 1994.

\bibitem{Box:1992}
G.~E.~P. Box and G.~C. Tiao, {\em Bayesian Inference in Statistical Analysis}.
\newblock John Wiley \& Sons, Chicester, UK, 1992.

\bibitem{JaynesBook}
E.~T. Jaynes, {\em Probability Theory. The logic of science}.
\newblock Cambridge University Press, Cambridge, UK, 2003.

\bibitem{Berger:2003}
J.~Berger, {\it Could fisher, jeffreys and neyman have agreed on testing?},
  {\em Statistical Science} {\bf 18} (2003), no.~1 1--12. rejoinder: {\em
  ibid.}, 28-32.

\bibitem{Trotta:2005ar}
R.~Trotta, {\it {Applications of Bayesian model selection to cosmological
  parameters}},  {\em Mon. Not. Roy. Astron. Soc.} {\bf 378} (2007) 72--82,
  [\href{http://xxx.lanl.gov/abs/astro-ph/0504022}{{\tt astro-ph/0504022}}].

\bibitem{Trotta:2008qt}
R.~Trotta, {\it {Bayes in the sky: Bayesian inference and model selection in
  cosmology}},  {\em Contemp. Phys.} {\bf 49} (2008) 71--104,
  [\href{http://xxx.lanl.gov/abs/0803.4089}{{\tt arXiv:0803.4089}}].

\bibitem{Gordon:2007xm}
C.~Gordon and R.~Trotta, {\it {Bayesian Calibrated Significance Levels Applied
  to the Spectral Tilt and Hemispherical Asymmetry}},  {\em Mon. Not. Roy.
  Astron. Soc.} {\bf 382} (2007) 1859--1863,
  [\href{http://xxx.lanl.gov/abs/0706.3014}{{\tt arXiv:0706.3014}}].

\bibitem{March:2010ex}
M.~March, G.~Starkman, R.~Trotta, and P.~Vaudrevange, {\it {Should we doubt the
  cosmological constant?}},  {\em Mon.Not.Roy.Astron.Soc.} {\bf 410} (2011)
  2488--2496, [\href{http://xxx.lanl.gov/abs/1005.3655}{{\tt
  arXiv:1005.3655}}].

\bibitem{Berger:1987}
J.~O. Berger and T.~Sellke, {\it {Testing a Point Null Hypothesis: The
  Irreconcilability of P Values and Evidence}},  {\em J. Am. Stat. Ass} {\bf
  82} (1987) 112--122.

\bibitem{Berger:2004}
J.~O. Berger and M.~Bayarri, {\it {The interplay of Bayesian and Frequentist
  analysis}},  {\em Stat. Science} {\bf 19} (2004) 58--80.

\bibitem{Johnson:2013}
V.~E. Johnson, {\it {Revised standards for statistical evidence}},  {\em PNAS}
  (2013) Published online before print: doi: 10.1073/pnas.1313476110.

\bibitem{Cousins2013}
R.~D. Cousins, {\it {The Jeffreys-Lindley Paradox and Discovery Criteria in
  High Energy Physics}},  \href{http://xxx.lanl.gov/abs/1310.3791}{{\tt
  arXiv:1310.3791}}.

\bibitem{Kullback:1951}
S.~Kullback and R.~Leibler, {\it {}},  {\em Ann. Math. Stat.} {\bf 22} (1951)
  79--86.

\bibitem{Kunz:2006mc}
M.~Kunz, R.~Trotta, and D.~Parkinson, {\it {Measuring the effective complexity
  of cosmological models}},  {\em Phys. Rev.} {\bf D74} (2006) 023503,
  [\href{http://xxx.lanl.gov/abs/astro-ph/0602378}{{\tt astro-ph/0602378}}].

\bibitem{Spiegelhalter}
D.~Spiegelhalter et~al., {\it {Bayesian Measures of Model Complexity and Fit}},
   {\em J. Roy. Stat. Soc. B} {\bf 64} (2002) 583--639.

\bibitem{Feroz:2011bj}
F.~Feroz, K.~Cranmer, M.~Hobson, R.~Ruiz~de Austri, and R.~Trotta, {\it
  {Challenges of Profile Likelihood Evaluation in Multi-Dimensional SUSY
  Scans}},  {\em JHEP} {\bf 1106} (2011) 042,
  [\href{http://xxx.lanl.gov/abs/1101.3296}{{\tt arXiv:1101.3296}}].

\bibitem{Feroz:2007kg}
F.~{Feroz} and M.~P. {Hobson}, {\it {Multimodal nested sampling: an efficient
  and robust alternative to Markov Chain Monte Carlo methods for astronomical
  data analyses}},  {\em Mon. Not. R. Astron. Soc.} {\bf 384} (Feb., 2008)
  449--463, [\href{http://xxx.lanl.gov/abs/0704.3704}{{\tt arXiv:0704.3704}}].

\bibitem{Feroz:2008xx}
F.~{Feroz}, M.~P. {Hobson}, and M.~{Bridges}, {\it {MULTINEST: an efficient and
  robust Bayesian inference tool for cosmology and particle physics}},  {\em
  Mon. Not. R. Astron. Soc.} {\bf 398} (Oct., 2009) 1601--1614,
  [\href{http://xxx.lanl.gov/abs/0809.3437}{{\tt arXiv:0809.3437}}].

\bibitem{Martin:2006rs}
J.~Martin and C.~Ringeval, {\it {Inflation after WMAP3: Confronting the
  slow-roll and exact power spectra to CMB data}},  {\em JCAP} {\bf 0608}
  (2006) 009, [\href{http://xxx.lanl.gov/abs/astro-ph/0605367}{{\tt
  astro-ph/0605367}}].

\bibitem{Martin:2010kz}
J.~Martin and C.~Ringeval, {\it {First CMB Constraints on the Inflationary
  Reheating Temperature}},  {\em Phys. Rev.} {\bf D82} (2010) 023511,
  [\href{http://xxx.lanl.gov/abs/1004.5525}{{\tt arXiv:1004.5525}}].

\bibitem{Martin:2010hh}
J.~Martin, C.~Ringeval, and R.~Trotta, {\it {Hunting Down the Best Model of
  Inflation with Bayesian Evidence}},  {\em Phys.Rev.} {\bf D83} (2011) 063524,
  [\href{http://xxx.lanl.gov/abs/1009.4157}{{\tt arXiv:1009.4157}}].

\bibitem{Easther:2011yq}
R.~Easther and H.~V. Peiris, {\it {Bayesian Analysis of Inflation II: Model
  Selection and Constraints on Reheating}},  {\em Phys.Rev.} {\bf D85} (2012)
  103533, [\href{http://xxx.lanl.gov/abs/1112.0326}{{\tt arXiv:1112.0326}}].

\bibitem{Norena:2012rs}
J.~Norena, C.~Wagner, L.~Verde, H.~V. Peiris, and R.~Easther, {\it {Bayesian
  Analysis of Inflation III: Slow Roll Reconstruction Using Model Selection}},
  {\em Phys.Rev.} {\bf D86} (2012) 023505,
  [\href{http://xxx.lanl.gov/abs/1202.0304}{{\tt arXiv:1202.0304}}].

\bibitem{Ringeval:2007am}
C.~Ringeval, {\it {The exact numerical treatment of inflationary models}},
  {\em Lect. Notes Phys.} {\bf 738} (2008) 243--273,
  [\href{http://xxx.lanl.gov/abs/astro-ph/0703486}{{\tt astro-ph/0703486}}].

\bibitem{Martin:2007ue}
J.~Martin and J.~Yokoyama, {\it {Generation of Large-Scale Magnetic Fields in
  Single-Field Inflation}},  {\em JCAP} {\bf 0801} (2008) 025,
  [\href{http://xxx.lanl.gov/abs/0711.4307}{{\tt arXiv:0711.4307}}].

\bibitem{Demozzi:2012wh}
V.~Demozzi and C.~Ringeval, {\it {Reheating constraints in inflationary
  magnetogenesis}},  {\em JCAP} {\bf 1205} (2012) 009,
  [\href{http://xxx.lanl.gov/abs/1202.3022}{{\tt arXiv:1202.3022}}].

\bibitem{Kuroyanagi:2013ns}
S.~Kuroyanagi, C.~Ringeval, and T.~Takahashi, {\it {Early Universe Tomography
  with CMB and Gravitational Waves}},  {\em Phys.Rev.} {\bf D87} (2013) 083502,
  [\href{http://xxx.lanl.gov/abs/1301.1778}{{\tt arXiv:1301.1778}}].

\bibitem{Ringeval:2013hfa}
C.~Ringeval, T.~Suyama, and J.~Yokoyama, {\it {Magneto-reheating constraints
  from curvature perturbations}},  {\em JCAP} {\bf 1309} (2013) 020,
  [\href{http://xxx.lanl.gov/abs/1302.6013}{{\tt arXiv:1302.6013}}].

\bibitem{Planck:2013kta}
{\bf Planck collaboration} Collaboration, P.~Ade et~al., {\it {Planck 2013
  results. XV. CMB power spectra and likelihood}},
  \href{http://xxx.lanl.gov/abs/1303.5075}{{\tt arXiv:1303.5075}}.

\bibitem{Schwarz:2001vv}
D.~J. Schwarz, C.~A. Terrero-Escalante, and A.~A. Garcia, {\it Higher order
  corrections to primordial spectra from cosmological inflation},  {\em Phys.
  Lett.} {\bf B517} (2001) 243--249,
  [\href{http://xxx.lanl.gov/abs/astro-ph/0106020}{{\tt astro-ph/0106020}}].

\bibitem{Schwarz:2004tz}
D.~J. Schwarz and C.~A. Terrero-Escalante, {\it Primordial fluctuations and
  cosmological inflation after wmap 1.0},  {\em JCAP} {\bf 0408} (2004) 003,
  [\href{http://xxx.lanl.gov/abs/hep-ph/0403129}{{\tt hep-ph/0403129}}].

\bibitem{Martin:2013uma}
J.~Martin, C.~Ringeval, and V.~Vennin, {\it {K-inflationary Power Spectra at
  Second Order}},  {\em JCAP} {\bf 1306} (2013) 021,
  [\href{http://xxx.lanl.gov/abs/1303.2120}{{\tt arXiv:1303.2120}}].

\bibitem{Ade:2013ktc}
{\bf Planck Collaboration} Collaboration, P.~Ade et~al., {\it {Planck 2013
  results. I. Overview of products and scientific results}},
  \href{http://xxx.lanl.gov/abs/1303.5062}{{\tt arXiv:1303.5062}}.

\bibitem{Lewis:2002ah}
A.~Lewis and S.~Bridle, {\it Cosmological parameters from cmb and other data: a
  monte- carlo approach},  {\em Phys. Rev.} {\bf D66} (2002) 103511,
  [\href{http://xxx.lanl.gov/abs/astro-ph/0205436}{{\tt astro-ph/0205436}}].

\bibitem{Lewis:1999bs}
A.~Lewis, A.~Challinor, and A.~Lasenby, {\it Efficient computation of cmb
  anisotropies in closed frw models},  {\em Astrophys. J.} {\bf 538} (2000)
  473--476, [\href{http://xxx.lanl.gov/abs/astro-ph/9911177}{{\tt
  astro-ph/9911177}}].

\bibitem{Shepard:1968}
D.~Shepard, {\it A two-dimensional interpolation function for
  irregularly-spaced data},  in {\em Proceedings of the 1968 23rd ACM national
  conference}, ACM '68, (New York, NY, USA), pp.~517--524, ACM, 1968.

\bibitem{Thacker:2010}
W.~I. Thacker, J.~Zhang, L.~T. Watson, J.~B. Birch, M.~A. Iyer, and M.~W.
  Berry, {\it Algorithm 905: Sheppack: Modified shepard algorithm for
  interpolation of scattered multivariate data},  {\em ACM Trans. Math. Softw.}
  {\bf 37} (Sept., 2010) 34:1--34:20.

\bibitem{Mazumdar:2013gya}
A.~Mazumdar and B.~Zaldivar, {\it {Quantifying the reheating temperature of the
  universe}},  \href{http://xxx.lanl.gov/abs/1310.5143}{{\tt arXiv:1310.5143}}.

\bibitem{Starobinsky:1980te}
A.~A. Starobinsky, {\it {A New Type of Isotropic Cosmological Models Without
  Singularity}},  {\em Phys.Lett.} {\bf B91} (1980) 99--102.

\bibitem{Ellis:2013nxa}
J.~Ellis, D.~V. Nanopoulos, and K.~A. Olive, {\it {Starobinsky-like
  Inflationary Models as Avatars of No-Scale Supergravity}},  {\em JCAP} {\bf
  1310} (2013) 009, [\href{http://xxx.lanl.gov/abs/1307.3537}{{\tt
  arXiv:1307.3537}}].

\bibitem{Ellis:2013xoa}
J.~Ellis, D.~V. Nanopoulos, and K.~A. Olive, {\it {No-Scale Supergravity
  Realization of the Starobinsky Model of Inflation}},  {\em Phys.Rev.Lett.}
  {\bf 111} (2013) 111301, [\href{http://xxx.lanl.gov/abs/1305.1247}{{\tt
  arXiv:1305.1247}}].

\bibitem{Conlon:2005jm}
J.~P. Conlon and F.~Quevedo, {\it {Kahler moduli inflation}},  {\em JHEP} {\bf
  0601} (2006) 146, [\href{http://xxx.lanl.gov/abs/hep-th/0509012}{{\tt
  hep-th/0509012}}].

\bibitem{Krippendorf:2009zza}
S.~Krippendorf and F.~Quevedo, {\it {Metastable SUSY Breaking, de Sitter Moduli
  Stabilisation and Kahler Moduli Inflation}},  {\em JHEP} {\bf 0911} (2009)
  039, [\href{http://xxx.lanl.gov/abs/0901.0683}{{\tt arXiv:0901.0683}}].

\bibitem{Burgess:2013sla}
C.~Burgess, M.~Cicoli, and F.~Quevedo, {\it {String Inflation After Planck
  2013}},  {\em JCAP} {\bf 1311} (2013) 003,
  [\href{http://xxx.lanl.gov/abs/1306.3512}{{\tt arXiv:1306.3512}}].

\bibitem{Ijjas:2013vea}
A.~Ijjas, P.~J. Steinhardt, and A.~Loeb, {\it {Inflationary paradigm in trouble
  after Planck2013}},  {\em Phys.Lett.} {\bf B723} (2013) 261--266,
  [\href{http://xxx.lanl.gov/abs/1304.2785}{{\tt arXiv:1304.2785}}].

\bibitem{Germani:2010gm}
C.~Germani and A.~Kehagias, {\it {New Model of Inflation with Non-minimal
  Derivative Coupling of Standard Model Higgs Boson to Gravity}},  {\em
  Phys.Rev.Lett.} {\bf 105} (2010) 011302,
  [\href{http://xxx.lanl.gov/abs/1003.2635}{{\tt arXiv:1003.2635}}].

\bibitem{Germani:2010hd}
C.~Germani and A.~Kehagias, {\it {UV-Protected Inflation}},  {\em
  Phys.Rev.Lett.} {\bf 106} (2011) 161302,
  [\href{http://xxx.lanl.gov/abs/1012.0853}{{\tt arXiv:1012.0853}}].

\bibitem{Germani:2011ua}
C.~Germani and Y.~Watanabe, {\it {UV-protected (Natural) Inflation: Primordial
  Fluctuations and non-Gaussian Features}},  {\em JCAP} {\bf 1107} (2011) 031,
  [\href{http://xxx.lanl.gov/abs/1106.0502}{{\tt arXiv:1106.0502}}].

\bibitem{Futamase:1987ua}
T.~Futamase and K.-i. Maeda, {\it {Chaotic Inflationary Scenario in Models
  Having Nonminimal Coupling With Curvature}},  {\em Phys.Rev.} {\bf D39}
  (1989) 399--404.

\bibitem{Einhorn:2009bh}
M.~B. Einhorn and D.~T. Jones, {\it {Inflation with Non-minimal Gravitational
  Couplings in Supergravity}},  {\em JHEP} {\bf 1003} (2010) 026,
  [\href{http://xxx.lanl.gov/abs/0912.2718}{{\tt arXiv:0912.2718}}].

\bibitem{Buchmuller:2012ex}
W.~Buchmüller, V.~Domcke, and K.~Schmitz, {\it {Superconformal D-Term
  Inflation}},  {\em JCAP} {\bf 1304} (2013) 019,
  [\href{http://xxx.lanl.gov/abs/1210.4105}{{\tt arXiv:1210.4105}}].

\bibitem{Kallosh:2013hoa}
R.~Kallosh and A.~Linde, {\it {Universality Class in Conformal Inflation}},
  {\em JCAP} {\bf 1307} (2013) 002,
  [\href{http://xxx.lanl.gov/abs/1306.5220}{{\tt arXiv:1306.5220}}].

\bibitem{Kallosh:2013pby}
R.~Kallosh and A.~Linde, {\it {Superconformal generalization of the chaotic
  inflation model $\frac{\lambda}{4} \phi^{4} - \frac{\xi}{2} \phi^{2}R$}},
  {\em JCAP} {\bf 1306} (2013) 027,
  [\href{http://xxx.lanl.gov/abs/1306.3211}{{\tt arXiv:1306.3211}}].

\bibitem{Kallosh:2013lkr}
R.~Kallosh and A.~Linde, {\it {Superconformal generalizations of the
  Starobinsky model}},  {\em JCAP} {\bf 1306} (2013) 028,
  [\href{http://xxx.lanl.gov/abs/1306.3214}{{\tt arXiv:1306.3214}}].

\bibitem{Ferrara:2013rsa}
S.~Ferrara, R.~Kallosh, A.~Linde, and M.~Porrati, {\it {Minimal Supergravity
  Models of Inflation}},  \href{http://xxx.lanl.gov/abs/1307.7696}{{\tt
  arXiv:1307.7696}}.

\bibitem{Kallosh:2013daa}
R.~Kallosh and A.~Linde, {\it {Multi-field Conformal Cosmological Attractors}},
   \href{http://xxx.lanl.gov/abs/1309.2015}{{\tt arXiv:1309.2015}}.

\bibitem{Kallosh:2013tua}
R.~Kallosh, A.~Linde, and D.~Roest, {\it {A universal attractor for inflation
  at strong coupling}},  \href{http://xxx.lanl.gov/abs/1310.3950}{{\tt
  arXiv:1310.3950}}.

\bibitem{Silverstein:2008sg}
E.~Silverstein and A.~Westphal, {\it {Monodromy in the CMB: Gravity Waves and
  String Inflation}},  {\em Phys. Rev.} {\bf D78} (2008) 106003,
  [\href{http://xxx.lanl.gov/abs/0803.3085}{{\tt arXiv:0803.3085}}].

\bibitem{delaMacorra:1995qh}
A.~de~la Macorra and S.~Lola, {\it {Inflation in S dual superstring models}},
  {\em Phys.Lett.} {\bf B373} (1996) 299--305,
  [\href{http://xxx.lanl.gov/abs/hep-ph/9511470}{{\tt hep-ph/9511470}}].

\bibitem{Dudas:2012vv}
E.~Dudas, N.~Kitazawa, S.~Patil, and A.~Sagnotti, {\it {CMB Imprints of a
  Pre-Inflationary Climbing Phase}},  {\em JCAP} {\bf 1205} (2012) 012,
  [\href{http://xxx.lanl.gov/abs/1202.6630}{{\tt arXiv:1202.6630}}].

\bibitem{Pajer:2008uy}
E.~Pajer, {\it {Inflation at the Tip}},  {\em JCAP} {\bf 0804} (2008) 031,
  [\href{http://xxx.lanl.gov/abs/0802.2916}{{\tt arXiv:0802.2916}}].

\bibitem{ArkaniHamed:2003mz}
N.~Arkani-Hamed, H.-C. Cheng, P.~Creminelli, and L.~Randall, {\it
  {Pseudonatural inflation}},  {\em JCAP} {\bf 0307} (2003) 003,
  [\href{http://xxx.lanl.gov/abs/hep-th/0302034}{{\tt hep-th/0302034}}].

\bibitem{Boubekeur:2005zm}
L.~Boubekeur and D.~Lyth, {\it {Hilltop inflation}},  {\em JCAP} {\bf 0507}
  (2005) 010, [\href{http://xxx.lanl.gov/abs/hep-ph/0502047}{{\tt
  hep-ph/0502047}}]. Latex, 20 pages, 5 figures. Minor changes, references
  added.

\bibitem{Channuie:2012bv}
P.~Channuie, J.~J. Jorgensen, and F.~Sannino, {\it {Composite Inflation from
  Super Yang-Mills, Orientifold and One-Flavor QCD}},  {\em Phys.Rev.} {\bf
  D86} (2012) 125035, [\href{http://xxx.lanl.gov/abs/1209.6362}{{\tt
  arXiv:1209.6362}}].

\bibitem{Kachru:2003aw}
S.~Kachru, R.~Kallosh, A.~D. Linde, and S.~P. Trivedi, {\it {De Sitter vacua in
  string theory}},  {\em Phys.Rev.} {\bf D68} (2003) 046005,
  [\href{http://xxx.lanl.gov/abs/hep-th/0301240}{{\tt hep-th/0301240}}].

\bibitem{Kachru:2003sx}
S.~Kachru, R.~Kallosh, A.~D. Linde, J.~M. Maldacena, L.~P. McAllister, et~al.,
  {\it {Towards inflation in string theory}},  {\em JCAP} {\bf 0310} (2003)
  013, [\href{http://xxx.lanl.gov/abs/hep-th/0308055}{{\tt hep-th/0308055}}].

\bibitem{Lorenz:2007ze}
L.~Lorenz, J.~Martin, and C.~Ringeval, {\it {Brane inflation and the WMAP data:
  a Bayesian analysis}},  {\em JCAP} {\bf 0804} (2008) 001,
  [\href{http://xxx.lanl.gov/abs/0709.3758}{{\tt arXiv:0709.3758}}].

\bibitem{Clesse:2008pf}
S.~Clesse and J.~Rocher, {\it {Avoiding the blue spectrum and the fine-tuning
  of initial conditions in hybrid inflation}},  {\em Phys.Rev.} {\bf D79}
  (2009) 103507, [\href{http://xxx.lanl.gov/abs/0809.4355}{{\tt
  arXiv:0809.4355}}].

\bibitem{Clesse:2009ur}
S.~Clesse, C.~Ringeval, and J.~Rocher, {\it {Fractal initial conditions and
  natural parameter values in hybrid inflation}},  {\em Phys. Rev.} {\bf D80}
  (2009) 123534, [\href{http://xxx.lanl.gov/abs/0909.0402}{{\tt
  arXiv:0909.0402}}].

\bibitem{Clesse:2010iz}
S.~Clesse, {\it {Hybrid inflation along waterfall trajectories}},  {\em
  Phys.Rev.} {\bf D83} (2011) 063518,
  [\href{http://xxx.lanl.gov/abs/1006.4522}{{\tt arXiv:1006.4522}}].

\bibitem{Clesse:2012dw}
S.~Clesse and B.~Garbrecht, {\it {Slow Roll during the Waterfall Regime: The
  Small Coupling Window for SUSY Hybrid Inflation}},  {\em Phys.Rev.} {\bf D86}
  (2012) 023525, [\href{http://xxx.lanl.gov/abs/1204.3540}{{\tt
  arXiv:1204.3540}}].

\bibitem{Clesse:2013jra}
S.~Clesse, B.~Garbrecht, and Y.~Zhu, {\it {Non-Gaussianities and Curvature
  Perturbations from Hybrid Inflation}},
  \href{http://xxx.lanl.gov/abs/1304.7042}{{\tt arXiv:1304.7042}}.

\bibitem{Kinney:1997hm}
W.~H. Kinney and A.~Riotto, {\it {Dynamical supersymmetric inflation}},  {\em
  Astropart.Phys.} {\bf 10} (1999) 387--395,
  [\href{http://xxx.lanl.gov/abs/hep-ph/9704388}{{\tt hep-ph/9704388}}].

\bibitem{Kinney:1998dv}
W.~H. Kinney and A.~Riotto, {\it {A Signature of inflation from dynamical
  supersymmetry breaking}},  {\em Phys.Lett.} {\bf B435} (1998) 272--276,
  [\href{http://xxx.lanl.gov/abs/hep-ph/9802443}{{\tt hep-ph/9802443}}].

\bibitem{Bezrukov:2011mv}
F.~Bezrukov, P.~Channuie, J.~J. Joergensen, and F.~Sannino, {\it {Composite
  Inflation Setup and Glueball Inflation}},  {\em Phys.Rev.} {\bf D86} (2012)
  063513, [\href{http://xxx.lanl.gov/abs/1112.4054}{{\tt arXiv:1112.4054}}].

\bibitem{Barrow:1995xb}
J.~D. Barrow and P.~Parsons, {\it {Inflationary models with logarithmic
  potentials}},  {\em Phys.Rev.} {\bf D52} (1995) 5576--5587,
  [\href{http://xxx.lanl.gov/abs/astro-ph/9506049}{{\tt astro-ph/9506049}}].

\end{thebibliography}\endgroup

\end{document}